\documentclass[10pt,sigconf,authorversion]{acmart}
\pdfoutput=1

\usepackage{amsmath}
\usepackage{amsfonts}
\usepackage{amssymb}
\usepackage{multirow}
\usepackage{enumitem}
\usepackage{listings}
\usepackage{booktabs} %
\usepackage{xspace}
\usepackage{alltt}
\usepackage{xcolor}
\usepackage{microtype}
\usepackage{mathtools}
\usepackage{subcaption}
\usepackage{balance}
\usepackage{tikz}
\usepackage[noend]{algpseudocode}
\usepackage{algorithmicx,algorithm}
\usepackage{comment}
\usepackage{microtype}
\usepackage{pgffor}
\usepackage{mwe}
\usepackage{todo}

\newlength{\listingindent}                %
\usepackage[LGRgreek]{mathastext}

\hypersetup{
  colorlinks=false,
  linkcolor=darkred,
  citecolor=darkgreen,
  urlcolor=darkblue
}

\usepackage[noabbrev,capitalise]{cleveref}

\definecolor{light-gray}{gray}{0.95}
\definecolor{mid-gray}{gray}{0.85}
\definecolor{darkred}{rgb}{0.7,0.25,0.25}
\definecolor{darkgreen}{rgb}{0.15,0.55,0.15}
\definecolor{darkblue}{rgb}{0.1,0.1,0.5}
\definecolor{blue}{rgb}{0.19,0.58,1}

\newcommand{\grey}[1]{\textcolor{mid-gray}{#1}}
\newcommand{\red}[1]{\textcolor{red}{#1}}

\newcommand{\eat}[1]{}
\newcommand{\ititle}[1]{\vspace{2pt}\noindent\emph{#1}}
\newcommand{\stitle}[1]{\vspace{2pt}\noindent\textbf{#1}}

\lstdefinestyle{SQLStyle}{
  language=SQL,
  showspaces=false,
  basicstyle=\ttfamily\scriptsize,
  commentstyle=\color{gray},
  mathescape=true,
  numbers=none,
  escapeinside={^}{^},
  captionpos=b,
  float=tp,
  floatplacement=tbp,
  belowskip=-0.05em,
   mathescape=false
}

\usepackage[utf8]{inputenc}

\newcommand{\prob}[0]{Query 2.0 Debugging Problem\xspace}
\newcommand{\sys}[0]{\textsf{Rain}\xspace}

\newcommand{\auc}[0]{$AUC_{CR}$}

\newcommand{\loss}[0]{\textsf{Loss}\xspace}
\newcommand{\infloss}[0]{\textsf{InfLoss}\xspace}
\newcommand{\naive}[0]{\textsf{TwoStep}\xspace}
\newcommand{\holistic}[0]{\textsf{Holistic}\xspace}

\newcommand{\qrels}{\textit{queried relations}\xspace}
\newcommand{\qrecs}{\textit{queried records}\xspace}
\newcommand{\qdat}{\textit{queried data}\xspace}
\newcommand{\qdats}{\textit{queried dataset}\xspace}

\newcommand{\trecs}{\textit{training records}\xspace}
\newcommand{\ctrecs}{\textit{Training records}\xspace}

\newcommand{\trec}{\textit{training record}\xspace}

\newcommand{\norm}[1]{\left\lVert#1\right\rVert}
\newcommand*\circled[1]{\tikz[baseline=(char.base)]{
            \node[shape=circle,draw,inner sep=0.5pt] (char) {#1};}}

\DeclareMathOperator*{\argmin}{arg\,min}

\newcommand{\qfunc}{$q(\pmb{\theta}_{\epsilon}^*)$\xspace}

\newcommand{\apptechport}{appendix\xspace}

\AtBeginDocument{%
  \providecommand\BibTeX{{%
    \normalfont B\kern-0.5em{\scshape i\kern-0.25em b}\kern-0.8em\TeX}}}

\copyrightyear{2020}
\acmYear{2020}
\setcopyright{acmlicensed}
\acmConference[SIGMOD'20]{Proceedings of the 2020 ACM SIGMOD International Conference on Management of Data}{June 14--19, 2020}{Portland, OR, USA}
\acmBooktitle{Proceedings of the 2020 ACM SIGMOD International Conference on Management of Data (SIGMOD'20), June 14--19, 2020, Portland, OR, USA}
\acmPrice{15.00}
\acmDOI{10.1145/3318464.3389696}
\acmISBN{978-1-4503-6735-6/20/06}
\begin{document}

\title[Complaint-driven Training Data Debugging for Query 2.0]{Complaint-driven Training Data Debugging \\ for~Query~2.0}

\author{Weiyuan Wu}
\affiliation{%
  \institution{Simon Fraser University}
  \streetaddress{8888 University Dr.}
  \city{Burnaby}
  \state{BC}
  \country{Canada}
  \postcode{V5A 1S6}
}
\email{youngw@sfu.ca}

\author{Lampros Flokas}
\affiliation{%
  \institution{Columbia University}
  \streetaddress{116th St & Broadway}
  \city{New York}
  \state{NY}
  \postcode{10027}
}
\email{lamflokas@cs.columbia.edu}

\author{Eugene Wu}
\affiliation{%
  \institution{Columbia University}
  \streetaddress{116th St & Broadway}
  \city{New York}
  \state{NY}
  \postcode{10027}
}
\email{ewu@cs.columbia.edu}

\author{Jiannan Wang}
\affiliation{%
  \institution{Simon Fraser University}
  \streetaddress{8888 University Dr.}
  \city{Burnaby}
  \state{BC}
  \country{Canada}
  \postcode{V5A 1S6}
}
\email{jnwang@sfu.ca}

\renewcommand{\shortauthors}{Wu, et al.}

\begin{abstract}
  
As the need for machine learning (ML) increases rapidly across all industry sectors, there is a significant interest among commercial database providers to support ``Query 2.0'', which integrates model inference into SQL queries. Debugging Query 2.0 is very challenging since an unexpected query result may be caused by the bugs in training data (e.g., wrong labels, corrupted features). In response, we propose \sys, a complaint-driven training data debugging system. \sys allows users to specify complaints over the query's intermediate or final output, and aims to return a minimum set of training examples so that if they were removed, the complaints would be resolved. To the best of our knowledge, we are the first to study this problem. A naive solution requires retraining an exponential number of ML models. We propose two novel heuristic approaches based on influence functions which both require linear retraining steps.  We provide an in-depth analytical and empirical analysis of the two approaches and conduct extensive experiments to evaluate their effectiveness using four real-world datasets.  Results show that \sys achieves the highest recall@k among all the baselines while still returns results interactively.  
\end{abstract}

\maketitle

\section{Introduction}

Database researchers have long advocated the value of integrating model inference within the DBMS: data used for model inference is already in the DBMS, it brings the code (models) to the data, and it provides a familiar relational user interface. Early libraries such as MADLib~\cite{madlib} provide this functionality by leveraging user-defined functions and type extensions in the DBMS.  The recent and tremendous success of ML in recommendation, ranking, predictions, and structured extraction over the past decade have led commercial data management systems~\cite{sqlflow,Agrawal2019CloudyWH,madlib,bigqueryml} to increasingly providing first-class support for in-DBMS inference: Google's BigQuery ML~\cite{bigqueryml} integrates native TensorFlow support, and SQLServer supports ONNX~\cite{onnx} models.  These developments point towards mainstream adoption of this new querying paradigm that we call Query 2.0\footnote{In analogy to Machine Learning as ``Software 2.0''~\cite{dawnblog,karpathy,overton}}.

\begin{figure}
    \centering
    \includegraphics[width=\columnwidth]{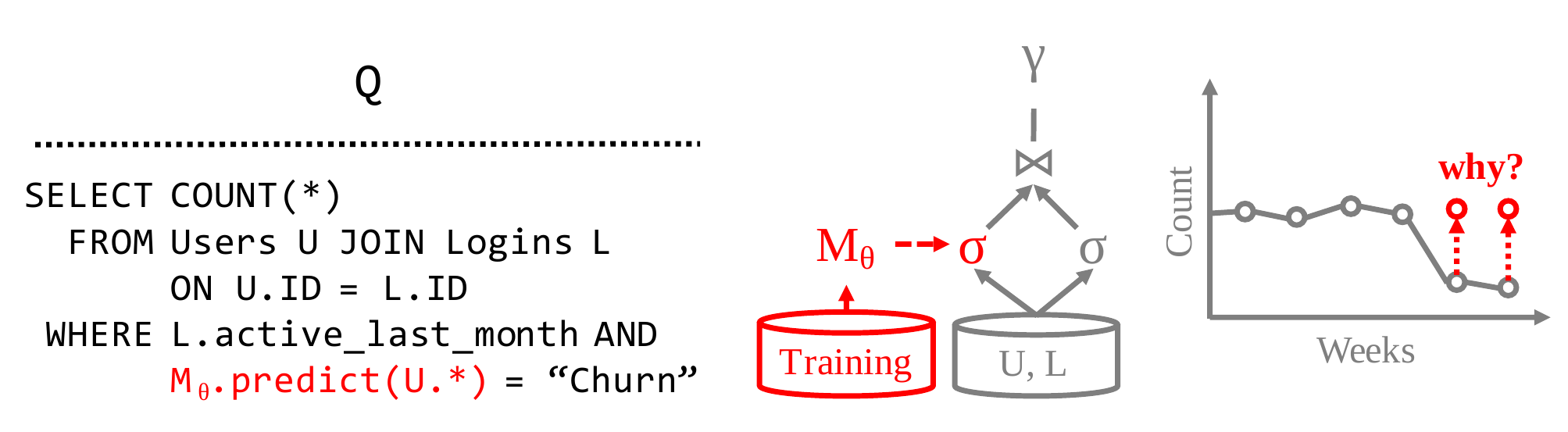}
    \caption{CompanyX cohort query, its workflow, and output visualization where the user specifies surprising output values.  
             Training and model inference steps in \red{red}. }
    \label{fig:intro-qplan}
\end{figure}

Many companies already leverage Query 2.0 in their core business.  CompanyX\footnote{Name anonymized.} customers can define user cohorts using traditional and model-based predicates (details in \Cref{sec:usecase}).  For example, \Cref{fig:intro-qplan} finds and counts the number of active users in the previous month (\texttt{active\_last\_month}) that are likely to churn ($M_{\pmb{\theta}}$\texttt{.predict()}).   The latter predicate uses the model $M_{\pmb{\theta}}$ to estimate whether the user will churn.  Cohorts are used for email campaigns, downstream analyses, and client monitoring.  In fact, 100\% of the company's user segmentation logic are performed within the DBMS.  Beyond CompanyX, both industry~\cite{logicblox,sqlflow} and research~\cite{subjectivedb,mlbase,systemml,lu2018accelerating,Jankov2019DeclarativeRC} are advocating for Query 2.0.

Unfortunately, Query 2.0 is considerably more challenging to debug than traditional relational queries because the results depend on not only the \qdat~\footnote{In machine learning literature \qdat is sometimes called inference data or serving data.} (e.g., \texttt{U,L}), but also the {\it training data} that are used to fit the predictive models used in the query.  Training data are a major factor in determining a model's accuracy, and when a model makes incorrect predictions, it is challenging to even identify the erroneous \trecs~\cite{dawnblog}.  Thus, \textbf{even if the query and \qdat are correct, errors in the training data can cause incorrect query results.}

As one example, CompanyX tracks users on e-commerce websites and scrapes the pages for data to estimate user retention.  They regularly retrain their model $M_{\pmb{\theta}}$. However, systematic errors, such as changing the name of a product category or adding a new check-out step, can cause $M_{\pmb{\theta}}$ to suddenly underestimate user churn likelihoods.  Customers will see a surprising cohort size drop in the monitoring chart  (\Cref{fig:intro-qplan}) and {\it complain}\footnote{Perhaps angrily.}.   Despite assertions and error checking in their workflow systems, CompanyX engineers still spend considerable time to find the training errors.  Ideally, a debugging system can help them quickly identify examples of the \trecs that were responsible for the customer complaint.

Query debugging is not new, and there are existing explanation and debugging approaches for relational queries or machine learning models.  SQL explanation~\cite{scorpion,tiresias,roy,bailis} uses user complaints of query results to identify \qrecs  or predicates, and can fix the complaint through intervention (deleting those records).  However in the context of Query 2.0, these methods would only identify errors in the \qdat (e.g., \texttt{U,L} in \Cref{fig:intro-qplan}), rather than in $M_{\pmb{\theta}}$'s training data.

On the other hand, case-based ML explanation algorithms~\cite{fisher,duti} use labeled mispredictions to identify training points that, if removed, would fix the mispredictions.  This is akin to specifying complaints over the intermediate outputs of the query (specifically, the outputs of the $M_{\pmb{\theta}}\texttt{.predict()}$ predicate).   Unfortunately, finding and labeling the mispredictions can take considerable effort.  Further, users such as CompanyX's customers only see the final chart.

To this end, we present \sys, a system to facilitate {\it complaint-driven data debugging for Query 2.0}. Given that the query and the \qdat are correct, \sys detects label errors in the training data. Users simply report errors in intermediate or final query results as {\it complaints}, which specify whether an output value should be higher, lower, or equal to an alternative value, or if an output tuple should not exist.  \sys  returns a subset of \trecs that, if the models are retrained without those records, would most likely address the complaints.  This problem combines aspects of integer programming, bi-level optimization, and combinatorial search over all subsets of \trecs---each is challenging in isolation, and together poses novel challenges faced neither by SQL nor ML explanation approaches.

To address these challenges, this paper describes and evaluates two techniques that bring together SQL and ML explanation techniques.  Both iteratively identify \trecs that, if removed, are most likely to fix user complaints.  \naive uses a two-step approach: it models the output of model inference as a view, and uses an existing SQL explanation method to identify records in the view that are responsible for user complaints.  Those records are marked as mispredictions and then used as input to a case-based ML explanation algorithm.  This method works well when SQL explanation can correctly identify the model mispredictions (or the user directly labels them).  However, it can work poorly when there are many satisfying solutions for the complaints in the SQL explanation step; we call this complaint \textbf{ambiguity}, and provide theoretical intuition and empirical evidence that it causes \naive to incorrectly identify erroneous training points.    

To address these limitations, the \holistic approach models the entire pipeline---the query plan, model training, and user complaints---as a single optimization problem.   This directly measures the effect of each \trec on the user complaints, without needing to guess mispredictions correctly.  We also provide theoretical intuition for when and why \holistic should be more effective than existing approaches that do not account for SQL queries nor user complaints.    To summarize, our contributions include:
\begin{itemize}[leftmargin=*, topsep=0mm, itemsep=0mm]
    \item A formalization of complaint-driven training data debugging for Query 2.0, along with motivating use cases. 
    
    \item The design and implementation of \sys, a solution framework that integrates elements of existing SQL and case-based ML explanation algorithms. \sys supports SPJA queries that use differentiable models such as linear models and neural networks.  
    
    \item \naive, which sequentially combines existing ILP-based SQL explanation approaches and ML influence analysis techniques.  Our theoretical analysis shows that \naive is sensitive to the ILP's solution space, and we empirically validate this in the experiments.
    
    \item \holistic, which combine user complaints, the query, and model training in a single problem that avoids the ambiguity issues in \naive.
    
    \item An extensive evaluation of \sys against existing explanation baselines.  We use a range of datasets containing relational, textual, and image data.  We validate our theoretical analyses: \naive is susceptible to performance degradation when ambiguity is high, and that approaches that do not use complaints are misled when there are considerable systematic training set errors.  We find that \holistic's accuracy dominates the other approaches---including settings where alternative approaches cannot find {\it any} erroneous \trecs---and iteratively returns \trecs in interactive time. %
\end{itemize}

\vspace{1em}

The remainder of this paper is organized as follows.  Section~\ref{sec:usecase} presents example use cases. Section~\ref{sec:probdef} formally defines the Query 2.0 debugging problem and discusses the computational challenge. We propose two novel approaches to solve the problem. Section~\ref{sec:background} presents their main ideas and Section~\ref{sec:sys} describes the overall system architecture and details. Experimental results are presented in Section~\ref{sec:exp}, followed by related work (Section~\ref{sec:rw}) and conclusion (Section~\ref{sec:conclusion}). 
\section{Use case}\label{sec:usecase}

\sys helps identify systematic errors in training datasets that cause model mispredictions that, later on, introduce errors in downstream analyses.  These errors can come from errors in manual labeling, procedural labelling~\cite{Ratner2017SnorkelFT}, or automated data generation processes~\cite{Colyer2019PuttingML}.  This section presents illustrative use cases that can benefit from complaint-based debugging.

\subsection{Example Use Cases}

\stitle{E-commerce Marketing: }
CompanyX specializes in retail marketing. One of its core services manages email marketing campaigns for its customers. 10-20 ML models predict different user characteristics (e.g., will a user churn, product affinity).  
Customers see model predictions as attributes in views, and can use them, or raw user profile data, to create predicates to define user cohorts that are used in email campaigns and tracked over time (e.g., \Cref{fig:intro-qplan}).  

For development simplicity, CompanyX uses Google BigQuery for model training, cohort creation, and monitoring; the queries are instrumented at different points to be visualized or monitored purposes.  For example, customer-facing metrics dashboards visualize user cohort sizes over time, and customers can set alerts for when the cohort's size drops or increases very rapidly, or exceeds some threshold.    

CompanyX collects training data by scraping their customers' e-commerce websites.  However, changes to the website---such as adding a new check-out step, or changing a product category---can introduce systematic training errors that degrade the re-trained models, and ultimately trigger customer monitoring alerts and lead to customer questions.  Pipeline monitoring is not enough to pinpoint the relevant training records, and their engineers are challenged to find and characterize the culprit training records.

\stitle{Entity Resolution:}
A data scientist scrapes and trains a boolean classification model to use for entity resolution (e.g., given two business records, the model can determine whether they refer to the same real-world entity).  However, when she uses it as the join condition over two business listings ($Listings_1\bowtie_{M_\theta.predict(*) = 1} Listings_2$), she finds that the dining business categories have zero matches.  She is sure that should not be the case and wants to understand why the classifier is incorrect. 

\stitle{Image Analysis:}
An engineer collects an image dataset and wants to train a hot-dog classifier. To create labels, she decides to use distant supervision~\cite{Zeng2015DistantSF}, and writes a programmatic labelling function.  She uses the classifier to label a hot-dog, and a non-hot-dog dataset, equi-joins the two datasets on the predicted label, and plots the resulting count.  She is surprised that there are many join results when there should not have been any, and complains that the count should be~$0$.

\subsection{Desired Criteria}
Ultimately, manual pipeline and training data analysis is time-consuming and difficult.  The above use cases highlight desired criteria that motivate complaint-driven data debugging for Query 2.0.  First, is the ability to express data errors at different points in the query pipeline.  This is important because users may only have access to specific output or intermediate results, or only have the time/expertise to comment on aggregated query results rather than manually label individual model predictions.  

For example in \Cref{fig:intro-qplan}, the user may specify errors in the final query result, but an ML engineer may collect a sample of the model predictions in the output of $M_{\pmb{\theta}}$\texttt{.predict(U.*)} and identify errors there as well.  Similarly, another customer may find errors in a separate query that uses $M_{\pmb{\theta}}$.  The system should be able to use all pieces of information to identify the erroneous training records.

Second, users want to describe {\it how} data are incorrect and what their expectations of what correct data should look like.  This requires a flexible complaint specification, rather than labeling mispredictions.  For instance, when viewing \Cref{fig:intro-qplan}'s chart, the customer may state that the right-most erroneous points should be the value of the red points, or perhaps that they should not exist at all.

\section{Problem Definition}\label{sec:probdef}
This section formalizes the Query 2.0 debugging problem that we will study in this work. Also, we are going to discuss the computational hurdles in solving the problem efficiently.  

\subsection{Defining Query 2.0}

Query 2.0 consists of a SQL query that embeds one or more ML models. This work focuses on Select-Project-Join-Aggregate (SPJA) SQL queries which have zero or more inner joins and embed a single classification ML model. In contrast to classification models, which assign probabilities to each class, regression does not always have probabilistic interpretations to the outputs.  Supporting those models, the full SQL standard, and multiple models is left to future work.  Note however, that the query can use the same model in multiple expressions.  

Specifically, we support SP, SPJ, SPJA queries, such as:
\lstset{style=SQLStyle, basicstyle=\ttfamily\small}
\begin{lstlisting}
  SELECT agg(^$\cdot$^), ^$\cdots$^ FROM ^$R_1, R_2 \cdots R_n$^ 
  WHERE ^$C_1$^ AND ^$\cdots$^ AND ^$C_m$^
  GROUP BY ^$G_1, G_2 \cdots G_k$^ 
\end{lstlisting}

\noindent where agg can be COUNT, SUM, or AVG, and each $C_i$ is either a filter condition or a join condition. Conjunctive and disjunctive predicates are supported as well.  A model $M$ can appear in the SELECTION, WHERE, or GROUP BY clause (Table~\ref{tab:querytwo-example}).

{\renewcommand{\arraystretch}{1.5}
\begin{table}[tb]
    \centering \scriptsize
    \begin{tabular}{c l}
       $Q_1:$  &  \textsf{SELECT AVG(M.predict(R)) FROM~R} \\
       $Q_2:$  & \textsf{SELECT COUNT(*) FROM R WHERE M.predict($R$)} \\ 
       $Q_3:$  & \textsf{SELECT * FROM $R_1$ and $R_2$ WHERE M.predict($R_1$) = M.predict($R_2$)} \\ 
       $Q_4:$  & \textsf{SELECT * FROM $R_1$ and $R_2$ WHERE M.predict($R_1$+$R_2$)} \\ 
       $Q_5:$  & \textsf{SELECT COUNT(*) FROM $R$ GROUP BY M.predict($R$)} \\ %
    \end{tabular}
    \vspace{1em}
    \caption{Query 2.0 examples. Model prediction can be embedded in an aggregation/projections ($Q_1$), filters ($Q_2$), join conditions ($Q_3$, $Q_4$), and group bys ($Q_5$).}
    \label{tab:querytwo-example}\vspace{-2em}
\end{table}
}

\begin{itemize}[leftmargin=*]\itemsep0.25em
    \item SELECTION: model prediction appears in an aggregation function, denoted by agg(M.predict). For example, if $M$ estimates customer salary, then $Q_1$ returns the average estimated salary.  
    \item WHERE: model prediction appears in a filter condition or a join condition. For example, if $M$ predicts if a customer will churn or not, then $Q_2$ returns the number of customers that may churn.  If $M$ extracts the user type, then $Q_3$ returns pairs of customers from two datasets that are the same user type (note that $Q_3$ is a SPJ query).  Finally, if $M$ estimates if two records are the same entity, then $Q_4$ finds pairs of records that are the same entity.
    \item GROUP BY: model prediction appears in the GROUP BY clause. For example, if $M$  predicts the sentiment of a customer comment, then $Q_5$ returns the number of comments for each sentiment class (positive, neutral, or negative).  
\end{itemize}

\noindent Let $T$ be the training set for model $M$ and $\mathcal{D} = \{R_1, R_2, \cdots, R_n\}$ denotes a database containing \qrels. The trained model $M$ will make predictions using data from $\mathcal{D}$. Given a query $Q$, we denote its output result over $\mathcal{D}$ by $Q(\mathcal{D};M(T))$. If the context is clear, the notation is simplified as $Q(\mathcal{D};T)$.

\subsection{Complaint Models} 
A user may have a {\it complaint} about the query output $Q(\mathcal{D};T)$. We consider two types: {\it value complaints} and {\it tuple complaints}.

A value complaint lets the user ask why an output attribute value in $Q(\mathcal{D};T)$ is not equal to (larger than or smaller than) another value. In \Cref{fig:intro-qplan}, the user can specify why the two right-most low points in the visualization are not equal to (or larger than) the corresponding red points.   %

A tuple complaint lets the user ask why an output tuple in $Q(\mathcal{D};T)$ appears in the output. This can be because a tuple should have been filtered by a predicate that compares with a model prediction, or because an aggregated group exists when it should not.   For example, the user may ask why a pair of loyal customers are in the join output of $Q_3$ in Table~\ref{tab:querytwo-example}. 

Definition~\ref{d:complaint} presents a formal definition of complaints.

\begin{definition}[Complaint]\label{d:complaint}
    A complaint $c(t)$ is expressed as a boolean constraint over a tuple $t$ in the output relation $Q(\mathcal{D};T)$. The complaint can take two forms. The first is a {\it Value Complaint} over an attribute value $t[a]$, where $op\in\{=,\le,\ge\}$ and $v$ may take any value in the attribute's domain (if $t[a]$ is discrete, then $\le,\ge$ do not apply):
    \begin{equation}
    c_{\textsf{value}}\big(t, Q(\mathcal{D};T)\big)=
    \begin{cases}
      True, & \text{if}\ \textsf{ t[a] op v} \\
      False, & \text{otherwise}
    \end{cases}
  \end{equation}
    The second is a {\it Tuple Complaint} over the tuple $t$ which states that $t$ should not be in the output relation:
    \begin{equation}
    c_{\textsf{tuple}}\big(t, Q(\mathcal{D};T)\big)=
    \begin{cases}
      True, & \text{if}\ \textsf{ $t \notin Q(\mathcal{D};T)$} \\
      False, & \text{otherwise}
    \end{cases}
  \end{equation}
\end{definition}

\stitle{Multiple Complaints:} 
The user may express multiple complaints against the result of $Q(\mathcal{D};T)$ or even against intermediate results of the query.  In addition, if the user executed other queries using the same model $M(T)$, then complaints against those queries may also be used to identify training set errors. For ease of presentation, the text will focus on the single complaint case. However, the proposed approaches support multiple complaints, and we evaluate them in the experiments. 

\subsection{Problem Statement}
Given a Query 2.0 query, there can be several ways to account for a user's complaint $c$ by making changes to the training set $T$. For example, one might modify training examples in $T$, augment it with new training examples or even delete training examples. While all the above interventions make sense in different scenarios, for simplicity in this work we will focus on deletions of training examples from $T$. Given the definitions of the previous subsection, we are ready to define the Query 2.0 debugging problem.
\begin{definition}[\prob]
Given a training dataset $T$, a database $\mathcal{D}$ containing \qrels, a query $Q$, and a complaint $c$, the goal is to identify the minimum set of \trecs such that if they were deleted, the complaint would be resolved:
\begin{align*}
    & \underset{{\Delta \subseteq T}}{\text{minimize}}  & & |\Delta| \hspace{1cm} \\
    & \text{subject to} & & c\big(t, Q(\mathcal{D}; T \setminus \Delta)\big) = True
\end{align*}
\label{p:sqlml}
\end{definition}
\noindent A brute force solution is to enumerate every possible set of deletions, and for each set, to retrain the model, update the query result, and evaluate the complaint. However, this needs to retrain up to $2^{|T|}$ models. The key is to reduce the number of models retrained. In the following, we propose two novel heuristic approaches which both reduce the number from exponential to linear. %

\section{Background and Preliminaries}\label{sec:background}
In this section, we first introduce the concept of {\it influence functions} in ML explanation and then present the main ideas of our approaches.

\subsection{Influence Functions}

Influence functions provide a powerful way to estimate how the model parameters change by adding/deleting/updating a training point {\it without retraining the model}. For example, suppose one wants to know to delete which training point will lead to the best model parameters (i.e., the minimum model loss). A brute-force approach needs to enumerate every training point and retrain $|T|$ models. As will be shown below, influence functions do not involve any retraining.

Let a training set $T$ include $n$ pairs of feature vectors and labels $\pmb{z}_i = (\pmb{x}_i,y_i)$. An ML model parametrized by $\pmb{\theta}$ is trained with the following loss function ($\ell$ is the \trec loss):
\begin{equation*}
    L(\pmb{\theta}) = \frac{1}{n}\sum_{i=1}^n \ell( \pmb{z}_i, \pmb{\theta}) 
\end{equation*}
A strongly convex function $L$ has a unique solution\footnote{Influence functions have been extended to non-convex models, and we evaluate a neural network model in our \apptechport.}:
\begin{equation*}
    \pmb{\theta}^* = \argmin_{\pmb{\theta}} L(\pmb{\theta})
\end{equation*}
Adding a new training sample $\pmb{z}$ with weight $\epsilon$ to the training loss  leads to new set of optimal parameters $\pmb{\theta}_{\epsilon}^*$:
\begin{equation}
    \label{eq:optobj}
    \pmb{\theta}_{\epsilon}^* = \argmin_{\pmb{\theta}} \; \{ L(\pmb{\theta}) + \epsilon \cdot \ell( \pmb{z}, \pmb{\theta}) \}
\end{equation}
In general, we are interested in a closed-form expression of $\pmb{\theta}_{\epsilon}^*$ for $\epsilon=\pm \frac{1}{n}$, which can estimate the effect of adding or removing a training point without retraining. Unfortunately, such a closed-form expression does not generally exist. The Influence Function approach quantifies the case when $\epsilon\approx 0$. By the first order optimality condition, since $\pmb{\theta}_{\epsilon}^*$ minimizes the objective of \cref{eq:optobj}
\begin{equation*}
     \nabla_{\pmb{\theta}} L(\pmb{\theta}_{\epsilon}^*) + \epsilon \cdot \nabla_{\pmb{\theta}} \ell( \pmb{z}, \pmb{\theta}_{\epsilon}^*) = \pmb{0}
\end{equation*}
The derivative of the equation above with respect to $\epsilon$, taking into account that $\pmb{\theta}_{\epsilon}^*$ is a function of $\epsilon$, yields
\begin{equation*}
     \nabla_{\pmb{\theta}}^2 L(\pmb{\theta}_{\epsilon}^*) \frac{d \pmb{\theta}_{\epsilon}^*}{d\epsilon} + \epsilon \cdot \nabla_{\pmb{\theta}}^2 \ell( \pmb{z}, \pmb{\theta}_{\epsilon}^*)\frac{d \pmb{\theta}_{\epsilon}^*}{d\epsilon} + \nabla_{\pmb{\theta}} \ell( \pmb{z}, \pmb{\theta}_{\epsilon}^*) = \pmb{0}
\end{equation*}
Recent work has shown that using the derivative where $\epsilon=0$ is a good approximation of the change in model parameters for  ($\pmb{\theta}_{-1/n}^*$ or $\pmb{\theta}_{1/n}^*$)~\cite{swissarmy,influencegroup}. Substituting $H_{\pmb{\theta}^*}=\nabla^2_{\theta} L(\pmb{\theta^*})$, where $H_{\pmb{\theta}^*}$ is the Hessian of the loss function $L(\pmb{\theta})$, and simple algebra derives the following when $\epsilon=0$:
\begin{equation*}
    \left. \frac{d \pmb{\theta}_{\epsilon}^*}{d\epsilon}\right|_{\substack{\epsilon=0}} = - H_{\pmb{\theta}^*}^{-1} \nabla_{\pmb{\theta}} \ell(\pmb{z},\pmb{\theta}^*)
\end{equation*}
Note $\epsilon$ is dropped from $\pmb{\theta}_{\epsilon}^*$ because it is set to $0$.

In our problem, we wish to approximate the effect of training points on user complaints.  To do so, we will construct a differentiable function \qfunc that represents user complaints by encoding the SQL query, ML model, and user complaints.  \Cref{sec:solution} and \Cref{sec:howtocomp} describe two encoding procedures. Given \qfunc, the effect of a training point is straightforward using the chain rule:
\begin{equation}\label{eq:inffunc}
    \left. \frac{d q(\pmb{\theta}_{\epsilon}^*) }{d\epsilon}\right|_{\substack{\epsilon=0}} = - \nabla_{\pmb{\theta}} q(\pmb{\theta}^*) \ H_{\pmb{\theta}^*}^{-1} \nabla_{\pmb{\theta}} \ell( \pmb{z}, \pmb{\theta}^*)
\end{equation}

Computing $H_{\pmb{\theta}^*}^{-1} \nabla_{\pmb{\theta}} \ell( \pmb{z},\pmb{\theta}^*)$ can become a significant bottleneck as a naive implementation requires $\mathcal{O}(d^2)$ space and $\mathcal{O}(d^3)$ time. The authors of \cite{influence} leverage prior work \cite{hessianfree} so that the total time and space complexity scales linearly in the dimension $d$. The calculation is posed as a linear system of equations, and approximately solved using the conjugate gradient algorithm. Instead of inverting the Hessian, the conjugate gradient relies on Hessian vector products that can be efficiently computed via backpropagation.

\begin{figure*}[!t]
    \centering
    \includegraphics[width=0.95\textwidth]{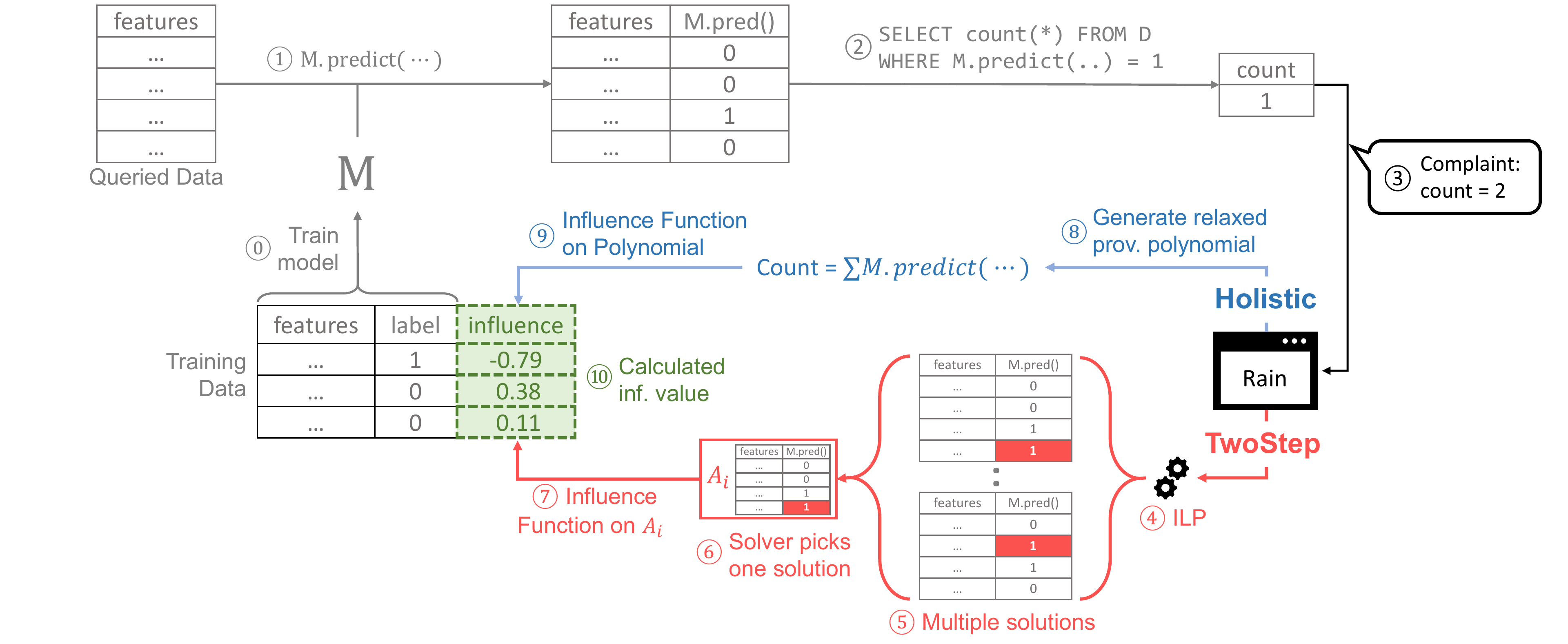}
    \caption{\sys architecture.}
    \label{fig:arch}
\end{figure*}

\subsection{Main Ideas of Our Approaches}\label{sec:mainideas}

Unfortunately, influence functions cannot be directly applied to solve the \prob since we need to calculate the impact of deletions of training points on a Query 2.0 query output, and SQL queries are not naturally differentiable. 

We use two novel ideas to address this challenge.   \naive first calculates the impact of deletions of training points on model parameters and then calculates the impact of the changes of model parameters on the query result.   \holistic encodes a Query 2.0 query (both SQL and model parts) into a single differentiable function and then directly calculates the impact of deletions of training points on the query result. 

We developed \sys, a Query 2.0 debugging system that implements \naive and \holistic approaches. The next section will describe the system details.

\subsection{Why are Complaints Important?}
Influence analysis can already be used to detect training errors based on the model loss without the need for complaints~\cite{influence}. The high sensitivity of the loss on a \trec can be interpreted as a corrupted \trec. Thus why are complaints important?

The main reason is that models can overfit to systematic training errors, and cause loss-based rankings to rank such errors arbitrarily low in terms of loss sensitivity. For example, changes in the checkout code might cause CompanyX to not log successful transactions for some customers; the trained model may then assume that similar customers will churn. 

In contrast, SQL queries and complaints provide a vocabulary to specify systematic errors. This vocabulary generalizes existing work that labels individual mispredictions~\cite{duti,fisher} or specifies undesirable prediction output distributions~\cite{DBLP:conf/icml/AgarwalBD0W18}. Our experiments show that even a single aggregation complaint can identify systematic training errors more effectively than hundreds of labeled mispredictions.

\section{The \sys System}\label{sec:sys}

This section describes the overall architecture of \sys, which uses either \naive (\Cref{sec:howtocomp}) or \holistic (\cref{sec:solution}) to solve the \prob.

\subsection{Architecture Overview}

\sys (\Cref{fig:arch}) consists of a query processor that supports training machine learning models (step \circled{0}), performing model inference (step \circled{1}) and executing SQL queries based on the model outputs (step \circled{2}).  The user examines the output or intermediate result set of a query $Q$, and specifies a set of complaints $C$ (step \circled{3} complaints that the result should be 2 instead of 1). The optimizer uses a simple heuristic to choose between the two methods. As we will discuss in \Cref{sec:howtocomp},  \naive is preferable when there is a unique way to fix the querying set predictions that resolves $C$. For all other cases, \holistic is used.  

\naive turns the complaints into a discrete ILP problem and uses an off-the-shelf solver (step \circled{4}) to label a subset of the model inferences with their (estimated) correct predictions. If multiple satisfying solutions exist (step \circled{5}), solvers will opaquely output one of the solutions dependent on the specific implementation (step \circled{6}). The solution is encoded as an influence function to estimate how each \trec ``fixes'' the mispredictions (step \circled{7}).  \holistic encodes the query and model training as a single relaxed provenance polynomial (step \circled{8}) that serves as an influence function to estimate how much each \trec ``fixes'' the complaints (step \circled{9}). For both approaches, \sys finds the top $k$ \trecs by influence (step \circled{10}). 

Both approaches will first rerun $Q$ (step \circled{2}) in a ``debug mode'' to generate fine-grained lineage metadata that encodes the optimization problem. \sys then runs a train-rank-fix scheme, where each iteration (re)trains the model (step \circled{0}), reruns the query (step \circled{1}-\circled{2}), finds and deletes the top \trecs by influence (step \circled{4}-\circled{10}), and repeats.  The result is a sequence of \trecs that comprise the output explanation $\Delta$. Assuming each iteration selects the top-$k$ \trecs, then \sys executes $\frac{|\Delta|}{k}$ iterations.

\subsection{\naive Approach}\label{sec:howtocomp}
Query 2.0 plans consist of relational pipelines and model inference.  Since there are existing solutions to address each in isolation, the naive approach combines them into a \naive solution.  This section describes this approach, and provides intuition on its strengths and limitations.  

\subsubsection{Approach Details}
We now describe the SQL and Influence Analysis steps of \naive.  

\stitle{SQL Step: } 
At a high level, \naive replaces each model inference expression, such as $M_\theta$\texttt{.predict(U.*)} in \Cref{fig:intro-qplan}, with a materialized {\it prediction view} containing the input's primary key and the prediction result.  Let $V_M$ be the prediction view for model $M$, and $D_T$ be the database containing the views.  $Q(D;T)$ can be rewritten as $Q_m(D_T)$ to instead refer to the model views rather than perform model inference directly.    For instance, the query in \Cref{fig:intro-qplan} would be rewritten as follows:
\lstset{style=SQLStyle, basicstyle=\ttfamily\small}
\begin{lstlisting}
 SELECT COUNT(*) FROM Users U, Logins L, ^$\mathbf{V_m}$^
  WHERE U.ID = L.ID AND U.ID = ^$V_m$^.ID 
        AND L.active_last_month
        AND ^$\mathbf{V_m}$.\textbf{prediction = "Churn"}^
\end{lstlisting}
\vspace{1em}
\noindent We build on Tiresias~\cite{tiresias}, which takes as input a set of complaints, along with attributes in \qrels that can be changed to fix those complaints.  It translates the complaints and query into an ILP, where marked attributes are replaced with free variables that the solver (e.g., Gurobi~\cite{gurobi}, CPLEX~\cite{ilog2014cplex}) assigns.  We mark the predicted attribute in the prediction views, and the objective minimizes the number of prediction changes.

The translation to an ILP relies on database provenance concepts. Each potential output of $Q_m$ defines a function over the prediction view that evaluates to $1$ if the tuple exists in the query output for the given prediction view or $0$ if not. In addition, each aggregation output value of $Q_m$ defines a function over the prediction view that returns the aggregation value. Prior provenance work~\cite{semirings,aggprov} shows how to translate the supported queries into symbolic representations of these functions also known as provenance polynomials, which Tiresias encodes as ILP constraints.  

We illustrate the reduction for the example in \cref{fig:intro-qplan}:
\begin{example}
Let the query plan for \Cref{fig:intro-qplan} first filter and join $L$ with $U$, and then apply the churn filter before the aggregation. Let $K$ be the number of the remaining rows after the join and filter on $L$, and $\pmb{r} \in \{0,1 \}^K$ be the binary model predictions over these rows.  $r_i=1$ means the user is predicted to churn, and the query result is $\sum_{i=1}^K r_i$.  If the user complains that the query output should be $X$, then the generated ILP is as follows, where $t_i\ne r_i$ means that record $i$ should be labeled as a misprediction:%
    {\begin{align}\label{eq:q-plan-ilp}
    & \underset{\pmb{t} \in \{0,1 \}^K}{\text{minimize}}  & & \sum_{i=1}^K | t_i - r_i| \hspace{1cm} \\ \nonumber
    & \text{subject to} & & \sum_{i=1}^K t_i = X 
    \end{align}}%
\end{example}

Rain goes beyond this simple example and supports the queries and complaints described in \cref{sec:probdef}.

\stitle{Influence Analysis Step: }
The previous step assigns each record $\pmb{x}_i$ a (possibly ``corrected'') label $t_i$: $\{\pmb{x}_i, t_i \}_{i=1}^K$.   Let $p_{t_i}(\pmb{x}_i, \pmb{\theta})$ be the probability that model $M_{\pmb{\theta}}$ predicts $\pmb{x}_i$ to be class $t_i$, where $\pmb{\theta}$ is the vector of the ML model parameters. We construct function $q(\pmb{\theta}) = - \sum_{i=1}^K p_{t_i}(\pmb{x}_i, \pmb{\theta})$ that is used as input to an influence analysis framework~\cite{influence,sgdcleansing,duti,fisher}. These frameworks return a ranking of training points that, if removed, are most likely to change the predictions of $\pmb{x}_i$ to $t_i$; this indirectly addresses the user's complaint.   

For example, suppose we use the influence analysis framework of \cite{influence}.  \naive uses \cref{eq:inffunc} to score every \trec. The initially trained model has optimal parameters $\pmb{\theta}^*$. The training loss Hessian $H_{\pmb{\theta}^*}$ and the training loss gradient of each \trec $\nabla_{\pmb{\theta}} \ell(\pmb{z},\pmb{\theta}^*)$ are evaluated at $\pmb{\theta}^*$. The function $q$ constructed by \naive is then substituted to encode the user's complaint. \ctrecs with large positive scores imply that their removal would decrease $q$ the most, implicitly addressing the complaint. \naive ranks these records at the top.

In most settings, the number of records not marked as a misprediction ($t_i=r_i$) is considerably larger than those marked as mispredictions ($t_i\ne r_i$), and encoding all of them slows down the influence analysis step.    In our experiments, we only encode the marked mispredictions into $q(\cdot)$ in \Cref{eq:inffunc}, and empirically find that they result in comparable rankings as when encoding all records.

\subsubsection{Limitations and Analysis}\label{sec:naiveanalysis}
Although \naive is simple, there are several limitations due to the nature of the ILP formulation of the SQL step.  First, the ILP problem can be \textit{ambiguous} and is not guaranteed to identify the correct solution.  Second, \naive depends on the user submitting a correct complaint.  We discuss both limitations in this subsection.

\stitle{Ambiguity: }
The generated ILP may not always have a unique solution. For example, \cref{fig:arch} shows how the ILP of a complaint on a $\texttt{COUNT}$ aggregation can have multiple solutions $A_1, A_2, \dots, A_n$ (step \circled{4}). We call such complaints {\it ambiguous}. Picking a solution $A_i$ in step \circled{5} that makes incorrect prediction fixes can negatively affect the influence step \circled{6}. Intuitively, a complaint with more ILP solutions should lead to worse rankings because, among all solutions that minimize the ILP problem, only a few minimize \Cref{p:sqlml}. We identify two sources of ambiguity.  

The first are aggregations. In \cref{fig:arch}, flipping any single prediction $0$ is a valid and minimal solution, but only one solution is correct. The same argument extends to all the aggregates supported by \sys as all of them are symmetric with respect to their inputs.

The second are join and selection predicates. Consider a join $A\bowtie_{A.a = B.b} B$, where $A.a$ and $B.b$ are both estimated by a model $M$.  If the user specifies that a join result should not exist, then one has to choose between changing $A.a$ or $B.b$. More generally, selection predicates that involve two or more model predictions can also be ambiguous.

Our \apptechport lists specific settings where ambiguity provably causes \naive to rank the true training errors arbitrarily low, thus forbidding us sampling multiple solutions from ILP to avoid bad results for the whole problem. Unfortunately, formally quantifying its effect in the general case is challenging because partially correct solutions $A_i$ can still yield high quality rankings depending on the model and the corrupted \trecs.

Our experiments vary the level of ambiguity and empirically suggest that \naive performs better when the number of solutions of the SQL step is smaller.

\stitle{Complaint Sensitivity:}
The second limitation is due to the discrete formulation of the ILP: identifying correct assignments depends on the correctness of the complaint. For example, if the user selected a slightly incorrect $X$ in \Cref{eq:q-plan-ilp}, the satisfying assignments can be considerably different than the true mispredictions.  Unfortunately, if the user finds surprising points in a visualization, she may have an intuition that the point should be higher or lower, but is unlikely to know its exact correct value.  We see this sensitivity in our experiments.

\subsection{\holistic Approach}\label{sec:solution}
In this section, we present the \holistic approach that addresses many of the limitations of TwoStep. The key insight is to connect \trecs with the user complaints by modeling the query probabilistically and interpreting the confidence of model predictions as probabilities.   This lets us leverage prior work in probabilistic databases~\cite{probdb,probdbsensitivity} to represent Query 2.0 statements as a differentiable function that is amendable to influence analysis. Note that although provenance and influence analysis alone build on prior work, integrating them for the purpose of complaint-driven training data debugging is the key novelty.  

\subsubsection{Relaxation Approach}

As noted above, the symbolic SQL query representations are not naturally differentiable due to discrete inputs (values in the prediction views), and thus are incompatible with an influence analysis framework.   In contrast to \naive, \holistic leverages techniques from probabilistic databases~\cite{probdbsensitivity,probdb} to relax these functions of discrete inputs into continuous variable functions.

Revisiting \cref{eq:q-plan-ilp}, \holistic substitutes the count of churn predictions with the expectation of the count. For example, let $r_i(\pmb{\theta})$ be the boolean churn prediction and $p_i(\pmb{\theta})$ be the churn probability assigned by $M_{\pmb{\theta}}$, \holistic substitutes:
\begin{equation*}
    \sum_{i=1}^K r_i(\pmb{\theta}) \rightarrow \sum_{i=1}^K p_i(\pmb{\theta}).
\end{equation*}

Unfortunately, expectations of provenance polynomials are not always straightforward to compute. Even calculating the expectation of a k-DNF formula is \#P-complete~\cite{probdbsensitivity}. To sidestep the computational difficulty of exact probabilistic relaxation, we propose a tractable alternative under the simplifying assumption that variables and sub-expressions are independent. We first replace discrete predictions in the provenance polynomial with their corresponding probabilities (similar to $r_i(\pmb{\theta})\rightarrow p_i(\pmb{\theta})$ above). We then replace boolean operators (\texttt{AND}, \texttt{OR}, \texttt{NOT}) with continuous alternatives
{\small\begin{equation*}
\begin{aligned}
    x &\texttt{ AND }&& y &&\rightarrow x \cdot y\\
    x &\texttt{ OR }&&  y &&\rightarrow 1 - (1-x) \cdot (1-y)\\
    &\texttt{ NOT }&&  x &&\rightarrow (1-x). \\
\end{aligned}
\end{equation*}}
Observe that the first two formulas above can be mapped to the probability formulas for the \texttt{AND} and \texttt{OR} of two independent random variables. Our relaxation applies this rule even when $x$ and $y$ are complex expressions that share random variables and thus may not be independent. When each variable appears only once in the provenance polynomial as discussed in \cite{probdbsensitivity}, our approach yields the actual expectation.

Our relaxation focuses on tractability. Alternative differentiable relaxations of logical constraints based on probabilistic interpretations are axiomatically principled~\cite{semanticloss} albeit generally intractable. Comparing relaxation approaches is a promising direction for future work.

\subsubsection{Translating complaints to influence functions}
To adapt the above into an influence analysis framework, we translate user complaints over relaxed provenance polynomials into a differentiable function $q(\pmb{\theta})$ that we want to minimize.  We will first assume one equality complaint $t_i[a] = X$ on a single value, and then relax these assumptions to support multiple, more general complaints. 

Let $r_q(\pmb{\theta})$ be the relaxed provenance polynomial for $t_i[a]$.  We adapt it to the complaint by defining $q(\pmb{\theta}) =  \left( r_q(\pmb{\theta}) - X \right)^2$.  Minimizing $q(\pmb{\theta})$ forces $t_i[a]$ to be close to $X$.  Akin to \Cref{sec:howtocomp}, this function is now compatible with modern influence analysis frameworks~\cite{influence,sgdcleansing,duti}.

We support tuple complaints by taking the relaxed tuple polynomial $r_q(\pmb{\theta})$ for tuple $t$, and defining $q(\pmb{\theta}) = (r_q(\pmb{\theta}) - 0)^2$.  Inequality value complaints like $t[a] >= X$ are supported within the train-rank-fix scheme of the system.  While the complaint is false, we model it as an equality complaint; iterations where the inequality is satisfied can ignore the complaint until it is once again violated.  Finally, to support multiple complaints, we sum their $q$ functions.

\section{Experiments}\label{sec:exp}
Our experiments seek to understand the trade-offs of \sys as compared to existing SQL-only and ML-only explanation methods, and to understand when complaint-based data debugging can be effective.  We then study how ambiguity, increasing the number of complaints, and errors in the complaints affect \sys and the baselines.  The majority of our experiments are performed using linear models. Our \apptechport also uses neural network models.

\subsection{Experimental Settings}
We now describe the experimental settings.  We use a range of SPJA queries summarized in \Cref{tab:exp-qs}.

{\renewcommand{\arraystretch}{1.5}
\begin{table}[t]
    \centering \scriptsize
    \begin{tabular}{rl}
       $Q_1$  &  \textsf{SELECT COUNT(*) FROM DBLP WHERE predict(*)='match'} \\
       $Q_2$  & \textsf{SELECT COUNT(*) FROM Enron} \\
               & \textsf{WHERE predict(*)=`spam' AND text LIKE `\%\texttt{\bf word}\%'} \\ 
       $Q_3$  & \textsf{SELECT * FROM MNIST L, MNIST R WHERE predict(L) = predict(R)} \\ 
       $Q_4$  & \textsf{SELECT COUNT(*) FROM MNIST L, MNIST R WHERE predict(L) = predict(R)} \\ 
       $Q_5$  & \textsf{SELECT COUNT(*) FROM MNIST WHERE predict(*)=1} \\ %
       $Q_6$  & \textsf{SELECT AVG(predict(*)) FROM Adult GROUP BY gender} \\
       $Q_7$  & \textsf{SELECT AVG(predict(*)) FROM Adult GROUP BY agedecade} \\
    \end{tabular}
    \caption{Summary of queries used in the experiments.  $predict(\cdot)$ is shorthand for $M_\theta.predict(\cdot)$.}
    \label{tab:exp-qs}
    \vspace{-.3in}
\end{table}
}

\subsubsection{Approaches}
We evaluate 3 baselines and the two approaches in this paper.  Each approach returns a ranked list of training points using a train-rank-fix scheme. Each iteration trains the model, and then selects and removes the top-10 ranked \trecs. Thus, removed records affect future iterations and potentially improves the results. %

For the baselines, \textbf{\loss} ranks from the highest training loss to lowest, it is the most convenient approach because it is naturally computed during training; \textbf{\infloss} uses the model-based influence analysis~\cite{influence} to rank a training point higher if removing it increases its individual training loss the most. This is the state of the art approach of using the influence analysis framework for training set debugging without requiring additional labels. We compare these against \textbf{\naive} (\Cref{sec:howtocomp}) and \textbf{\holistic} (\Cref{sec:solution}). 

\subsubsection{Datasets}  We use record, text, and image-based datasets.  In each experiment, we will systematically corrupt the labels of $\mathcal{K}$ \trecs.

\stitle{DBLP-GOOG} publication entity resolution dataset used in \cite{magellandata}. Each publication entry contains four attributes: title, author list, venue, and year.  It contains two bibliographical sources---DBLP and Google Scholar---and the logistic regression model classifies a pair of DBLP, Scholar entries as same or not.  We represent each pair using 17 features from \cite{dblpgoog}. The dataset is split in a training and querying set and a logistic regression model is trained.

\stitle{ADULT} income dataset \cite{uci}, also known as the ``Census Income'' dataset. The task of this dataset is to predict based on census data whether a person makes more than 50K\$ per year. Following the code of the author's of \cite{discrimination}, we take three features of the dataset, namely age, education and gender and turn them in 18 binary variables. This process creates a lot of training examples with identical features (but not necessarily identical labels). Creating large groups of training examples with identical features is a necessary preprocessing step for many approaches of countering bias in learning \cite{capuchin}. In \Cref{sec:multi-query}, we shall see that it also introduces complications in training bug detection.

\stitle{ENRON} spam classification dataset~\cite{enron}. It contains 5172 emails received and sent by ENRON employees. The logistic regression model classifies each email as spam or not spam. Each email is represented as a bag of words.

\stitle{MNIST} digits recognition dataset \cite{mnist} contains 70000 hand-written images of 0-9 digits, each consisting of a $28 \times 28$ grid of pixels. The task is given an input image to output the digit depicted. We will experiment on this dataset using both logistic regression and neural architectures trained on 10000 training examples. 

\subsubsection{Training Errors: }  Our experiments generate systematic training set errors by corrupting training labels.  To do so, we choose records that match a predicate, and change the labels for a subset of the matching records.  For example, for some of the MNIST image experiments, we select images of the digit $1$, and change varying subsets of those images to be labeled $7$.  We describe the predicate and subset size in the corresponding experiments.

\subsubsection{Complaints: } The majority of our experiments specify equality value complaints for outputs of aggregation queries, tuple deletion complaints for outputs of join and non-aggregation queries.  The complaints are generated from the ground truth.  In \Cref{sec:exp-effort}, we execute two queries on the same query dataset and submit complaints for both queries; we also simulate misspecified equality value complaints that overestimate or underestimate the correct value, or where the value is completely incorrect.  

\subsubsection{Metrics: }  We report recall $r_k$ as the percentage of correctly identified \trecs in the top-$k$ returned records, where $k\in[0,\mathcal{K}]$ increases to the number of actual corruptions $\mathcal{K}$. Unlike ML model evaluation, we note that for a given $k$, precision can be derived from recall.

Comparing curves across experiments can be challenging, thus we take inspiration from the area under the curve measure for precision-recall curves ($AUC_{PR}$) to introduce an area under the curve measure for our corruption-recall curves.  We call it $AUC_{CR}$, and compute it as the normalized average of the recalls across all $k$ values: $AUC=\frac{2}{\mathcal{K}}\sum_{k=1}^\mathcal{K} r_k$ where $r_k$ are the recall percentages. We also report running time when appropriate.  

\sloppy

\subsubsection{Implementation: } All our experiments are implemented in Tensorflow~\cite{tensorflow} and run on a google cloud n1-highmem-32 machine (32 vCPU, 208GB memory) with 4 NVIDIA V100 GPU. All models are implemented in Keras, and trained using the L-BFGS algorithm in Tensorflow. As noted in \cref{sec:background}, we use the conjugate gradient algorithm to efficiently calculate $\nabla_{\pmb{\theta}} q(\pmb{\theta}^*) H_{\pmb{\theta}^*}^{-1}$. 

\fussy

\subsection{Baseline Comparison: SPA Queries}\label{sec:useful-complaint}

We first evaluate the efficacy of complaint-based methods as compared to the baselines for detecting systematic errors in training records.  We use a \texttt{COUNT(*)} query, and a single value complaint with the correct equality value.  We first report detailed results for systematic corruptions of the DBLP dataset, where we flip a percentage of the \texttt{match} training labels to be \texttt{notmatch}.  The percentage varies from 30\% to 70\% of the \texttt{match} training records, affecting 7\% to 17\% of the training labels accordingly. We run $Q_1$ from \Cref{tab:exp-qs}, and complain that the count is incorrect.

\begin{figure}[h]
    \centering
    \includegraphics[width=\columnwidth]{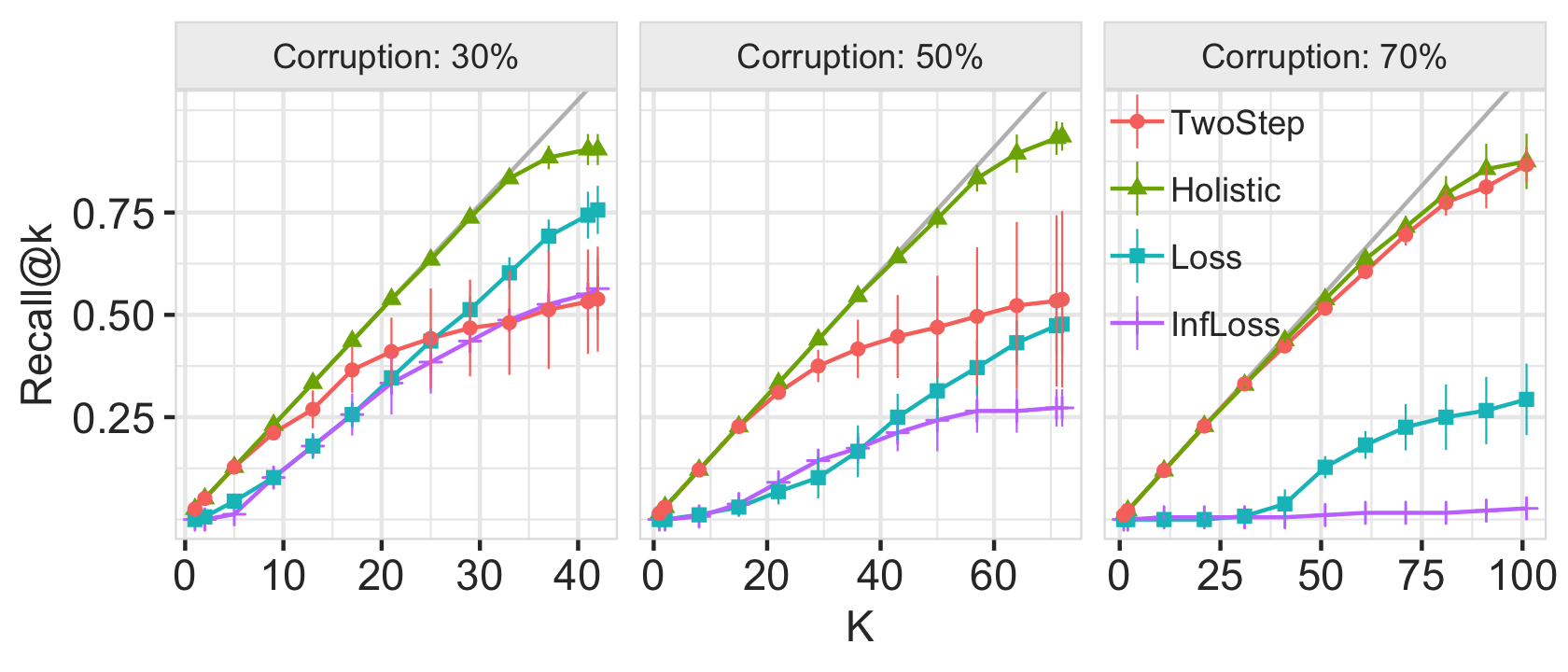}
    \caption{Recall curves when varying corruption rate for DBLP (\grey{grey} line is perfect recall).  Loss-based approaches perform poorly as corruption rate increases, while \naive improves at very high corruption rates (70\%).  \holistic dominates the other approaches.}
    \label{fig:cr-dblp}
\end{figure}

\begin{figure}[htpb]
    \centering
    \begin{minipage}[t]{.48\linewidth}
        \centering
        \includegraphics[width=\columnwidth]{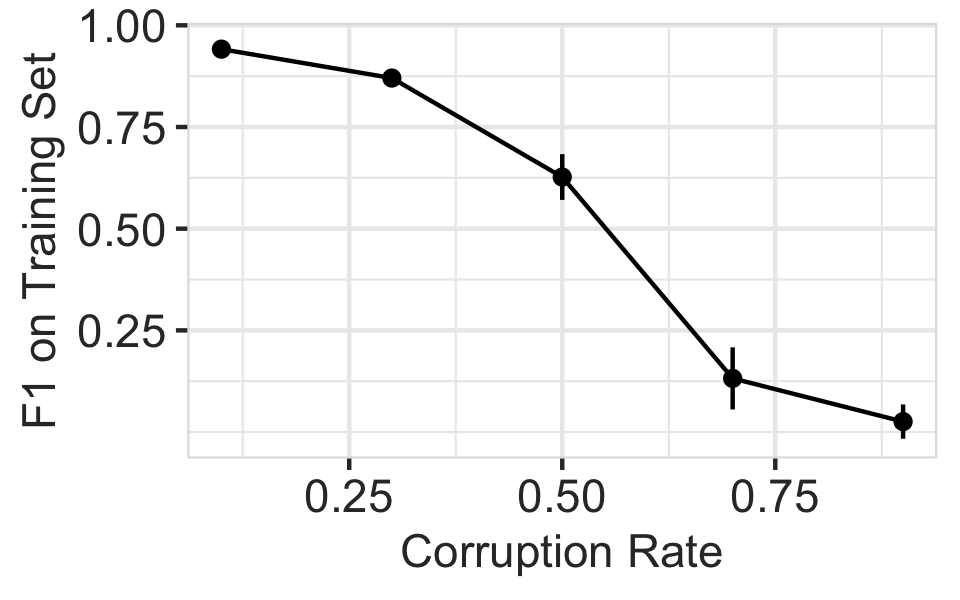}
        \caption{F1 vs corruption rate on DBLP}
        \label{fig:f1-dblp}
    \end{minipage}\hfill
    \begin{minipage}[t]{.48\linewidth}
        \centering
        \includegraphics[width=\columnwidth]{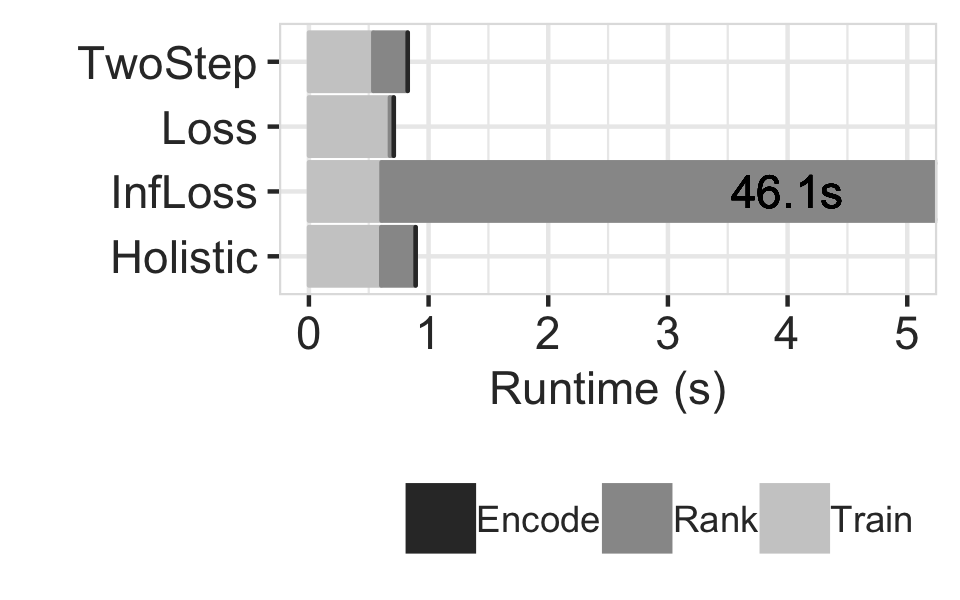}
        \caption{Per-iteration runtime on DBLP, 50\% corruption.  \infloss takes 46.1s.}
        \label{fig:time-dblp}
    \end{minipage}
\end{figure}

\Cref{fig:cr-dblp} shows the recall curves for low (30\%), medium (50\%), and high (70\%) corruption rates, where the \grey{grey} line is a reference for perfect recall.  Both loss-based approaches (\loss, \infloss) degrade substantially as the corruption rate increases because the model begins to overfit to the training corruptions instead. This is corroborated by \Cref{fig:f1-dblp}. There we observe the F1 score of the model, the geometric mean of the model precision and recall, on the querying set as the corruption rate increases. For small corruption rates, the model treats the few corruptions as outliers and it does not fit them leading to robust performance. However, this changes for corruption rates larger than $50\%$ where performance starts to drop drastically indicating that the model has started fitting to the corrupted data. \naive initially performs poorly, but improves as the systematic errors dominate the training set ($70\%$) and reduce the complaint ambiguity.  In contrast, \holistic is nearly perfect, and is robust to the different corruption rates. For reference, the \auc of the approaches for medium corruption are shown as the first row in \Cref{tab:enron}.

\Cref{fig:time-dblp} shows the runtime for each train-rank-fix iteration.  We report three values, based on the terms in \Cref{eq:inffunc}. Train refers to model retraining to compute the model parameters $\pmb{\theta}^*$; 
Encode refers to the cost of computing the influence function   $-\nabla_{\pmb{\theta}} q(\pmb{\theta}^*)$; 
Rank refers to evaluating $\nabla_{\pmb{\theta}} q(\pmb{\theta}^*) H_{\pmb{\theta}^*}^{-1}$, which is dominated by calculating the Hessian vector products required by the conjugate gradient approach of \cite{hessianfree}. \loss is the fastest because it simply uses the training loss and avoids costly influence estimation; \infloss has similar or worse recall curves than \loss, but is by far the slowest because it computes a unique influence function for each training record.  \holistic and \naive are comparable, and dominated by the ranking cost.

We next evaluate the ENRON dataset using $Q_2$, where the search word in the \textsf{LIKE} predicate is either `http' or `deal'.  The corruptions simulate rule-based labeling functions.   For the `http' query, we label all training emails containing `http' as spam (13\% of emails, of which 76\% already labeled spam).  The label corruption method is similar for the `deal' query (18\% of emails, 2.7\% labeled spam).  \Cref{tab:enron} summarizes the results: \infloss, \loss and \naive perform poorly. It is worth pointing out that \infloss takes 2 days to produce the results. \holistic performs much better for `deal' because 17.5\% more training labels were flipped, in contrast to only 3.14\% for `http'.

{\small
\begin{table}[htpb]
    \begin{tabular}{cccccc}
         Dataset &    & \infloss  & Loss & \naive & \holistic \\ 
         \hline
         DBLP   && 0.30    & 0.35 & 0.71   & \red{\textbf{0.99}}\\
         ENRON & '\%\texttt{http}\%' &   0.05    & 0.02  & 0.04     & \red{\textbf{0.12}}      \\  
         ENRON & '\%\texttt{deal}\%' &   0.17    & 0.02  & 0.07     & \red{\textbf{0.40}}      \\  
    \end{tabular}
    \caption{AUC for DBLP with medium corruption, and ENRON with different search words.}
    \label{tab:enron}
\end{table}
}

{\it\ititle{Takeaways:}
    Loss-based approaches are sensitive to the number of systematic errors in the training set---at large corruption rates, the model can overfit to the errors and lead to poor debugging quality.  In contrast, complaints help ensure training records are ranked according to their effects on the complaints.  We find that \infloss takes over 40s per iteration, yet performs poorly under systematic errors.  
    For these reasons, we do not evaluate \infloss in subsequent experiments, but keep \loss to serve as a comparison point.
}
\subsection{Baseline Comparison: SPJA Queries}

\begin{figure*}[htpb]
    \centering
    \begin{subfigure}[t]{.24\textwidth}
        \centering
        \includegraphics[width=\columnwidth]{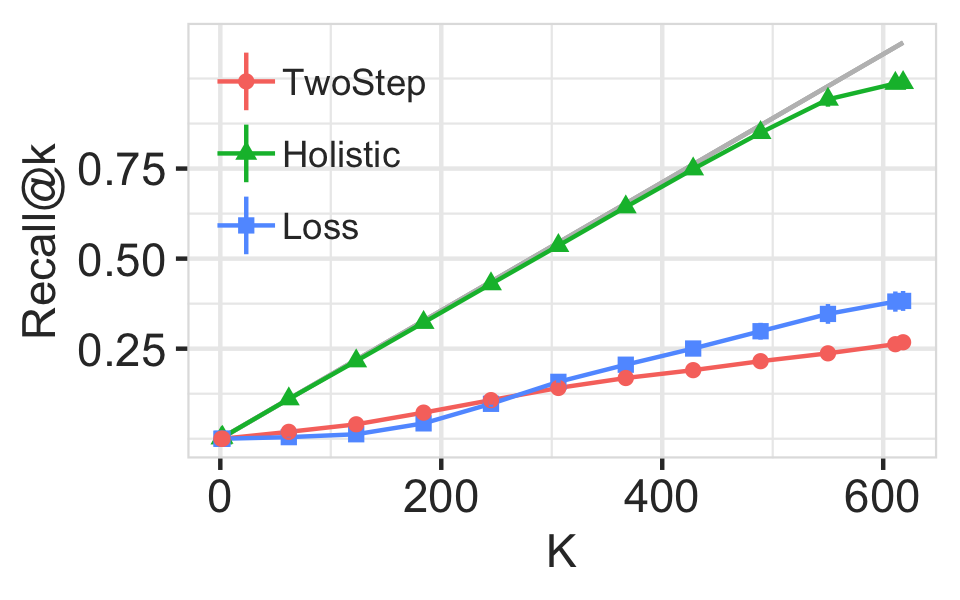}
        \caption{\centering Recall for point complaints (50\% corruption).}
        \label{fig:rc-mnist-joincount-row-0.5}
    \end{subfigure}%
    \begin{subfigure}[t]{.24\textwidth}
        \centering
        \includegraphics[width=\columnwidth]{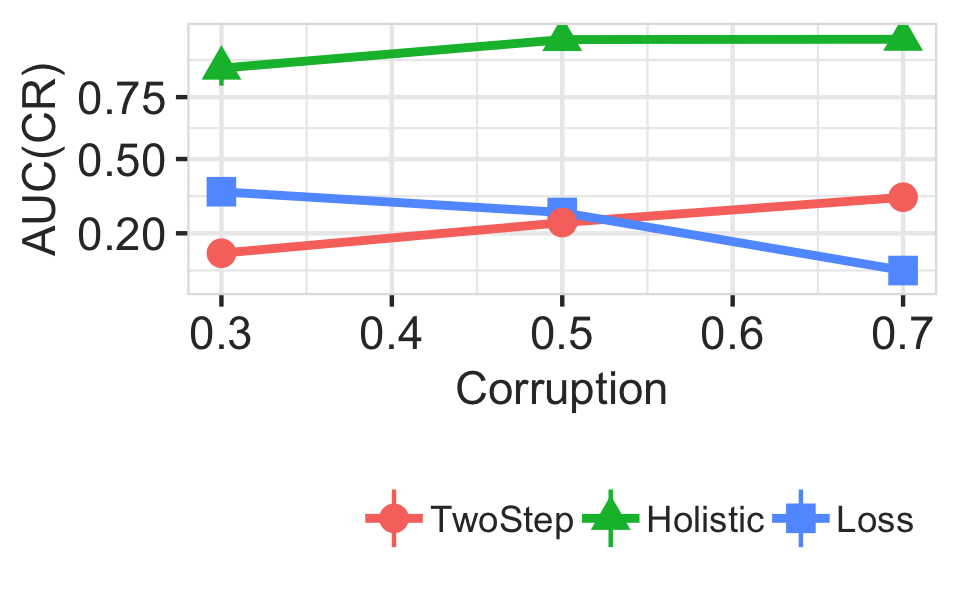}
        \caption{\auc for point complaints.}
        \label{fig:auc-mnist-joincount-row-0.5}
    \end{subfigure}%
    \begin{subfigure}[t]{.24\textwidth}
        \centering
        \includegraphics[width=\columnwidth]{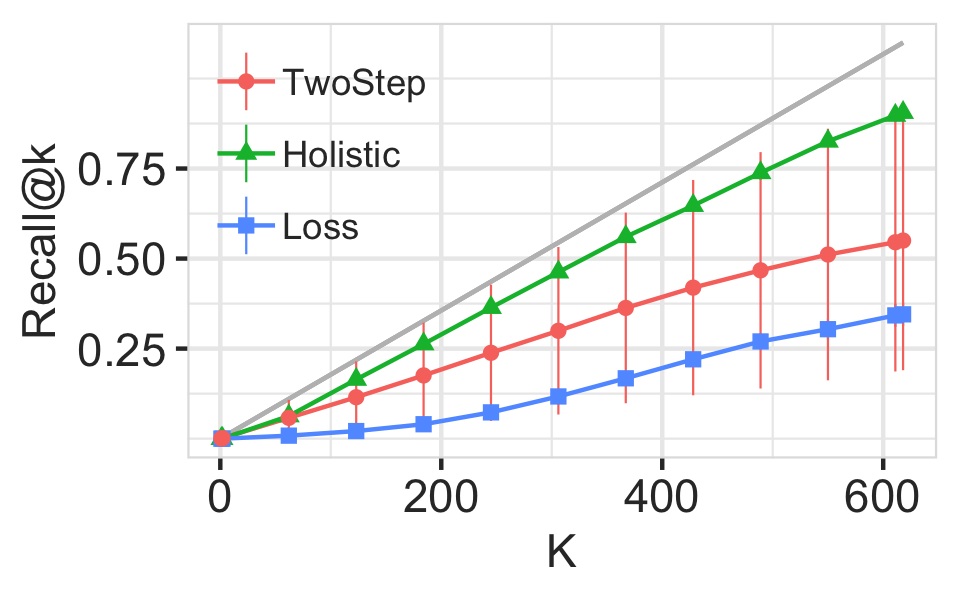}
        \caption{\centering Recall for COUNT complaint (50\% corruption).}
        \label{fig:rc-mnist-joincount-0.5}
    \end{subfigure}%
    \begin{subfigure}[t]{.24\textwidth}
        \centering
        \includegraphics[width=\columnwidth]{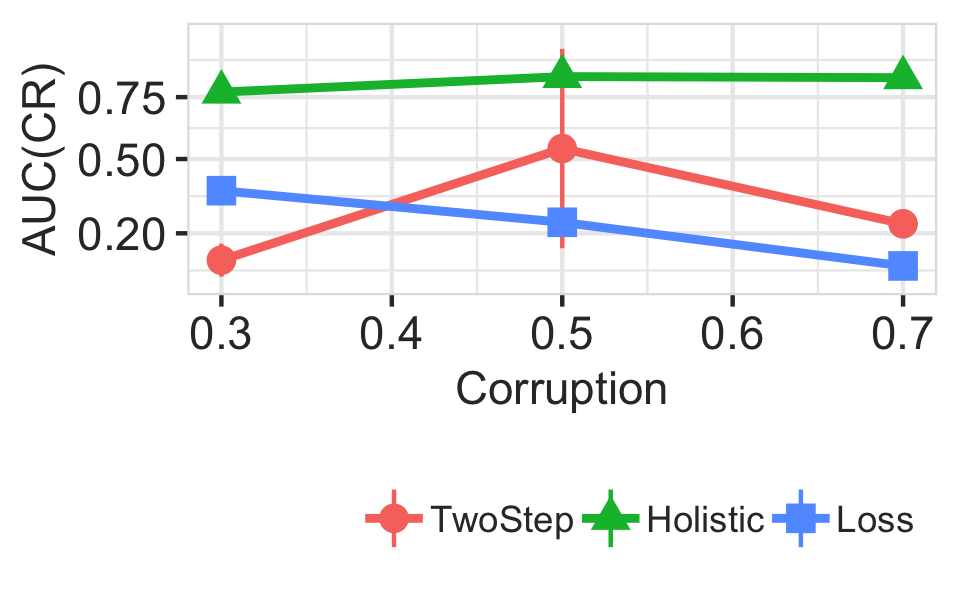}
        \caption{\auc for COUNT complaint.}
        \label{fig:auc-mnist-joincount-0.5}
    \end{subfigure}%
    \caption{MNIST complaints on individual join rows (a-b), or COUNT of join results (c-d). }
    \label{fig:mnist-join}
\end{figure*}

This section uses the MNIST dataset to evaluate complaint-based debugging against the baselines for SPJA queries containing joins.  The first two experiments join two image subsets that do not overlap in their digits, and thus expect no results of the join operation.  We introduce corruptions by flipping a random subset of digit $1$ images to be labeled $7$ instead.  We corrupt 30\% (low), 50\% (medium), and 70\% (high) of the labels, impacting 3\%, 5\% and 7\% of the total training labels accordingly. We chose MNIST to make the problem more ambiguous: the model is a 10-digit classifier, thus there are 10 ways ($1=1$, $2=2$, e.t.c.) to incorrectly satisfy the join condition, but 90 ways to incorrectly fix it (all other label combinations). We thus expect \naive to perform poorly due to a large number of satisfying, but incorrect, ILP solutions.

We first use $Q_3$, which joins images of $1$ with images of $7$. We generate tuple complaints for join results where the left (or right) side of the join was correctly predicted, but the right (left) side was incorrect.  This results in 121, 550, and 931 complaints for the low, medium, and high corruption rates.  \Cref{fig:rc-mnist-joincount-row-0.5} shows that \naive and \loss perform poorly compared to \holistic, despite 550 complaints.  When varying the corruption rate in \Cref{fig:auc-mnist-joincount-row-0.5}, \naive improves slightly, but is still dominated by \holistic.  

Our second experiment runs a COUNT aggregation ($Q_4$) on $Q_3$'s results.  The left relation contains images with digits $1$ through $5$; the right relation contains digits $6-9,0$.  The complaint says that the result should be $0$---this is the same as a delete complaint on all join tuples, and states that all left tuples should not have the same prediction as any in the right relation.  As expected, the lower ambiguity improves the likelihood that \naive's ILP picks a good satisfying solution, but the large standard deviation shows that it is unstable (\Cref{fig:rc-mnist-joincount-row-0.5}).    \Cref{fig:auc-mnist-joincount-0.5} shows both \loss and \naive perform poorly across corruption rates; note that \naive is erratic between runs and doesn't show a clear trend.  

Our third experiment joins two image datasets that overlap.  We use the same relations as the previous experiment, and set the corruption rate to 50\%.   However, we move a subset of the $1$ digit images from the left relation to the right, which we call the {\it mix rate}.  For example, a mix rate of 25\% means that we move 25\% of the $1$ images ($296$ out of 1125) from the left relation to the right---the true output of $Q_4$ should be $829\times 296=245384$, whereas the incorrect output was $1044470$.  As noted in \Cref{sec:naiveanalysis}, this is far more ambiguous than the previous experiment.  As we vary the mix rate between $5\%$, $25\%$, $35\%$, the \auc for \loss is stable at $\approx0.24$, whereas \holistic is initially high then decreases slightly (\auc$=0.78,0.57,0.48$, respectively).  \naive does not solve the ILP within $30$ minutes, thus we cannot report its results.  

\smallskip
{\it\ititle{Takeaways: }
    Overall, \holistic achieves the highest recall on SPA and SPJA queries as compared to the baselines as well as \naive. \naive is sensitive to the ILP solver as well as the level of ambiguity, which we will evaluate in the next subsection.  
}

\subsection{Effects of Ambiguity}

The previous experiments suggested the effects of high ambiguity on the different approaches.  In this experiment, we use the same setup as $Q_3$ in the SPJ experiment, and carefully vary the amount of complaint ambiguity.  In the previous experiment, the complaint only specifies that the join output record should not exist, but does not prescribe how to fix it.   Here, we will replace a subset $\alpha$ of those complaints with unambiguous complaints.  Specifically, for a complaint over a join output record $(l\in L, r\in R)$, we replace it with value complaints on the output of the model predictions $predict(l)$ and $predict(r)$.   We corrupt 30\%  of the $1$ digits as in the previous experiment.

\begin{figure}[htpb]
    \centering
    \includegraphics[width=\columnwidth]{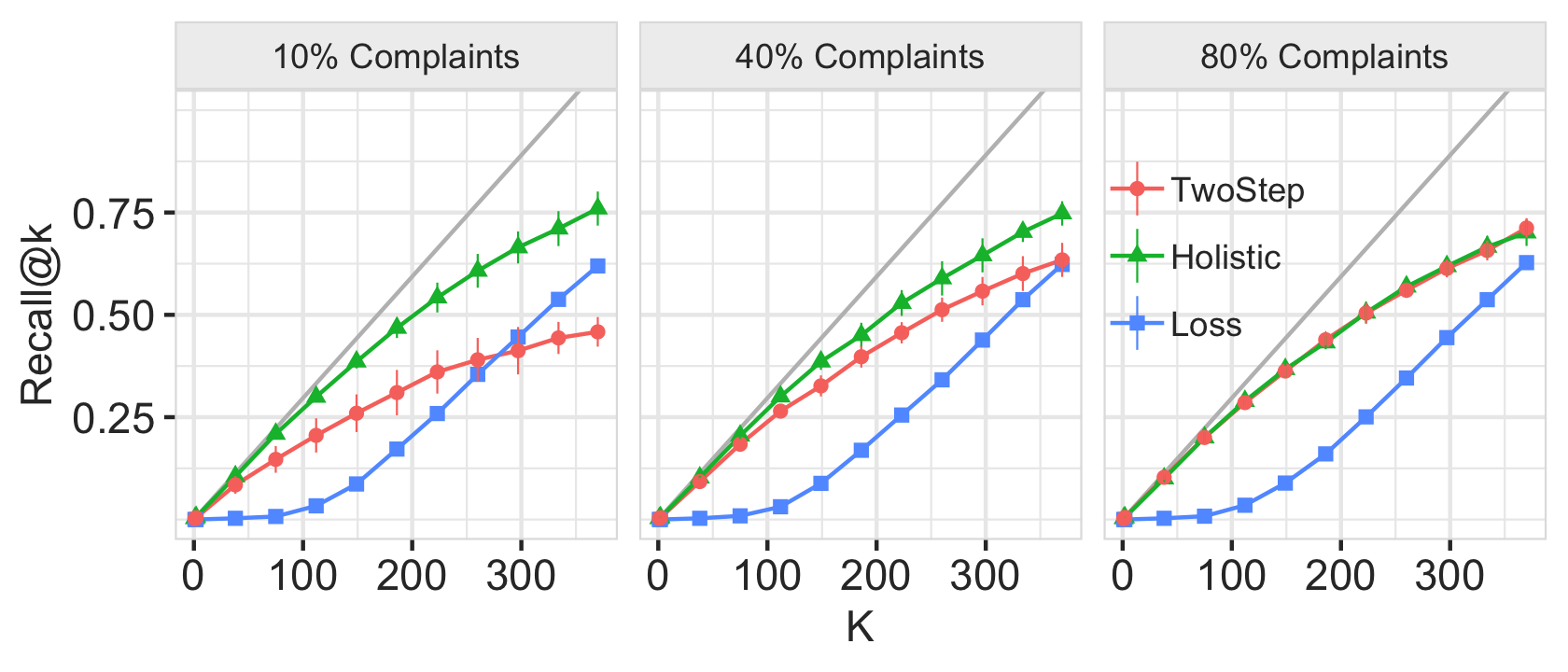}
    \caption{Varying ambiguity of the MNIST point complaints experiment.  Each facet varies the percentage of join result complaints that are replaced with direct complaints over the model mispredictions. }
    \label{fig:mnist-joinrow-ambiguity-0.3}
\end{figure}

\Cref{fig:mnist-joinrow-ambiguity-0.3} shows that \holistic dominates the approaches at high ambiguity (10\% complaints), however at low ambiguity (80\%), \naive is competitive with \holistic. In addition, this experiment illustrates how \sys can make use of complaints from different parts of the query plan. Specifically, we can view the join record complaints as complaints on the output of $Q_3$ and the unambiguous complaints on the predictions of $L$ and $R$ as complaints that target the provenance of $Q_3$.

{\it\ititle{Takeaways:} \naive is sensitive to ambiguity. \naive converges to \holistic when ambiguity is reduced by,  for example, directly labelling many model mispredictions.  
}

\subsection{Multi-Query Complaints}\label{sec:multi-query}

\begin{figure}
    \centering
    \includegraphics[width=.9\columnwidth]{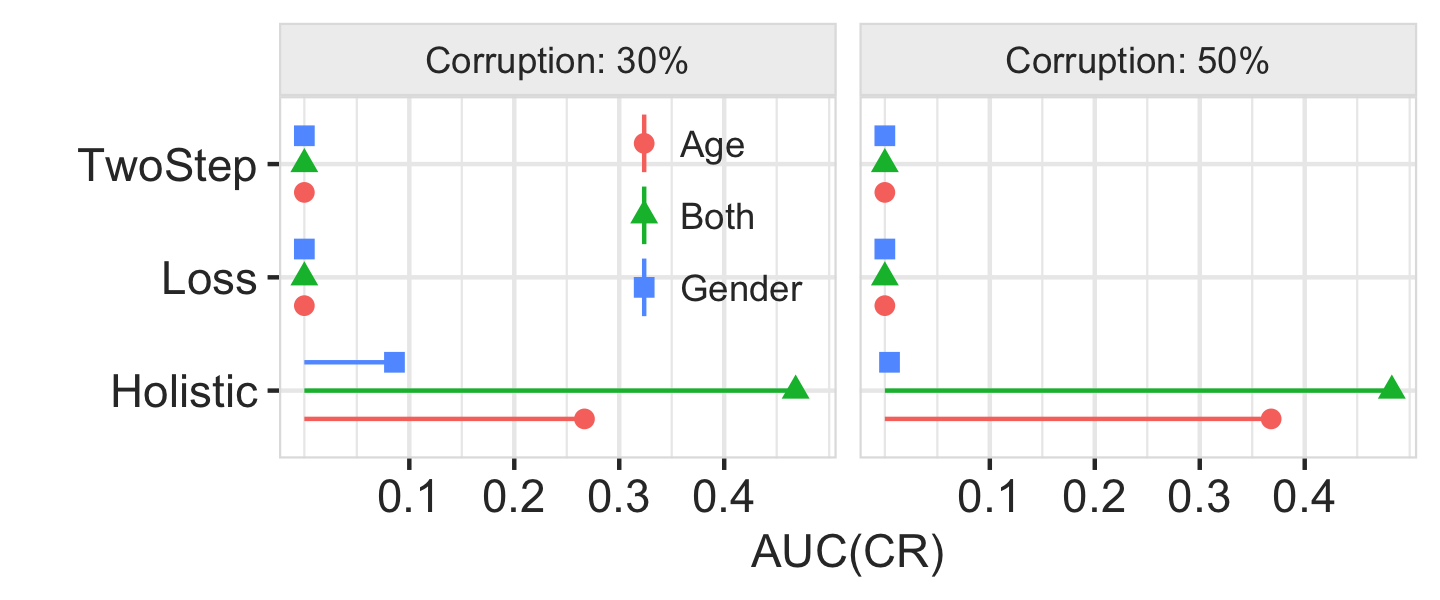}
    \caption{\holistic can benefit from combining complaints of multiple queries.}
    \label{fig:mnist-adult}
\end{figure}
So far, we have evaluated \sys using a single query and on a single attribute.  In this experiment, we use the multi-attribute Adult dataset, and illustrate that complaints over {\it different} queries (that use the same model) can be combined to more effectively identify training set errors.   We execute $Q_6$ and $Q_7$ from \Cref{tab:exp-qs}. $Q_6$ groups the dataset by {\it Gender} and creates a value complaint for the male average value. $Q_7$ aggregates the dataset by {\it Age} (bucketed into decades), and creates a value complaint for the 40-50 age group's average value. To corrupt the training set, we select records that satisfy the conjunction of low income, male, and 40-50 years old, and flip $\alpha\%$ of their labels from low income (y=0) to high income (y=1). $8.2\%$ of the training set matches this predicate. We set $\alpha\in\{30\%,50\%\}$ thus affecting the labels of $2.4\%$ and $4.1\%$ training points respectively.

\Cref{fig:mnist-adult} shows that \naive, \loss, and \holistic when given each complaint in isolation, and when given both.  \naive and \loss are unable to find any erroneous training records. One of the reasons is that the preprocessing step borrowed from ~\cite{discrimination} only uses three attributes to construct their features.  This results in many duplicate training points (118/6512 points are unique).  Thus, considerably more iterations for \naive and \loss are spent proposing and removing duplicates.  Further, \naive's SQL step is agnostic to the model and training set, and fails to leverage this information when solving the ILP.

\holistic is, to a lesser degree, affected by the duplicates for the {\it Gender} complaint. This is because {\it Gender} is less selective than {\it Age}: in the training set, only $23.1\%$ of males are between 40 and 50 but $71.3\%$ of people between 40 and 50 are males.   \holistic benefits considerably from using both complaints because they serve to narrow the possible training errors to those within the corrupted subspace.

{\it\ititle{Takeaways:}
    Users often run multiple queries over the same dataset.  We find that \holistic is able to leverage complaints across multiple queries. In contrast, techniques that are oblivious to the complaints (\loss) or oblivious to the model and training (\naive) perform poorly.  
}

\subsection{Do Complaints Reduce Debugging Effort?}
\label{sec:exp-effort}

One of the potential benefits of a complaint-based debugging approach is that users can specify a few aggregate but potentially ambiguous complaints, rather than label many individual, unambiguous, model predictions.  In addition, it is desirable that complaints are robust to mis-specifications.  For example, if a result value is 20 but should be 49, then a value complaint that is $50$, or $60$, or $45$ should not greatly affect the returned training records.   We now evaluate both of these questions in sequence.  We use the MNIST dataset with corruptions that flip 10\% of the training images with the digit $1$ to be labeled $7$.

\begin{figure}
    \centering
     \includegraphics[width=.6\columnwidth]{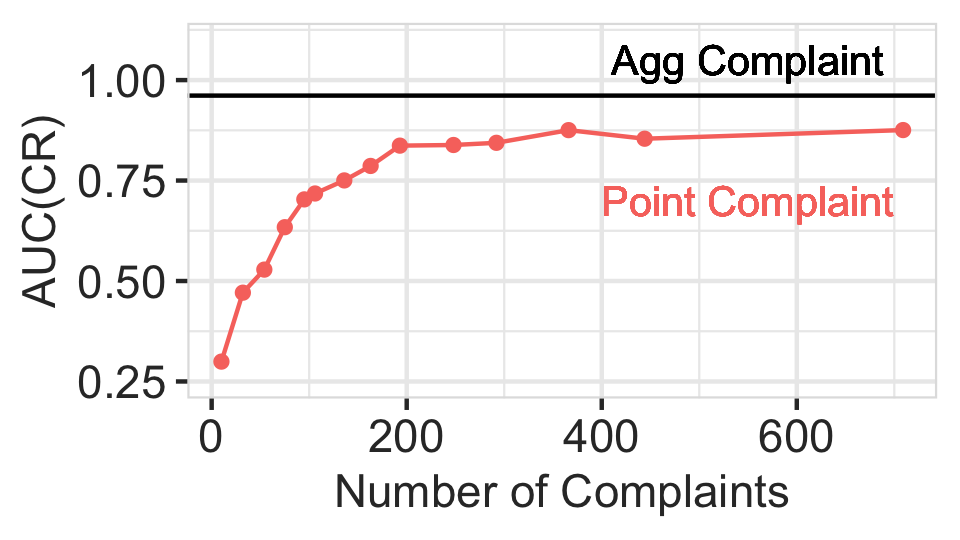}
    \caption{Comparison of one aggregate complaint (black) and increasing the number of point complaints (\red{red})}
    \label{fig:mnist-avp}
\end{figure}
First, we compare aggregate-level and prediction-level complaints.  \textit{Agg Complaint} is a single value complaint over $Q_5$, which counts the number of $1$ digits; \textit{Point} \textit{Complaints} varies the number of complaints of model mispredictions from 1 to 709, and is equivalent to state-of-the-art influence analysis~\cite{influence}).   \Cref{fig:mnist-avp} shows that the aggregate complaint is enough to achieve \auc$\approx1$, whereas \naive requires over $200$ point complaints to reach \auc$\approx0.87$.  This suggests that, from an user perspective, aggregate-level complaints can require less effort.

A potential drawback of aggregate-level complaints is that they may be sensitive to mis-specification.  To evaluate this, we introduce three types of errors to the user's value complaint.  The errors vary in the user-specified $X$ in the equality complaint $t[a] = X$, as compared to the ground truth $X^*$.  {\it Overshoot} overcompensates for the error by setting $X = 1.2 \times X^*$, meaning if the query result was $10$ and the ground truth was $100$, then $X$ is set to $120$.  {\it Partial} under-estimates the error but correctly identifies the direction the query result should move---$X$ is set to the average of the query result and the ground truth (e.g., $55$ in the preceding example).  {\it Wrong} overcompensates in the  incorrect direction, and sets $X=0.8 \times t[a]$. 

\begin{figure}[htpb]
    \centering
    \includegraphics[width=\columnwidth]{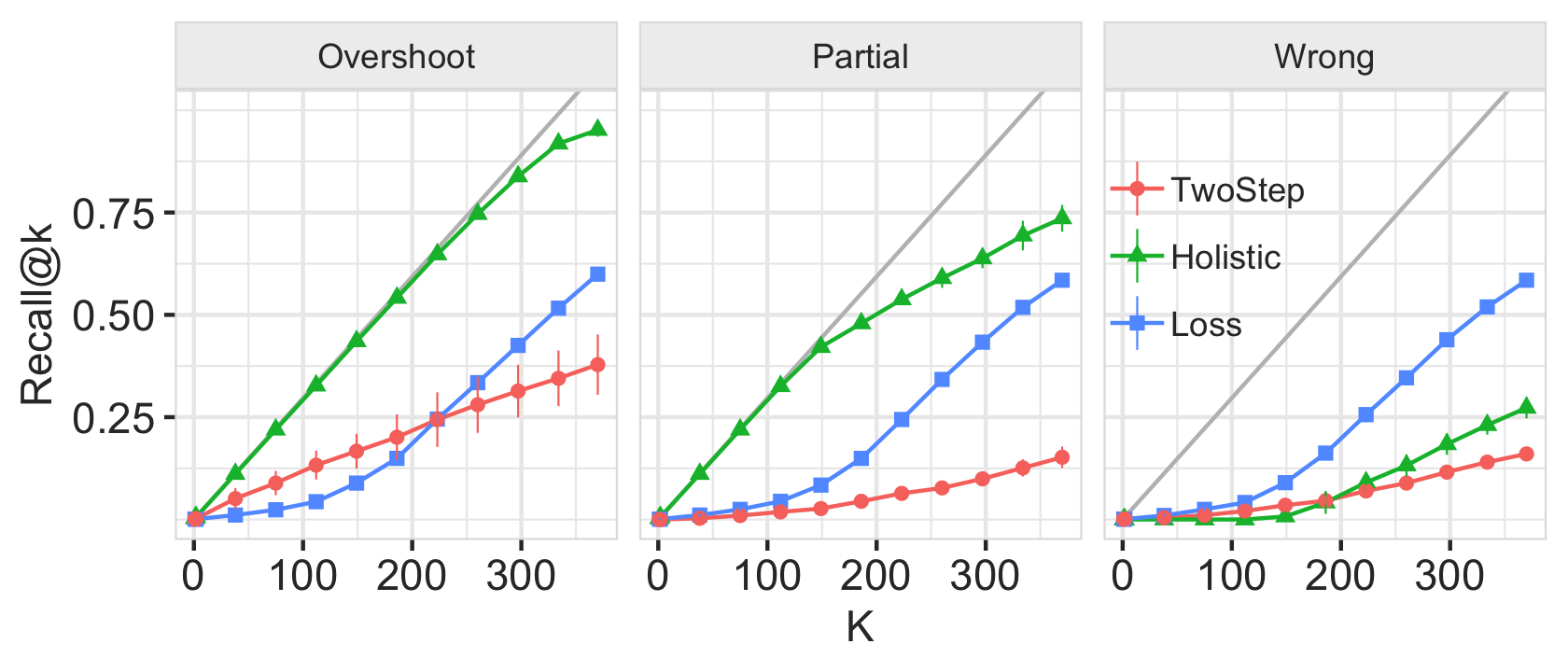}
    \caption{How errors in complaints affect each approach. }
    \label{fig:mnist-fuzzy-complaint}
\end{figure}

\Cref{fig:mnist-fuzzy-complaint} shows that \holistic is relatively robust to misspecified complaints, as long as they point in the correct direction of error.  Specifically, the \holistic {\it Partial} curve degrades around $K=150$ because the complaint has been satisfied. \holistic performs poorly when the complaint direction is {\it Wrong} because it tries to identify \trecs that if removed reduce the count whereas the true corruptions do the opposite. \naive similarly degrades, whereas \loss is insensitive because it does not rely on complaints at all.  

{\it\ititle{Takeaways: }
    Complaint-based approaches allow users to provide few ambiguous complaints over aggregated results, and still accurately identify training set errors.  \holistic is robust to misspecifications as long as the direction of the complaint is correct.
}

\section{Related Work}\label{sec:rw}

\sys provides complaint-driven data debugging for relational queries that use machine learning inference.  This is most closely related to SQL explanations in the DB community, ML explanations in the ML community.  It is also related to data cleaning for machine learning, as well as debugging ML pipelines in general.

\stitle{SQL Explanation:}
SQL explanation seeks to explain errors in a query result.  Errors may be specified as incorrect values, how values should change, tuples that should not exist, or tuples that should exist.   These errors can be explained as subsets of the \qrels~\cite{probdbsensitivity}, predicates over the \qrels that should be deleted~\cite{scorpion,roy,bailis,Roy2014AFA}, values of the \qrels that should be changed~\cite{tiresias}, changes to the query~\cite{Chapman1965WhyN,Tan2017ReverseEA,Tran2010HowTC}, or changes to past queries~\cite{qfix}.    This line of work is generally related to causal interventions in queries~\cite{Meliou2014CausalityAE,Salimi2018HypDBAD} and reverse data management~\cite{Meliou2011ReverseDM}.  

Provenance~\cite{probdbsensitivity,semirings,probdb} returns the \qrecs, and how they were combined, for a given output record.  This is a form of explanation, serves as the starting point for many of the SQL explanation approaches above, including this work.   From our perspective, \sys traces user complaints back through the query,  and by using influence analysis, back to the training records.

\stitle{ML Explanation: }
Gilpin et. al~\cite{mlexplanationsurvey} provide an excellent survey of ML Interpretation.  A major aspect of ML explanation is in understanding why a model makes a specific prediction for a data point, and techniques include surrogate models~\cite{lime}, saliency maps~\cite{Simonyan2013DeepIC,Zeiler2013VisualizingAU,deeplift}, decision sets~\cite{decissionsets}, rule summaries~\cite{anchors,Krishnan2017PALMML}, hidden unit analysis~\cite{Sellam2018DeepBaseDI}, sensitivity analysis~\cite{Kantchelian2015EvasionAH,Nilforoshan2017LeveragingQP}, and general feature attribution methods~\cite{integratedgradients,shap}.     

Most related are case-based explanations that identify training records that affected a set of mispredictions.  Of these, influence analysis methods are prominent. Techniques such as DUTI~\cite{duti} model this task as a bi-level optimization problem that may require several rounds of model retraining to identify a single training point. Influence Functions~\cite{influence} avoid retraining by approximating this influence locally.  

The limitation of these approaches is that they assume that the user has identified model mispredictions. In contrast, \sys focuses on query result complaints that have been affected by model mispredictions.  However, it may not be directly known which of the \qrecs have been mispredicted.

\stitle{Debugging ML Pipelines: }
Relational plans, such as those studied in this work, can be viewed as a restricted form of general data analysis pipelines.  Within this context, data errors are a major issue in ML pipelines~\cite{Polyzotis2017DataMC}.   Systems such as Data X-Ray~\cite{Wang2015DataXA} help debug large-scale data pipelines by summarizing data errors that share a common cause.  Data validation~\cite{Breck2019DataVF,Shawi2019AutomatedML} and model assertions~\cite{Kang2018ModelAF} help catch errors before deployment.  This work relies on record-level provenance to address complaints; provenance is increasingly viewed as an integral part of any modern ML pipeline~\cite{dawnblog,Agrawal2019CloudyWH}.

\stitle{Data Cleaning:} 
Machine learning relies on training data.  Data cleaning is both used to clean errors in training datasets to improve ML models, and leverage ML to identify errors.  Traditional data cleaning is largely based on constraints~\cite{Rahm2000DataCP,chu2013holistic}.  In contrast, recent work leverages knowledge of downstream ML models~\cite{activeclean,boostclean,alphaclean,Li2019CleanMLAB}, integrates cleaning signals from heterogeneous sources~\cite{Rekatsinas2017HoloCleanHD}, and leverages machine learning to perform error detection~\cite{heidari2019holodetect,Mahdavi2019RahaAC,abedjan2016detecting}.

In addition to general methods for addressing noisy labels~\cite{frenay2013classification}, techniques such as Snorkel help identify conflicting and noisy labels from different labeling sources~\cite{Ratner2017SnorkelFT}, while other work leverages oracles~\cite{dolatshah2018cleaning}.

Unlike the existing studies in data cleaning, our work is focused on detecting training set errors w.r.t. the user complaints expressed as Query 2.0.

\section{Conclusions and Discussion}\label{sec:conclusion}

Leveraging model inference within query execution (which we call Query 2.0) is rapidly gaining wide-spread adoption.  However, query results are now susceptible to errors in the model's training data.  Although there exist techniques to individually debug outliers of SQL queries, and prediction errors in ML models, techniques to address the combination of the two do not exist.

To this end, \sys helps users identify training set errors by leveraging not only the model and data, but also {\it user complaints} about final or intermediate query results.  \sys integrates these together to find the training records that will most address the user's complaints.  To do so, we introduce two approaches.  \naive splits the query into SQL-only and ML prediction-only subplans that can be solved using existing SQL and ML explanation techniques.  \holistic is an optimization that integrates both steps to directly estimate each training record's influence on user complaints.  Our experiments show that \holistic more accurately identifies systematic training set errors as compared to existing ML explanation techniques, across relational, image, and text datasets; linear and neural network models; and different SPJA queries.  

\stitle{Other Interventions: }
The type of intervention for fixing the training data is not restricted to only the deletion. Existing techniques like \cite{DBLP:conf/cvpr/TanakaIYA18} advocates doing label fixing while training and others like \cite{activeclean} proposes both feature and label fixing. \sys chooses deletion based intervention for two reasons: 1. Deletion based intervention is a natural and wide used in SQL explanation \cite{scorpion}. \sys uses deletion as the first step towards this broader Query 2.0 Debugging problem, 2. There can be many choices to fix the labels, even more for features. It is unclear how to find the correct fix. We leave other interventions as the future work.

\stitle{Systematic Debugging: } Combining separate analysis methods in a piece-wise manner, such as \naive, can perform poorly.  This is both because errors from one step will propagate and affect subsequent steps, and because information cannot be shared between steps. \holistic suggests that it is important to consider the entire pipeline and user specifications in a holistic manner.

Stepping back, there is an increasing need for system-wide debugging of data analytic pipelines that use model inference.   This paper advocates for a complaint-driven approach towards pipeline debugging.  Different users---customers, engineers, data scientists, and ML experts---have differing access, perspective, and expertise of the data that flows through these analytic pipelines~\cite{googletfxpaper,Polyzotis2017DataMC}.  
We plan to extend this work beyond SPJA queries to general relational and non-relational workflows, to improve the runtime of the system, and to study interventions beyond training record deletion.

\section*{Acknowledgements}
This work was supported in part by Mitacs through an Accelerate Grant, NSERC through a discovery grant and a CRD grant as well as NSF 1527765 \& 1564049 \& 1845638, Google LCC and Amazon.com, Inc.. All opinions, findings, conclusions and recommendations in this paper are those of the authors and do not necessarily reflect the views of the funding agencies.

\bibliographystyle{ACM-Reference-Format}
\bibliography{ms}


\begin{thebibliography}{85}


\ifx \showCODEN    \undefined \def \showCODEN     #1{\unskip}     \fi
\ifx \showDOI      \undefined \def \showDOI       #1{#1}\fi
\ifx \showISBNx    \undefined \def \showISBNx     #1{\unskip}     \fi
\ifx \showISBNxiii \undefined \def \showISBNxiii  #1{\unskip}     \fi
\ifx \showISSN     \undefined \def \showISSN      #1{\unskip}     \fi
\ifx \showLCCN     \undefined \def \showLCCN      #1{\unskip}     \fi
\ifx \shownote     \undefined \def \shownote      #1{#1}          \fi
\ifx \showarticletitle \undefined \def \showarticletitle #1{#1}   \fi
\ifx \showURL      \undefined \def \showURL       {\relax}        \fi
\providecommand\bibfield[2]{#2}
\providecommand\bibinfo[2]{#2}
\providecommand\natexlab[1]{#1}
\providecommand\showeprint[2][]{arXiv:#2}

\bibitem[\protect\citeauthoryear{Abadi, Agarwal, Barham, Brevdo, Chen, Citro,
  Corrado, Davis, Dean, Devin, Ghemawat, Goodfellow, Harp, Irving, Isard, Jia,
  Jozefowicz, Kaiser, Kudlur, Levenberg, Man\'{e}, Monga, Moore, Murray, Olah,
  Schuster, Shlens, Steiner, Sutskever, Talwar, Tucker, Vanhoucke, Vasudevan,
  Vi\'{e}gas, Vinyals, Warden, Wattenberg, Wicke, Yu, and Zheng}{Abadi
  et~al\mbox{.}}{2015}]%
        {tensorflow}
\bibfield{author}{\bibinfo{person}{Mart\'{\i}n Abadi}, \bibinfo{person}{Ashish
  Agarwal}, \bibinfo{person}{Paul Barham}, \bibinfo{person}{Eugene Brevdo},
  \bibinfo{person}{Zhifeng Chen}, \bibinfo{person}{Craig Citro},
  \bibinfo{person}{Greg~S. Corrado}, \bibinfo{person}{Andy Davis},
  \bibinfo{person}{Jeffrey Dean}, \bibinfo{person}{Matthieu Devin},
  \bibinfo{person}{Sanjay Ghemawat}, \bibinfo{person}{Ian Goodfellow},
  \bibinfo{person}{Andrew Harp}, \bibinfo{person}{Geoffrey Irving},
  \bibinfo{person}{Michael Isard}, \bibinfo{person}{Yangqing Jia},
  \bibinfo{person}{Rafal Jozefowicz}, \bibinfo{person}{Lukasz Kaiser},
  \bibinfo{person}{Manjunath Kudlur}, \bibinfo{person}{Josh Levenberg},
  \bibinfo{person}{Dan Man\'{e}}, \bibinfo{person}{Rajat Monga},
  \bibinfo{person}{Sherry Moore}, \bibinfo{person}{Derek Murray},
  \bibinfo{person}{Chris Olah}, \bibinfo{person}{Mike Schuster},
  \bibinfo{person}{Jonathon Shlens}, \bibinfo{person}{Benoit Steiner},
  \bibinfo{person}{Ilya Sutskever}, \bibinfo{person}{Kunal Talwar},
  \bibinfo{person}{Paul Tucker}, \bibinfo{person}{Vincent Vanhoucke},
  \bibinfo{person}{Vijay Vasudevan}, \bibinfo{person}{Fernanda Vi\'{e}gas},
  \bibinfo{person}{Oriol Vinyals}, \bibinfo{person}{Pete Warden},
  \bibinfo{person}{Martin Wattenberg}, \bibinfo{person}{Martin Wicke},
  \bibinfo{person}{Yuan Yu}, {and} \bibinfo{person}{Xiaoqiang Zheng}.}
  \bibinfo{year}{2015}\natexlab{}.
\newblock \bibinfo{title}{{TensorFlow}: Large-Scale Machine Learning on
  Heterogeneous Systems}.
\newblock
\newblock
\urldef\tempurl%
\url{http://tensorflow.org/}
\showURL{%
\tempurl}
\newblock
\shownote{Software available from tensorflow.org.}


\bibitem[\protect\citeauthoryear{Abedjan, Chu, Deng, Fernandez, Ilyas, Ouzzani,
  Papotti, Stonebraker, and Tang}{Abedjan et~al\mbox{.}}{2016}]%
        {abedjan2016detecting}
\bibfield{author}{\bibinfo{person}{Ziawasch Abedjan}, \bibinfo{person}{Xu Chu},
  \bibinfo{person}{Dong Deng}, \bibinfo{person}{Raul~Castro Fernandez},
  \bibinfo{person}{Ihab~F. Ilyas}, \bibinfo{person}{Mourad Ouzzani},
  \bibinfo{person}{Paolo Papotti}, \bibinfo{person}{Michael Stonebraker}, {and}
  \bibinfo{person}{Nan Tang}.} \bibinfo{year}{2016}\natexlab{}.
\newblock \showarticletitle{Detecting Data Errors: Where Are We and What Needs
  to Be Done?}
\newblock \bibinfo{journal}{\emph{Proc. VLDB Endow.}} \bibinfo{volume}{9},
  \bibinfo{number}{12} (\bibinfo{date}{Aug.} \bibinfo{year}{2016}),
  \bibinfo{pages}{993–1004}.
\newblock
\showISSN{2150-8097}
\urldef\tempurl%
\url{https://doi.org/10.14778/2994509.2994518}
\showDOI{\tempurl}


\bibitem[\protect\citeauthoryear{Abuzaid, Kraft, Suri, Gan, Xu, Shenoy,
  Ananthanarayan, Sheu, Meijer, Wu, Naughton, Bailis, and Zaharia}{Abuzaid
  et~al\mbox{.}}{2018}]%
        {bailis}
\bibfield{author}{\bibinfo{person}{Firas Abuzaid}, \bibinfo{person}{Peter
  Kraft}, \bibinfo{person}{Sahaana Suri}, \bibinfo{person}{Edward Gan},
  \bibinfo{person}{Eric Xu}, \bibinfo{person}{Atul Shenoy},
  \bibinfo{person}{Asvin Ananthanarayan}, \bibinfo{person}{John Sheu},
  \bibinfo{person}{Erik Meijer}, \bibinfo{person}{Xi Wu}, \bibinfo{person}{Jeff
  Naughton}, \bibinfo{person}{Peter Bailis}, {and} \bibinfo{person}{Matei
  Zaharia}.} \bibinfo{year}{2018}\natexlab{}.
\newblock \showarticletitle{DIFF: A Relational Interface for Large-Scale Data
  Explanation}.
\newblock \bibinfo{journal}{\emph{Proc. VLDB Endow.}} \bibinfo{volume}{12},
  \bibinfo{number}{4} (\bibinfo{date}{Dec.} \bibinfo{year}{2018}),
  \bibinfo{pages}{419–432}.
\newblock
\showISSN{2150-8097}
\urldef\tempurl%
\url{https://doi.org/10.14778/3297753.3297761}
\showDOI{\tempurl}


\bibitem[\protect\citeauthoryear{Agarwal, Beygelzimer, Dudik, Langford, and
  Wallach}{Agarwal et~al\mbox{.}}{2018}]%
        {DBLP:conf/icml/AgarwalBD0W18}
\bibfield{author}{\bibinfo{person}{Alekh Agarwal}, \bibinfo{person}{Alina
  Beygelzimer}, \bibinfo{person}{Miroslav Dudik}, \bibinfo{person}{John
  Langford}, {and} \bibinfo{person}{Hanna Wallach}.}
  \bibinfo{year}{2018}\natexlab{}.
\newblock \showarticletitle{A Reductions Approach to Fair Classification}. In
  \bibinfo{booktitle}{\emph{Proceedings of the 35th International Conference on
  Machine Learning}}, Vol.~\bibinfo{volume}{80}. \bibinfo{publisher}{PMLR},
  \bibinfo{address}{Stockholmsmässan, Stockholm Sweden},
  \bibinfo{pages}{60--69}.
\newblock
\urldef\tempurl%
\url{http://proceedings.mlr.press/v80/agarwal18a.html}
\showURL{%
\tempurl}


\bibitem[\protect\citeauthoryear{Agrawal, Chatterjee, Curino, Floratou, Gowdal,
  Interlandi, Jindal, Karanasos, Krishnan, Kroth, Leeka, Park, Patel, Poppe,
  Psallidas, Ramakrishnan, Roy, Saur, Sen, Weimer, Wright, and Zhu}{Agrawal
  et~al\mbox{.}}{2019}]%
        {Agrawal2019CloudyWH}
\bibfield{author}{\bibinfo{person}{Ashvin Agrawal}, \bibinfo{person}{Rony
  Chatterjee}, \bibinfo{person}{Carlo Curino}, \bibinfo{person}{Avrilia
  Floratou}, \bibinfo{person}{Neha Gowdal}, \bibinfo{person}{Matteo
  Interlandi}, \bibinfo{person}{Alekh Jindal}, \bibinfo{person}{Konstantinos
  Karanasos}, \bibinfo{person}{Subru Krishnan}, \bibinfo{person}{Brian Kroth},
  \bibinfo{person}{Jyoti Leeka}, \bibinfo{person}{Kwanghyun Park},
  \bibinfo{person}{Hiren Patel}, \bibinfo{person}{Olga Poppe},
  \bibinfo{person}{Fotis Psallidas}, \bibinfo{person}{Raghu Ramakrishnan},
  \bibinfo{person}{Abhishek Roy}, \bibinfo{person}{Karla Saur},
  \bibinfo{person}{Rathijit Sen}, \bibinfo{person}{Markus Weimer},
  \bibinfo{person}{Travis Wright}, {and} \bibinfo{person}{Yiwen Zhu}.}
  \bibinfo{year}{2019}\natexlab{}.
\newblock \showarticletitle{Cloudy with high chance of {DBMS:} {A} 10-year
  prediction for Enterprise-Grade {ML}}.
\newblock \bibinfo{journal}{\emph{CoRR}} (\bibinfo{year}{2019}).
\newblock
\urldef\tempurl%
\url{http://arxiv.org/abs/1909.00084}
\showURL{%
\tempurl}


\bibitem[\protect\citeauthoryear{Amsterdamer, Deutch, and Tannen}{Amsterdamer
  et~al\mbox{.}}{2011}]%
        {aggprov}
\bibfield{author}{\bibinfo{person}{Yael Amsterdamer}, \bibinfo{person}{Daniel
  Deutch}, {and} \bibinfo{person}{Val Tannen}.}
  \bibinfo{year}{2011}\natexlab{}.
\newblock \showarticletitle{Provenance for Aggregate Queries}. In
  \bibinfo{booktitle}{\emph{Proceedings of the Thirtieth ACM
  SIGMOD-SIGACT-SIGART Symposium on Principles of Database Systems}} (Athens,
  Greece) \emph{(\bibinfo{series}{PODS ’11})}.
  \bibinfo{publisher}{Association for Computing Machinery},
  \bibinfo{address}{New York, NY, USA}, \bibinfo{pages}{153–164}.
\newblock
\showISBNx{9781450306607}
\urldef\tempurl%
\url{https://doi.org/10.1145/1989284.1989302}
\showDOI{\tempurl}


\bibitem[\protect\citeauthoryear{Baylor, Breck, Cheng, Fiedel, Foo, Haque,
  Haykal, Ispir, Jain, Koc, Koo, Lew, Mewald, Modi, Polyzotis, Ramesh, Roy,
  Whang, Wicke, Wilkiewicz, Zhang, and Zinkevich}{Baylor et~al\mbox{.}}{2017}]%
        {googletfxpaper}
\bibfield{author}{\bibinfo{person}{Denis Baylor}, \bibinfo{person}{Eric Breck},
  \bibinfo{person}{Heng-Tze Cheng}, \bibinfo{person}{Noah Fiedel},
  \bibinfo{person}{Chuan~Yu Foo}, \bibinfo{person}{Zakaria Haque},
  \bibinfo{person}{Salem Haykal}, \bibinfo{person}{Mustafa Ispir},
  \bibinfo{person}{Vihan Jain}, \bibinfo{person}{Levent Koc},
  \bibinfo{person}{Chiu~Yuen Koo}, \bibinfo{person}{Lukasz Lew},
  \bibinfo{person}{Clemens Mewald}, \bibinfo{person}{Akshay~Naresh Modi},
  \bibinfo{person}{Neoklis Polyzotis}, \bibinfo{person}{Sukriti Ramesh},
  \bibinfo{person}{Sudip Roy}, \bibinfo{person}{Steven~Euijong Whang},
  \bibinfo{person}{Martin Wicke}, \bibinfo{person}{Jarek Wilkiewicz},
  \bibinfo{person}{Xin Zhang}, {and} \bibinfo{person}{Martin Zinkevich}.}
  \bibinfo{year}{2017}\natexlab{}.
\newblock \showarticletitle{TFX: A TensorFlow-Based Production-Scale Machine
  Learning Platform}. In \bibinfo{booktitle}{\emph{Proceedings of the 23rd ACM
  SIGKDD International Conference on Knowledge Discovery and Data Mining}}
  (Halifax, NS, Canada) \emph{(\bibinfo{series}{KDD ’17})}.
  \bibinfo{publisher}{Association for Computing Machinery},
  \bibinfo{address}{New York, NY, USA}, \bibinfo{pages}{1387–1395}.
\newblock
\showISBNx{9781450348874}
\urldef\tempurl%
\url{https://doi.org/10.1145/3097983.3098021}
\showDOI{\tempurl}


\bibitem[\protect\citeauthoryear{Boehm, Dusenberry, Eriksson, Evfimievski,
  Manshadi, Pansare, Reinwald, Reiss, Sen, Surve, and Tatikonda}{Boehm
  et~al\mbox{.}}{2016}]%
        {systemml}
\bibfield{author}{\bibinfo{person}{Matthias Boehm}, \bibinfo{person}{Michael~W.
  Dusenberry}, \bibinfo{person}{Deron Eriksson}, \bibinfo{person}{Alexandre~V.
  Evfimievski}, \bibinfo{person}{Faraz~Makari Manshadi},
  \bibinfo{person}{Niketan Pansare}, \bibinfo{person}{Berthold Reinwald},
  \bibinfo{person}{Frederick~R. Reiss}, \bibinfo{person}{Prithviraj Sen},
  \bibinfo{person}{Arvind~C. Surve}, {and} \bibinfo{person}{Shirish
  Tatikonda}.} \bibinfo{year}{2016}\natexlab{}.
\newblock \showarticletitle{SystemML: Declarative Machine Learning on Spark}.
\newblock \bibinfo{journal}{\emph{Proc. VLDB Endow.}} \bibinfo{volume}{9},
  \bibinfo{number}{13} (\bibinfo{date}{Sept.} \bibinfo{year}{2016}),
  \bibinfo{pages}{1425–1436}.
\newblock
\showISSN{2150-8097}
\urldef\tempurl%
\url{https://doi.org/10.14778/3007263.3007279}
\showDOI{\tempurl}


\bibitem[\protect\citeauthoryear{Breck, Polyzotis, Roy, Whang, and
  Zinkevich}{Breck et~al\mbox{.}}{2019}]%
        {Breck2019DataVF}
\bibfield{author}{\bibinfo{person}{Eric Breck}, \bibinfo{person}{Neoklis
  Polyzotis}, \bibinfo{person}{Sudip Roy}, \bibinfo{person}{Steven~Euijong
  Whang}, {and} \bibinfo{person}{Martin Zinkevich}.}
  \bibinfo{year}{2019}\natexlab{}.
\newblock \bibinfo{title}{Data Validation for Machine Learning}.
\newblock
\newblock
\urldef\tempurl%
\url{https://mlsys.org/Conferences/2019/doc/2019/167.pdf}
\showURL{%
\tempurl}


\bibitem[\protect\citeauthoryear{Chapman and Jagadish}{Chapman and
  Jagadish}{2009}]%
        {Chapman1965WhyN}
\bibfield{author}{\bibinfo{person}{Adriane Chapman} {and}
  \bibinfo{person}{H.~V. Jagadish}.} \bibinfo{year}{2009}\natexlab{}.
\newblock \showarticletitle{Why Not?}. In \bibinfo{booktitle}{\emph{Proceedings
  of the 2009 ACM SIGMOD International Conference on Management of Data}}
  (Providence, Rhode Island, USA) \emph{(\bibinfo{series}{SIGMOD ’09})}.
  \bibinfo{publisher}{Association for Computing Machinery},
  \bibinfo{address}{New York, NY, USA}, \bibinfo{pages}{523–534}.
\newblock
\showISBNx{9781605585512}
\urldef\tempurl%
\url{https://doi.org/10.1145/1559845.1559901}
\showDOI{\tempurl}


\bibitem[\protect\citeauthoryear{Chu, Ilyas, and Papotti}{Chu
  et~al\mbox{.}}{2013}]%
        {chu2013holistic}
\bibfield{author}{\bibinfo{person}{Xu Chu}, \bibinfo{person}{Ihab~F Ilyas},
  {and} \bibinfo{person}{Paolo Papotti}.} \bibinfo{year}{2013}\natexlab{}.
\newblock \showarticletitle{Holistic data cleaning: Putting violations into
  context}. In \bibinfo{booktitle}{\emph{2013 IEEE 29th International
  Conference on Data Engineering (ICDE)}}. \bibinfo{publisher}{{IEEE}},
  \bibinfo{pages}{458--469}.
\newblock


\bibitem[\protect\citeauthoryear{Colyer}{Colyer}{2019}]%
        {Colyer2019PuttingML}
\bibfield{author}{\bibinfo{person}{Adrian Colyer}.}
  \bibinfo{year}{2019}\natexlab{}.
\newblock \showarticletitle{Putting Machine Learning into Production Systems}.
\newblock \bibinfo{journal}{\emph{Queue}} \bibinfo{volume}{17},
  \bibinfo{number}{4}, Article \bibinfo{articleno}{Pages 60}
  (\bibinfo{date}{Aug.} \bibinfo{year}{2019}), \bibinfo{numpages}{2}~pages.
\newblock
\showISSN{1542-7730}
\urldef\tempurl%
\url{https://doi.org/10.1145/3358955.3365847}
\showDOI{\tempurl}


\bibitem[\protect\citeauthoryear{Dalvi and Suciu}{Dalvi and Suciu}{2004}]%
        {probdb}
\bibfield{author}{\bibinfo{person}{Nilesh Dalvi} {and} \bibinfo{person}{Dan
  Suciu}.} \bibinfo{year}{2004}\natexlab{}.
\newblock \showarticletitle{Efficient Query Evaluation on Probabilistic
  Databases}. In \bibinfo{booktitle}{\emph{Proceedings of the Thirtieth
  International Conference on Very Large Data Bases - Volume 30}} (Toronto,
  Canada) \emph{(\bibinfo{series}{VLDB ’04})}. \bibinfo{publisher}{VLDB
  Endowment}, \bibinfo{pages}{864–875}.
\newblock
\showISBNx{0120884690}


\bibitem[\protect\citeauthoryear{Das, Doan, G.~C., Gokhale, Konda, Govind, and
  Paulsen}{Das et~al\mbox{.}}{[n.d.]}]%
        {magellandata}
\bibfield{author}{\bibinfo{person}{Sanjib Das}, \bibinfo{person}{AnHai Doan},
  \bibinfo{person}{Paul~Suganthan G.~C.}, \bibinfo{person}{Chaitanya Gokhale},
  \bibinfo{person}{Pradap Konda}, \bibinfo{person}{Yash Govind}, {and}
  \bibinfo{person}{Derek Paulsen}.} \bibinfo{year}{[n.d.]}\natexlab{}.
\newblock \bibinfo{title}{The Magellan Data Repository}.
\newblock
  \bibinfo{howpublished}{\url{https://sites.google.com/site/anhaidgroup/projects/data}}.
\newblock


\bibitem[\protect\citeauthoryear{Dolatshah, Teoh, Wang, and Pei}{Dolatshah
  et~al\mbox{.}}{2018}]%
        {dolatshah2018cleaning}
\bibfield{author}{\bibinfo{person}{Mohamad Dolatshah}, \bibinfo{person}{Mathew
  Teoh}, \bibinfo{person}{Jiannan Wang}, {and} \bibinfo{person}{Jian Pei}.}
  \bibinfo{year}{2018}\natexlab{}.
\newblock \showarticletitle{Cleaning Crowdsourced Labels Using Oracles for
  Statistical Classification}.
\newblock \bibinfo{journal}{\emph{Proc. VLDB Endow.}} \bibinfo{volume}{12},
  \bibinfo{number}{4} (\bibinfo{date}{Dec.} \bibinfo{year}{2018}),
  \bibinfo{pages}{376–389}.
\newblock
\showISSN{2150-8097}
\urldef\tempurl%
\url{https://doi.org/10.14778/3297753.3297758}
\showDOI{\tempurl}


\bibitem[\protect\citeauthoryear{du~Pin~Calmon, Wei, Vinzamuri, Ramamurthy, and
  Varshney}{du~Pin~Calmon et~al\mbox{.}}{2017}]%
        {discrimination}
\bibfield{author}{\bibinfo{person}{Fl{\'{a}}vio du Pin~Calmon},
  \bibinfo{person}{Dennis Wei}, \bibinfo{person}{Bhanukiran Vinzamuri},
  \bibinfo{person}{Karthikeyan~Natesan Ramamurthy}, {and}
  \bibinfo{person}{Kush~R. Varshney}.} \bibinfo{year}{2017}\natexlab{}.
\newblock \showarticletitle{Optimized Pre-Processing for Discrimination
  Prevention}. In \bibinfo{booktitle}{\emph{Advances in Neural Information
  Processing Systems 30}}. \bibinfo{pages}{3992--4001}.
\newblock
\urldef\tempurl%
\url{http://papers.nips.cc/paper/6988-optimized-pre-processing-for-discrimination-prevention}
\showURL{%
\tempurl}


\bibitem[\protect\citeauthoryear{Dua and Graff}{Dua and Graff}{2017}]%
        {uci}
\bibfield{author}{\bibinfo{person}{Dheeru Dua} {and} \bibinfo{person}{Casey
  Graff}.} \bibinfo{year}{2017}\natexlab{}.
\newblock \bibinfo{title}{{UCI} Machine Learning Repository}.
\newblock
\newblock
\urldef\tempurl%
\url{http://archive.ics.uci.edu/ml}
\showURL{%
\tempurl}


\bibitem[\protect\citeauthoryear{Exchange}{Exchange}{2019}]%
        {onnx}
\bibfield{author}{\bibinfo{person}{Open Neural~Network Exchange}.}
  \bibinfo{year}{2019}\natexlab{}.
\newblock \bibinfo{title}{ONNX}.
\newblock \bibinfo{howpublished}{\url{https://onnx.ai/}}.
\newblock
\newblock
\shownote{[Online; accessed 10-October-2019].}


\bibitem[\protect\citeauthoryear{Fr{\'e}nay and Verleysen}{Fr{\'e}nay and
  Verleysen}{2013}]%
        {frenay2013classification}
\bibfield{author}{\bibinfo{person}{Beno{\^\i}t Fr{\'e}nay} {and}
  \bibinfo{person}{Michel Verleysen}.} \bibinfo{year}{2013}\natexlab{}.
\newblock \showarticletitle{Classification in the presence of label noise: a
  survey}.
\newblock \bibinfo{journal}{\emph{IEEE transactions on neural networks and
  learning systems}} \bibinfo{volume}{25}, \bibinfo{number}{5}
  (\bibinfo{year}{2013}), \bibinfo{pages}{845--869}.
\newblock


\bibitem[\protect\citeauthoryear{Gilpin, Bau, Yuan, Bajwa, Specter, and
  Kagal}{Gilpin et~al\mbox{.}}{2018}]%
        {mlexplanationsurvey}
\bibfield{author}{\bibinfo{person}{Leilani~H. Gilpin}, \bibinfo{person}{David
  Bau}, \bibinfo{person}{Ben~Z. Yuan}, \bibinfo{person}{Ayesha Bajwa},
  \bibinfo{person}{Michael Specter}, {and} \bibinfo{person}{Lalana Kagal}.}
  \bibinfo{year}{2018}\natexlab{}.
\newblock \showarticletitle{Explaining Explanations: An Overview of
  Interpretability of Machine Learning}. In \bibinfo{booktitle}{\emph{5th
  {IEEE} International Conference on Data Science and Advanced Analytics,
  {DSAA} 2018, Turin, Italy, October 1-3, 2018}}. \bibinfo{publisher}{{IEEE}},
  \bibinfo{pages}{80--89}.
\newblock
\urldef\tempurl%
\url{https://doi.org/10.1109/DSAA.2018.00018}
\showDOI{\tempurl}


\bibitem[\protect\citeauthoryear{Giordano, Stephenson, Liu, Jordan, and
  Broderick}{Giordano et~al\mbox{.}}{2019}]%
        {swissarmy}
\bibfield{author}{\bibinfo{person}{Ryan Giordano}, \bibinfo{person}{William
  Stephenson}, \bibinfo{person}{Runjing Liu}, \bibinfo{person}{Michael Jordan},
  {and} \bibinfo{person}{Tamara Broderick}.} \bibinfo{year}{2019}\natexlab{}.
\newblock \showarticletitle{A Swiss Army Infinitesimal Jackknife}. In
  \bibinfo{booktitle}{\emph{Proceedings of Machine Learning Research}},
  Vol.~\bibinfo{volume}{89}. \bibinfo{publisher}{PMLR},
  \bibinfo{pages}{1139--1147}.
\newblock
\urldef\tempurl%
\url{http://proceedings.mlr.press/v89/giordano19a.html}
\showURL{%
\tempurl}


\bibitem[\protect\citeauthoryear{Green, Karvounarakis, and Tannen}{Green
  et~al\mbox{.}}{2007}]%
        {semirings}
\bibfield{author}{\bibinfo{person}{Todd~J. Green}, \bibinfo{person}{Grigoris
  Karvounarakis}, {and} \bibinfo{person}{Val Tannen}.}
  \bibinfo{year}{2007}\natexlab{}.
\newblock \showarticletitle{Provenance Semirings}. In
  \bibinfo{booktitle}{\emph{Proceedings of the Twenty-Sixth ACM
  SIGMOD-SIGACT-SIGART Symposium on Principles of Database Systems}} (Beijing,
  China) \emph{(\bibinfo{series}{PODS ’07})}. \bibinfo{publisher}{Association
  for Computing Machinery}, \bibinfo{address}{New York, NY, USA},
  \bibinfo{pages}{31–40}.
\newblock
\showISBNx{9781595936851}
\urldef\tempurl%
\url{https://doi.org/10.1145/1265530.1265535}
\showDOI{\tempurl}


\bibitem[\protect\citeauthoryear{Gurobi~Optimization}{Gurobi~Optimization}{2019}]%
        {gurobi}
\bibfield{author}{\bibinfo{person}{LLC Gurobi~Optimization}.}
  \bibinfo{year}{2019}\natexlab{}.
\newblock \bibinfo{title}{Gurobi Optimizer Reference Manual}.
\newblock
\newblock
\urldef\tempurl%
\url{http://www.gurobi.com}
\showURL{%
\tempurl}


\bibitem[\protect\citeauthoryear{Hara, Nitanda, and Maehara}{Hara
  et~al\mbox{.}}{2019}]%
        {sgdcleansing}
\bibfield{author}{\bibinfo{person}{Satoshi Hara}, \bibinfo{person}{Atsushi
  Nitanda}, {and} \bibinfo{person}{Takanori Maehara}.}
  \bibinfo{year}{2019}\natexlab{}.
\newblock \showarticletitle{Data Cleansing for Models Trained with SGD}.
\newblock In \bibinfo{booktitle}{\emph{Advances in Neural Information
  Processing Systems 32}}. \bibinfo{pages}{4213--4222}.
\newblock
\urldef\tempurl%
\url{http://papers.nips.cc/paper/8674-data-cleansing-for-models-trained-with-sgd.pdf}
\showURL{%
\tempurl}


\bibitem[\protect\citeauthoryear{Heidari, McGrath, Ilyas, and
  Rekatsinas}{Heidari et~al\mbox{.}}{2019}]%
        {heidari2019holodetect}
\bibfield{author}{\bibinfo{person}{Alireza Heidari}, \bibinfo{person}{Joshua
  McGrath}, \bibinfo{person}{Ihab~F. Ilyas}, {and} \bibinfo{person}{Theodoros
  Rekatsinas}.} \bibinfo{year}{2019}\natexlab{}.
\newblock \showarticletitle{HoloDetect: Few-Shot Learning for Error Detection}.
  In \bibinfo{booktitle}{\emph{Proceedings of the 2019 International Conference
  on Management of Data}} (Amsterdam, Netherlands)
  \emph{(\bibinfo{series}{SIGMOD ’19})}. \bibinfo{publisher}{Association for
  Computing Machinery}, \bibinfo{address}{New York, NY, USA},
  \bibinfo{pages}{829–846}.
\newblock
\showISBNx{9781450356435}
\urldef\tempurl%
\url{https://doi.org/10.1145/3299869.3319888}
\showDOI{\tempurl}


\bibitem[\protect\citeauthoryear{Hellerstein, R\'{e}, Schoppmann, Wang,
  Fratkin, Gorajek, Ng, Welton, Feng, Li, and Kumar}{Hellerstein
  et~al\mbox{.}}{2012}]%
        {madlib}
\bibfield{author}{\bibinfo{person}{Joseph~M. Hellerstein},
  \bibinfo{person}{Christoper R\'{e}}, \bibinfo{person}{Florian Schoppmann},
  \bibinfo{person}{Daisy~Zhe Wang}, \bibinfo{person}{Eugene Fratkin},
  \bibinfo{person}{Aleksander Gorajek}, \bibinfo{person}{Kee~Siong Ng},
  \bibinfo{person}{Caleb Welton}, \bibinfo{person}{Xixuan Feng},
  \bibinfo{person}{Kun Li}, {and} \bibinfo{person}{Arun Kumar}.}
  \bibinfo{year}{2012}\natexlab{}.
\newblock \showarticletitle{The MADlib Analytics Library: Or MAD Skills, the
  SQL}.
\newblock \bibinfo{journal}{\emph{Proc. VLDB Endow.}} \bibinfo{volume}{5},
  \bibinfo{number}{12} (\bibinfo{date}{Aug.} \bibinfo{year}{2012}),
  \bibinfo{pages}{1700–1711}.
\newblock
\showISSN{2150-8097}
\urldef\tempurl%
\url{https://doi.org/10.14778/2367502.2367510}
\showDOI{\tempurl}


\bibitem[\protect\citeauthoryear{ILOG}{ILOG}{2014}]%
        {ilog2014cplex}
\bibfield{author}{\bibinfo{person}{IBM ILOG}.} \bibinfo{year}{2014}\natexlab{}.
\newblock \showarticletitle{Cplex optimization studio}.
\newblock \bibinfo{journal}{\emph{URL: http://www-01. ibm.
  com/software/commerce/optimization/cplex-optimizer}} (\bibinfo{year}{2014}).
\newblock


\bibitem[\protect\citeauthoryear{Jankov, Luo, Yuan, Cai, Zou, Jermaine, and
  Gao}{Jankov et~al\mbox{.}}{2019}]%
        {Jankov2019DeclarativeRC}
\bibfield{author}{\bibinfo{person}{Dimitrije Jankov}, \bibinfo{person}{Shangyu
  Luo}, \bibinfo{person}{Binhang Yuan}, \bibinfo{person}{Zhuhua Cai},
  \bibinfo{person}{Jia Zou}, \bibinfo{person}{Chris Jermaine}, {and}
  \bibinfo{person}{Zekai~J. Gao}.} \bibinfo{year}{2019}\natexlab{}.
\newblock \showarticletitle{Declarative Recursive Computation on an RDBMS: Or,
  Why You Should Use a Database for Distributed Machine Learning}.
\newblock \bibinfo{journal}{\emph{Proc. VLDB Endow.}} \bibinfo{volume}{12},
  \bibinfo{number}{7} (\bibinfo{date}{March} \bibinfo{year}{2019}),
  \bibinfo{pages}{822–835}.
\newblock
\showISSN{2150-8097}
\urldef\tempurl%
\url{https://doi.org/10.14778/3317315.3317323}
\showDOI{\tempurl}


\bibitem[\protect\citeauthoryear{Kanagal, Li, and Deshpande}{Kanagal
  et~al\mbox{.}}{2011}]%
        {probdbsensitivity}
\bibfield{author}{\bibinfo{person}{Bhargav Kanagal}, \bibinfo{person}{Jian Li},
  {and} \bibinfo{person}{Amol Deshpande}.} \bibinfo{year}{2011}\natexlab{}.
\newblock \showarticletitle{Sensitivity Analysis and Explanations for Robust
  Query Evaluation in Probabilistic Databases}. In
  \bibinfo{booktitle}{\emph{Proceedings of the 2011 ACM SIGMOD International
  Conference on Management of Data}} (Athens, Greece)
  \emph{(\bibinfo{series}{SIGMOD ’11})}. \bibinfo{publisher}{Association for
  Computing Machinery}, \bibinfo{address}{New York, NY, USA},
  \bibinfo{pages}{841–852}.
\newblock
\showISBNx{9781450306614}
\urldef\tempurl%
\url{https://doi.org/10.1145/1989323.1989411}
\showDOI{\tempurl}


\bibitem[\protect\citeauthoryear{Kang, Raghavan, Bailis, and Zaharia}{Kang
  et~al\mbox{.}}{2020}]%
        {Kang2018ModelAF}
\bibfield{author}{\bibinfo{person}{Daniel Kang}, \bibinfo{person}{Deepti
  Raghavan}, \bibinfo{person}{Peter Bailis}, {and} \bibinfo{person}{Matei
  Zaharia}.} \bibinfo{year}{2020}\natexlab{}.
\newblock \showarticletitle{Model Assertions for Monitoring and Improving ML
  Models}.
\newblock In \bibinfo{booktitle}{\emph{Proceedings of Machine Learning and
  Systems 2020}}. \bibinfo{pages}{481--496}.
\newblock


\bibitem[\protect\citeauthoryear{Kantchelian, Tygar, and Joseph}{Kantchelian
  et~al\mbox{.}}{2016}]%
        {Kantchelian2015EvasionAH}
\bibfield{author}{\bibinfo{person}{Alex Kantchelian}, \bibinfo{person}{J.~D.
  Tygar}, {and} \bibinfo{person}{Anthony Joseph}.}
  \bibinfo{year}{2016}\natexlab{}.
\newblock \showarticletitle{Evasion and Hardening of Tree Ensemble
  Classifiers}. In \bibinfo{booktitle}{\emph{Proceedings of The 33rd
  International Conference on Machine Learning}}, Vol.~\bibinfo{volume}{48}.
  \bibinfo{publisher}{PMLR}, \bibinfo{address}{New York, New York, USA},
  \bibinfo{pages}{2387--2396}.
\newblock
\urldef\tempurl%
\url{http://proceedings.mlr.press/v48/kantchelian16.html}
\showURL{%
\tempurl}


\bibitem[\protect\citeauthoryear{Karpathy}{Karpathy}{2017}]%
        {karpathy}
\bibfield{author}{\bibinfo{person}{Andrej Karpathy}.}
  \bibinfo{year}{2017}\natexlab{}.
\newblock \bibinfo{title}{Software 2.0}.
\newblock
  \bibinfo{howpublished}{\url{https://medium.com/@karpathy/software-2-0-a64152b37c35}}.
\newblock
\newblock
\shownote{[Online; accessed 10-October-2019].}


\bibitem[\protect\citeauthoryear{Khanna, Kim, Ghosh, and Koyejo}{Khanna
  et~al\mbox{.}}{2019}]%
        {fisher}
\bibfield{author}{\bibinfo{person}{Rajiv Khanna}, \bibinfo{person}{Been Kim},
  \bibinfo{person}{Joydeep Ghosh}, {and} \bibinfo{person}{Sanmi Koyejo}.}
  \bibinfo{year}{2019}\natexlab{}.
\newblock \showarticletitle{Interpreting Black Box Predictions using Fisher
  Kernels}. In \bibinfo{booktitle}{\emph{Proceedings of Machine Learning
  Research}}, Vol.~\bibinfo{volume}{89}. \bibinfo{publisher}{PMLR},
  \bibinfo{pages}{3382--3390}.
\newblock
\urldef\tempurl%
\url{http://proceedings.mlr.press/v89/khanna19a.html}
\showURL{%
\tempurl}


\bibitem[\protect\citeauthoryear{Koh, Ang, Teo, and Liang}{Koh
  et~al\mbox{.}}{2019}]%
        {influencegroup}
\bibfield{author}{\bibinfo{person}{Pang~Wei Koh}, \bibinfo{person}{Kai{-}Siang
  Ang}, \bibinfo{person}{Hubert H.~K. Teo}, {and} \bibinfo{person}{Percy
  Liang}.} \bibinfo{year}{2019}\natexlab{}.
\newblock \showarticletitle{On the Accuracy of Influence Functions for
  Measuring Group Effects}. In \bibinfo{booktitle}{\emph{Advances in Neural
  Information Processing Systems 32}}. \bibinfo{pages}{5255--5265}.
\newblock
\urldef\tempurl%
\url{http://papers.nips.cc/paper/8767-on-the-accuracy-of-influence-functions-for-measuring-group-effects}
\showURL{%
\tempurl}


\bibitem[\protect\citeauthoryear{Koh and Liang}{Koh and Liang}{2017}]%
        {influence}
\bibfield{author}{\bibinfo{person}{Pang~Wei Koh} {and} \bibinfo{person}{Percy
  Liang}.} \bibinfo{year}{2017}\natexlab{}.
\newblock \showarticletitle{Understanding Black-box Predictions via Influence
  Functions}. In \bibinfo{booktitle}{\emph{Proceedings of the 34th
  International Conference on Machine Learning}}, Vol.~\bibinfo{volume}{70}.
  \bibinfo{publisher}{PMLR}, \bibinfo{address}{International Convention Centre,
  Sydney, Australia}, \bibinfo{pages}{1885--1894}.
\newblock
\urldef\tempurl%
\url{http://proceedings.mlr.press/v70/koh17a.html}
\showURL{%
\tempurl}


\bibitem[\protect\citeauthoryear{Konda, Das, Suganthan G.~C., Doan, Ardalan,
  Ballard, Li, Panahi, Zhang, Naughton, Prasad, Krishnan, Deep, and
  Raghavendra}{Konda et~al\mbox{.}}{2016}]%
        {dblpgoog}
\bibfield{author}{\bibinfo{person}{Pradap Konda}, \bibinfo{person}{Sanjib Das},
  \bibinfo{person}{Paul Suganthan G.~C.}, \bibinfo{person}{AnHai Doan},
  \bibinfo{person}{Adel Ardalan}, \bibinfo{person}{Jeffrey~R. Ballard},
  \bibinfo{person}{Han Li}, \bibinfo{person}{Fatemah Panahi},
  \bibinfo{person}{Haojun Zhang}, \bibinfo{person}{Jeff Naughton},
  \bibinfo{person}{Shishir Prasad}, \bibinfo{person}{Ganesh Krishnan},
  \bibinfo{person}{Rohit Deep}, {and} \bibinfo{person}{Vijay Raghavendra}.}
  \bibinfo{year}{2016}\natexlab{}.
\newblock \showarticletitle{Magellan: Toward Building Entity Matching
  Management Systems}.
\newblock \bibinfo{journal}{\emph{Proc. VLDB Endow.}} \bibinfo{volume}{9},
  \bibinfo{number}{12} (\bibinfo{date}{Aug.} \bibinfo{year}{2016}),
  \bibinfo{pages}{1197–1208}.
\newblock
\showISSN{2150-8097}
\urldef\tempurl%
\url{https://doi.org/10.14778/2994509.2994535}
\showDOI{\tempurl}


\bibitem[\protect\citeauthoryear{Kraska, Talwalkar, Duchi, Griffith, Franklin,
  and Jordan}{Kraska et~al\mbox{.}}{2013}]%
        {mlbase}
\bibfield{author}{\bibinfo{person}{Tim Kraska}, \bibinfo{person}{Ameet
  Talwalkar}, \bibinfo{person}{John~C. Duchi}, \bibinfo{person}{Rean Griffith},
  \bibinfo{person}{Michael~J. Franklin}, {and} \bibinfo{person}{Michael~I.
  Jordan}.} \bibinfo{year}{2013}\natexlab{}.
\newblock \showarticletitle{MLbase: {A} Distributed Machine-learning System}.
  In \bibinfo{booktitle}{\emph{{CIDR} 2013, Sixth Biennial Conference on
  Innovative Data Systems Research, Asilomar, CA, USA, January 6-9, 2013,
  Online Proceedings}}. \bibinfo{publisher}{www.cidrdb.org}.
\newblock
\urldef\tempurl%
\url{http://cidrdb.org/cidr2013/Papers/CIDR13\_Paper118.pdf}
\showURL{%
\tempurl}


\bibitem[\protect\citeauthoryear{Krishnan, Franklin, Goldberg, and Wu}{Krishnan
  et~al\mbox{.}}{2017}]%
        {boostclean}
\bibfield{author}{\bibinfo{person}{Sanjay Krishnan},
  \bibinfo{person}{Michael~J. Franklin}, \bibinfo{person}{Ken Goldberg}, {and}
  \bibinfo{person}{Eugene Wu}.} \bibinfo{year}{2017}\natexlab{}.
\newblock \showarticletitle{BoostClean: Automated Error Detection and Repair
  for Machine Learning}.
\newblock \bibinfo{journal}{\emph{CoRR}} (\bibinfo{year}{2017}).
\newblock
\urldef\tempurl%
\url{http://arxiv.org/abs/1711.01299}
\showURL{%
\tempurl}


\bibitem[\protect\citeauthoryear{Krishnan, Wang, Wu, Franklin, and
  Goldberg}{Krishnan et~al\mbox{.}}{2016}]%
        {activeclean}
\bibfield{author}{\bibinfo{person}{Sanjay Krishnan}, \bibinfo{person}{Jiannan
  Wang}, \bibinfo{person}{Eugene Wu}, \bibinfo{person}{Michael~J. Franklin},
  {and} \bibinfo{person}{Ken Goldberg}.} \bibinfo{year}{2016}\natexlab{}.
\newblock \showarticletitle{ActiveClean: Interactive Data Cleaning for
  Statistical Modeling}.
\newblock \bibinfo{journal}{\emph{Proc. VLDB Endow.}} \bibinfo{volume}{9},
  \bibinfo{number}{12} (\bibinfo{date}{Aug.} \bibinfo{year}{2016}),
  \bibinfo{pages}{948–959}.
\newblock
\showISSN{2150-8097}
\urldef\tempurl%
\url{https://doi.org/10.14778/2994509.2994514}
\showDOI{\tempurl}


\bibitem[\protect\citeauthoryear{Krishnan and Wu}{Krishnan and Wu}{2017}]%
        {Krishnan2017PALMML}
\bibfield{author}{\bibinfo{person}{Sanjay Krishnan} {and}
  \bibinfo{person}{Eugene Wu}.} \bibinfo{year}{2017}\natexlab{}.
\newblock \showarticletitle{PALM: Machine Learning Explanations For Iterative
  Debugging}. In \bibinfo{booktitle}{\emph{HILDA@SIGMOD}}.
\newblock


\bibitem[\protect\citeauthoryear{Krishnan and Wu}{Krishnan and Wu}{2019}]%
        {alphaclean}
\bibfield{author}{\bibinfo{person}{Sanjay Krishnan} {and}
  \bibinfo{person}{Eugene Wu}.} \bibinfo{year}{2019}\natexlab{}.
\newblock \showarticletitle{AlphaClean: Automatic Generation of Data Cleaning
  Pipelines}.
\newblock \bibinfo{journal}{\emph{CoRR}} (\bibinfo{year}{2019}).
\newblock
\urldef\tempurl%
\url{http://arxiv.org/abs/1904.11827}
\showURL{%
\tempurl}


\bibitem[\protect\citeauthoryear{Lakkaraju, Bach, and Leskovec}{Lakkaraju
  et~al\mbox{.}}{2016}]%
        {decissionsets}
\bibfield{author}{\bibinfo{person}{Himabindu Lakkaraju},
  \bibinfo{person}{Stephen~H. Bach}, {and} \bibinfo{person}{Jure Leskovec}.}
  \bibinfo{year}{2016}\natexlab{}.
\newblock \showarticletitle{Interpretable Decision Sets: {A} Joint Framework
  for Description and Prediction}. In \bibinfo{booktitle}{\emph{Proceedings of
  the 22nd {ACM} {SIGKDD} International Conference on Knowledge Discovery and
  Data Mining, San Francisco, CA, USA, August 13-17, 2016}}.
  \bibinfo{publisher}{{ACM}}, \bibinfo{pages}{1675--1684}.
\newblock
\urldef\tempurl%
\url{https://doi.org/10.1145/2939672.2939874}
\showDOI{\tempurl}


\bibitem[\protect\citeauthoryear{LeCun, Cortes, and J.C.~Burges}{LeCun
  et~al\mbox{.}}{2010}]%
        {mnist}
\bibfield{author}{\bibinfo{person}{Yann LeCun}, \bibinfo{person}{Corinna
  Cortes}, {and} \bibinfo{person}{Christopher J.C.~Burges}.}
  \bibinfo{year}{2010}\natexlab{}.
\newblock \bibinfo{title}{{MNIST} handwritten digit database}.
\newblock \bibinfo{howpublished}{http://yann.lecun.com/exdb/mnist/}.
\newblock
\urldef\tempurl%
\url{http://yann.lecun.com/exdb/mnist/}
\showURL{%
\tempurl}


\bibitem[\protect\citeauthoryear{Li, Rao, Blase, Zhang, Chu, and Zhang}{Li
  et~al\mbox{.}}{2019b}]%
        {Li2019CleanMLAB}
\bibfield{author}{\bibinfo{person}{Peng Li}, \bibinfo{person}{Xi Rao},
  \bibinfo{person}{Jennifer Blase}, \bibinfo{person}{Yue Zhang},
  \bibinfo{person}{Xu Chu}, {and} \bibinfo{person}{Ce Zhang}.}
  \bibinfo{year}{2019}\natexlab{b}.
\newblock \showarticletitle{CleanML: A Benchmark for Joint Data Cleaning and
  Machine Learning [Experiments and Analysis]}.
\newblock \bibinfo{journal}{\emph{CoRR}} (\bibinfo{year}{2019}).
\newblock
\urldef\tempurl%
\url{http://arxiv.org/abs/1904.09483}
\showURL{%
\tempurl}


\bibitem[\protect\citeauthoryear{Li, Feng, Li, Mumick, Halevy, Li, and Tan}{Li
  et~al\mbox{.}}{2019a}]%
        {subjectivedb}
\bibfield{author}{\bibinfo{person}{Yuliang Li}, \bibinfo{person}{Aaron Feng},
  \bibinfo{person}{Jinfeng Li}, \bibinfo{person}{Saran Mumick},
  \bibinfo{person}{Alon Halevy}, \bibinfo{person}{Vivian Li}, {and}
  \bibinfo{person}{Wang-Chiew Tan}.} \bibinfo{year}{2019}\natexlab{a}.
\newblock \showarticletitle{Subjective Databases}.
\newblock \bibinfo{journal}{\emph{Proc. VLDB Endow.}} \bibinfo{volume}{12},
  \bibinfo{number}{11} (\bibinfo{date}{July} \bibinfo{year}{2019}),
  \bibinfo{pages}{1330–1343}.
\newblock
\showISSN{2150-8097}
\urldef\tempurl%
\url{https://doi.org/10.14778/3342263.3342271}
\showDOI{\tempurl}


\bibitem[\protect\citeauthoryear{LLC}{LLC}{2019}]%
        {bigqueryml}
\bibfield{author}{\bibinfo{person}{Google LLC}.}
  \bibinfo{year}{2019}\natexlab{}.
\newblock \bibinfo{title}{Introduction to BigQuery ML}.
\newblock
  \bibinfo{howpublished}{\url{https://cloud.google.com/bigquery-ml/docs/bigqueryml-intro}}.
\newblock
\newblock
\shownote{[Online; accessed 10-October-2019].}


\bibitem[\protect\citeauthoryear{Logicblox}{Logicblox}{2019}]%
        {logicblox}
\bibfield{author}{\bibinfo{person}{Logicblox}.}
  \bibinfo{year}{2019}\natexlab{}.
\newblock \bibinfo{title}{LogicBlox – Next Generation Analytics
  Applications}.
\newblock \bibinfo{howpublished}{\url{https://logicblox.com}}.
\newblock
\newblock
\shownote{[Online; accessed 10-October-2019].}


\bibitem[\protect\citeauthoryear{Lu, Chowdhery, Kandula, and Chaudhuri}{Lu
  et~al\mbox{.}}{2018}]%
        {lu2018accelerating}
\bibfield{author}{\bibinfo{person}{Yao Lu}, \bibinfo{person}{Aakanksha
  Chowdhery}, \bibinfo{person}{Srikanth Kandula}, {and}
  \bibinfo{person}{Surajit Chaudhuri}.} \bibinfo{year}{2018}\natexlab{}.
\newblock \showarticletitle{Accelerating Machine Learning Inference with
  Probabilistic Predicates}. In \bibinfo{booktitle}{\emph{Proceedings of the
  2018 International Conference on Management of Data}} (Houston, TX, USA)
  \emph{(\bibinfo{series}{SIGMOD ’18})}. \bibinfo{publisher}{Association for
  Computing Machinery}, \bibinfo{address}{New York, NY, USA},
  \bibinfo{pages}{1493–1508}.
\newblock
\showISBNx{9781450347037}
\urldef\tempurl%
\url{https://doi.org/10.1145/3183713.3183751}
\showDOI{\tempurl}


\bibitem[\protect\citeauthoryear{Lundberg and Lee}{Lundberg and Lee}{2017}]%
        {shap}
\bibfield{author}{\bibinfo{person}{Scott~M. Lundberg} {and}
  \bibinfo{person}{Su{-}In Lee}.} \bibinfo{year}{2017}\natexlab{}.
\newblock \showarticletitle{A Unified Approach to Interpreting Model
  Predictions}. In \bibinfo{booktitle}{\emph{Advances in Neural Information
  Processing Systems 30}}. \bibinfo{pages}{4765--4774}.
\newblock
\urldef\tempurl%
\url{http://papers.nips.cc/paper/7062-a-unified-approach-to-interpreting-model-predictions}
\showURL{%
\tempurl}


\bibitem[\protect\citeauthoryear{Mahdavi, Abedjan, Castro~Fernandez, Madden,
  Ouzzani, Stonebraker, and Tang}{Mahdavi et~al\mbox{.}}{2019}]%
        {Mahdavi2019RahaAC}
\bibfield{author}{\bibinfo{person}{Mohammad Mahdavi}, \bibinfo{person}{Ziawasch
  Abedjan}, \bibinfo{person}{Raul Castro~Fernandez}, \bibinfo{person}{Samuel
  Madden}, \bibinfo{person}{Mourad Ouzzani}, \bibinfo{person}{Michael
  Stonebraker}, {and} \bibinfo{person}{Nan Tang}.}
  \bibinfo{year}{2019}\natexlab{}.
\newblock \showarticletitle{Raha: A Configuration-Free Error Detection System}.
  In \bibinfo{booktitle}{\emph{Proceedings of the 2019 International Conference
  on Management of Data}} (Amsterdam, Netherlands)
  \emph{(\bibinfo{series}{SIGMOD ’19})}. \bibinfo{publisher}{Association for
  Computing Machinery}, \bibinfo{address}{New York, NY, USA},
  \bibinfo{pages}{865–882}.
\newblock
\showISBNx{9781450356435}
\urldef\tempurl%
\url{https://doi.org/10.1145/3299869.3324956}
\showDOI{\tempurl}


\bibitem[\protect\citeauthoryear{Martens}{Martens}{2010}]%
        {hessianfree}
\bibfield{author}{\bibinfo{person}{James Martens}.}
  \bibinfo{year}{2010}\natexlab{}.
\newblock \showarticletitle{Deep learning via Hessian-free optimization}. In
  \bibinfo{booktitle}{\emph{Proceedings of the 27th International Conference on
  Machine Learning (ICML-10)}}. \bibinfo{publisher}{Omnipress},
  \bibinfo{address}{Haifa, Israel}, \bibinfo{pages}{735--742}.
\newblock
\urldef\tempurl%
\url{http://www.icml2010.org/papers/458.pdf}
\showURL{%
\tempurl}


\bibitem[\protect\citeauthoryear{Meliou, Gatterbauer, and Suciu}{Meliou
  et~al\mbox{.}}{2011}]%
        {Meliou2011ReverseDM}
\bibfield{author}{\bibinfo{person}{Alexandra Meliou}, \bibinfo{person}{Wolfgang
  Gatterbauer}, {and} \bibinfo{person}{Dan Suciu}.}
  \bibinfo{year}{2011}\natexlab{}.
\newblock \showarticletitle{Reverse Data Management}.
\newblock \bibinfo{journal}{\emph{Proc. VLDB Endow.}} \bibinfo{volume}{4},
  \bibinfo{number}{12} (\bibinfo{year}{2011}), \bibinfo{pages}{1490--1493}.
\newblock
\urldef\tempurl%
\url{http://www.vldb.org/pvldb/vol4/p1490-meliou.pdf}
\showURL{%
\tempurl}


\bibitem[\protect\citeauthoryear{Meliou, Roy, and Suciu}{Meliou
  et~al\mbox{.}}{2014}]%
        {Meliou2014CausalityAE}
\bibfield{author}{\bibinfo{person}{Alexandra Meliou}, \bibinfo{person}{Sudeepa
  Roy}, {and} \bibinfo{person}{Dan Suciu}.} \bibinfo{year}{2014}\natexlab{}.
\newblock \showarticletitle{Causality and Explanations in Databases}.
\newblock \bibinfo{journal}{\emph{Proc. VLDB Endow.}} \bibinfo{volume}{7},
  \bibinfo{number}{13} (\bibinfo{date}{Aug.} \bibinfo{year}{2014}),
  \bibinfo{pages}{1715–1716}.
\newblock
\showISSN{2150-8097}
\urldef\tempurl%
\url{https://doi.org/10.14778/2733004.2733070}
\showDOI{\tempurl}


\bibitem[\protect\citeauthoryear{Meliou and Suciu}{Meliou and Suciu}{2012}]%
        {tiresias}
\bibfield{author}{\bibinfo{person}{Alexandra Meliou} {and} \bibinfo{person}{Dan
  Suciu}.} \bibinfo{year}{2012}\natexlab{}.
\newblock \showarticletitle{Tiresias: The Database Oracle for How-to Queries}.
  In \bibinfo{booktitle}{\emph{Proceedings of the 2012 ACM SIGMOD International
  Conference on Management of Data}} (Scottsdale, Arizona, USA)
  \emph{(\bibinfo{series}{SIGMOD ’12})}. \bibinfo{publisher}{Association for
  Computing Machinery}, \bibinfo{address}{New York, NY, USA},
  \bibinfo{pages}{337–348}.
\newblock
\showISBNx{9781450312479}
\urldef\tempurl%
\url{https://doi.org/10.1145/2213836.2213875}
\showDOI{\tempurl}


\bibitem[\protect\citeauthoryear{Metsis, Androutsopoulos, and Paliouras}{Metsis
  et~al\mbox{.}}{2006}]%
        {enron}
\bibfield{author}{\bibinfo{person}{Vangelis Metsis}, \bibinfo{person}{Ion
  Androutsopoulos}, {and} \bibinfo{person}{Georgios Paliouras}.}
  \bibinfo{year}{2006}\natexlab{}.
\newblock \showarticletitle{Spam Filtering with Naive Bayes - Which Naive
  Bayes?}. In \bibinfo{booktitle}{\emph{{CEAS} 2006 - The Third Conference on
  Email and Anti-Spam, July 27-28, 2006, Mountain View, California, {USA}}}.
\newblock
\urldef\tempurl%
\url{http://www.ceas.cc/2006/listabs.html\#15.pdf}
\showURL{%
\tempurl}


\bibitem[\protect\citeauthoryear{Nilforoshan and Wu}{Nilforoshan and
  Wu}{2017}]%
        {Nilforoshan2017LeveragingQP}
\bibfield{author}{\bibinfo{person}{Hamed Nilforoshan} {and}
  \bibinfo{person}{Eugene Wu}.} \bibinfo{year}{2017}\natexlab{}.
\newblock \showarticletitle{Leveraging Quality Prediction Models for Automatic
  Writing Feedback}. In \bibinfo{booktitle}{\emph{ICWSM}}.
\newblock


\bibitem[\protect\citeauthoryear{Polyzotis, Roy, Whang, and
  Zinkevich}{Polyzotis et~al\mbox{.}}{2017}]%
        {Polyzotis2017DataMC}
\bibfield{author}{\bibinfo{person}{Neoklis Polyzotis}, \bibinfo{person}{Sudip
  Roy}, \bibinfo{person}{Steven~Euijong Whang}, {and} \bibinfo{person}{Martin
  Zinkevich}.} \bibinfo{year}{2017}\natexlab{}.
\newblock \showarticletitle{Data Management Challenges in Production Machine
  Learning}. In \bibinfo{booktitle}{\emph{Proceedings of the 2017 ACM
  International Conference on Management of Data}} (Chicago, Illinois, USA)
  \emph{(\bibinfo{series}{SIGMOD ’17})}. \bibinfo{publisher}{Association for
  Computing Machinery}, \bibinfo{address}{New York, NY, USA},
  \bibinfo{pages}{1723–1726}.
\newblock
\showISBNx{9781450341974}
\urldef\tempurl%
\url{https://doi.org/10.1145/3035918.3054782}
\showDOI{\tempurl}


\bibitem[\protect\citeauthoryear{Rahm and Do}{Rahm and Do}{2000}]%
        {Rahm2000DataCP}
\bibfield{author}{\bibinfo{person}{Erhard Rahm} {and} \bibinfo{person}{Hong~Hai
  Do}.} \bibinfo{year}{2000}\natexlab{}.
\newblock \showarticletitle{Data Cleaning: Problems and Current Approaches}.
\newblock \bibinfo{journal}{\emph{IEEE Data Eng. Bull.}}  \bibinfo{volume}{23}
  (\bibinfo{year}{2000}), \bibinfo{pages}{3--13}.
\newblock


\bibitem[\protect\citeauthoryear{Ratner, Bach, Ehrenberg, and R\'{e}}{Ratner
  et~al\mbox{.}}{2017}]%
        {Ratner2017SnorkelFT}
\bibfield{author}{\bibinfo{person}{Alexander~J. Ratner},
  \bibinfo{person}{Stephen~H. Bach}, \bibinfo{person}{Henry~R. Ehrenberg},
  {and} \bibinfo{person}{Chris R\'{e}}.} \bibinfo{year}{2017}\natexlab{}.
\newblock \showarticletitle{Snorkel: Fast Training Set Generation for
  Information Extraction}. In \bibinfo{booktitle}{\emph{Proceedings of the 2017
  ACM International Conference on Management of Data}} (Chicago, Illinois, USA)
  \emph{(\bibinfo{series}{SIGMOD ’17})}. \bibinfo{publisher}{Association for
  Computing Machinery}, \bibinfo{address}{New York, NY, USA},
  \bibinfo{pages}{1683–1686}.
\newblock
\showISBNx{9781450341974}
\urldef\tempurl%
\url{https://doi.org/10.1145/3035918.3056442}
\showDOI{\tempurl}


\bibitem[\protect\citeauthoryear{R{\'{e}}, Niu, Gudipati, and
  Srisuwananukorn}{R{\'{e}} et~al\mbox{.}}{2019}]%
        {overton}
\bibfield{author}{\bibinfo{person}{Christopher R{\'{e}}}, \bibinfo{person}{Feng
  Niu}, \bibinfo{person}{Pallavi Gudipati}, {and} \bibinfo{person}{Charles
  Srisuwananukorn}.} \bibinfo{year}{2019}\natexlab{}.
\newblock \showarticletitle{Overton: {A} Data System for Monitoring and
  Improving Machine-Learned Products}.
\newblock \bibinfo{journal}{\emph{CoRR}} (\bibinfo{year}{2019}).
\newblock
\urldef\tempurl%
\url{http://arxiv.org/abs/1909.05372}
\showURL{%
\tempurl}


\bibitem[\protect\citeauthoryear{Rekatsinas, Chu, Ilyas, and R{\'e}}{Rekatsinas
  et~al\mbox{.}}{2017}]%
        {Rekatsinas2017HoloCleanHD}
\bibfield{author}{\bibinfo{person}{Theodoros Rekatsinas}, \bibinfo{person}{Xu
  Chu}, \bibinfo{person}{Ihab~F. Ilyas}, {and} \bibinfo{person}{Christopher
  R{\'e}}.} \bibinfo{year}{2017}\natexlab{}.
\newblock \showarticletitle{HoloClean: Holistic Data Repairs with Probabilistic
  Inference}.
\newblock \bibinfo{journal}{\emph{Proc. VLDB Endow.}}  \bibinfo{volume}{10}
  (\bibinfo{year}{2017}), \bibinfo{pages}{1190--1201}.
\newblock


\bibitem[\protect\citeauthoryear{Ribeiro, Singh, and Guestrin}{Ribeiro
  et~al\mbox{.}}{2016}]%
        {lime}
\bibfield{author}{\bibinfo{person}{Marco~Tulio Ribeiro},
  \bibinfo{person}{Sameer Singh}, {and} \bibinfo{person}{Carlos Guestrin}.}
  \bibinfo{year}{2016}\natexlab{}.
\newblock \showarticletitle{“Why Should I Trust You?”: Explaining the
  Predictions of Any Classifier}. In \bibinfo{booktitle}{\emph{Proceedings of
  the 22nd ACM SIGKDD International Conference on Knowledge Discovery and Data
  Mining}} (San Francisco, California, USA) \emph{(\bibinfo{series}{KDD
  ’16})}. \bibinfo{publisher}{Association for Computing Machinery},
  \bibinfo{address}{New York, NY, USA}, \bibinfo{pages}{1135–1144}.
\newblock
\showISBNx{9781450342322}
\urldef\tempurl%
\url{https://doi.org/10.1145/2939672.2939778}
\showDOI{\tempurl}


\bibitem[\protect\citeauthoryear{Ribeiro, Singh, and Guestrin}{Ribeiro
  et~al\mbox{.}}{2018}]%
        {anchors}
\bibfield{author}{\bibinfo{person}{Marco~T{\'{u}}lio Ribeiro},
  \bibinfo{person}{Sameer Singh}, {and} \bibinfo{person}{Carlos Guestrin}.}
  \bibinfo{year}{2018}\natexlab{}.
\newblock \showarticletitle{Anchors: High-Precision Model-Agnostic
  Explanations}. In \bibinfo{booktitle}{\emph{Proceedings of the Thirty-Second
  {AAAI} Conference on Artificial Intelligence}}. \bibinfo{publisher}{{AAAI}
  Press}, \bibinfo{pages}{1527--1535}.
\newblock
\urldef\tempurl%
\url{https://www.aaai.org/ocs/index.php/AAAI/AAAI18/paper/view/16982}
\showURL{%
\tempurl}


\bibitem[\protect\citeauthoryear{Roy, Orr, and Suciu}{Roy
  et~al\mbox{.}}{2015}]%
        {roy}
\bibfield{author}{\bibinfo{person}{Sudeepa Roy}, \bibinfo{person}{Laurel Orr},
  {and} \bibinfo{person}{Dan Suciu}.} \bibinfo{year}{2015}\natexlab{}.
\newblock \showarticletitle{Explaining Query Answers with Explanation-Ready
  Databases}.
\newblock \bibinfo{journal}{\emph{Proc. VLDB Endow.}} \bibinfo{volume}{9},
  \bibinfo{number}{4} (\bibinfo{date}{Dec.} \bibinfo{year}{2015}),
  \bibinfo{pages}{348–359}.
\newblock
\showISSN{2150-8097}
\urldef\tempurl%
\url{https://doi.org/10.14778/2856318.2856329}
\showDOI{\tempurl}


\bibitem[\protect\citeauthoryear{Roy and Suciu}{Roy and Suciu}{2014}]%
        {Roy2014AFA}
\bibfield{author}{\bibinfo{person}{Sudeepa Roy} {and} \bibinfo{person}{Dan
  Suciu}.} \bibinfo{year}{2014}\natexlab{}.
\newblock \showarticletitle{A Formal Approach to Finding Explanations for
  Database Queries}. In \bibinfo{booktitle}{\emph{Proceedings of the 2014 ACM
  SIGMOD International Conference on Management of Data}} (Snowbird, Utah, USA)
  \emph{(\bibinfo{series}{SIGMOD ’14})}. \bibinfo{publisher}{Association for
  Computing Machinery}, \bibinfo{address}{New York, NY, USA},
  \bibinfo{pages}{1579–1590}.
\newblock
\showISBNx{9781450323765}
\urldef\tempurl%
\url{https://doi.org/10.1145/2588555.2588578}
\showDOI{\tempurl}


\bibitem[\protect\citeauthoryear{Salimi, Cole, Li, Gehrke, and Suciu}{Salimi
  et~al\mbox{.}}{2018}]%
        {Salimi2018HypDBAD}
\bibfield{author}{\bibinfo{person}{Babak Salimi}, \bibinfo{person}{Corey Cole},
  \bibinfo{person}{Peter Li}, \bibinfo{person}{Johannes Gehrke}, {and}
  \bibinfo{person}{Dan Suciu}.} \bibinfo{year}{2018}\natexlab{}.
\newblock \showarticletitle{HypDB: A Demonstration of Detecting, Explaining and
  Resolving Bias in OLAP Queries}.
\newblock \bibinfo{journal}{\emph{Proc. VLDB Endow.}} \bibinfo{volume}{11},
  \bibinfo{number}{12} (\bibinfo{date}{Aug.} \bibinfo{year}{2018}),
  \bibinfo{pages}{2062–2065}.
\newblock
\showISSN{2150-8097}
\urldef\tempurl%
\url{https://doi.org/10.14778/3229863.3236260}
\showDOI{\tempurl}


\bibitem[\protect\citeauthoryear{Salimi, Rodriguez, Howe, and Suciu}{Salimi
  et~al\mbox{.}}{2019}]%
        {capuchin}
\bibfield{author}{\bibinfo{person}{Babak Salimi}, \bibinfo{person}{Luke
  Rodriguez}, \bibinfo{person}{Bill Howe}, {and} \bibinfo{person}{Dan Suciu}.}
  \bibinfo{year}{2019}\natexlab{}.
\newblock \showarticletitle{Interventional Fairness: Causal Database Repair for
  Algorithmic Fairness}. In \bibinfo{booktitle}{\emph{Proceedings of the 2019
  International Conference on Management of Data}} (Amsterdam, Netherlands)
  \emph{(\bibinfo{series}{SIGMOD ’19})}. \bibinfo{publisher}{Association for
  Computing Machinery}, \bibinfo{address}{New York, NY, USA},
  \bibinfo{pages}{793–810}.
\newblock
\showISBNx{9781450356435}
\urldef\tempurl%
\url{https://doi.org/10.1145/3299869.3319901}
\showDOI{\tempurl}


\bibitem[\protect\citeauthoryear{Sellam, Lin, Huang, Yang, Vondrick, and
  Wu}{Sellam et~al\mbox{.}}{2019}]%
        {Sellam2018DeepBaseDI}
\bibfield{author}{\bibinfo{person}{Thibault Sellam}, \bibinfo{person}{Kevin
  Lin}, \bibinfo{person}{Ian Huang}, \bibinfo{person}{Michelle Yang},
  \bibinfo{person}{Carl Vondrick}, {and} \bibinfo{person}{Eugene Wu}.}
  \bibinfo{year}{2019}\natexlab{}.
\newblock \showarticletitle{DeepBase: Deep Inspection of Neural Networks}. In
  \bibinfo{booktitle}{\emph{Proceedings of the 2019 International Conference on
  Management of Data}} (Amsterdam, Netherlands) \emph{(\bibinfo{series}{SIGMOD
  ’19})}. \bibinfo{publisher}{Association for Computing Machinery},
  \bibinfo{address}{New York, NY, USA}, \bibinfo{pages}{1117–1134}.
\newblock
\showISBNx{9781450356435}
\urldef\tempurl%
\url{https://doi.org/10.1145/3299869.3300073}
\showDOI{\tempurl}


\bibitem[\protect\citeauthoryear{Shawi, Maher, and Sakr}{Shawi
  et~al\mbox{.}}{2019}]%
        {Shawi2019AutomatedML}
\bibfield{author}{\bibinfo{person}{Radwa~El Shawi}, \bibinfo{person}{Mohamed
  Maher}, {and} \bibinfo{person}{Sherif Sakr}.}
  \bibinfo{year}{2019}\natexlab{}.
\newblock \showarticletitle{Automated Machine Learning: State-of-The-Art and
  Open Challenges}.
\newblock \bibinfo{journal}{\emph{CoRR}} (\bibinfo{year}{2019}).
\newblock
\urldef\tempurl%
\url{http://arxiv.org/abs/1906.02287}
\showURL{%
\tempurl}


\bibitem[\protect\citeauthoryear{Shrikumar, Greenside, and Kundaje}{Shrikumar
  et~al\mbox{.}}{2017}]%
        {deeplift}
\bibfield{author}{\bibinfo{person}{Avanti Shrikumar}, \bibinfo{person}{Peyton
  Greenside}, {and} \bibinfo{person}{Anshul Kundaje}.}
  \bibinfo{year}{2017}\natexlab{}.
\newblock \showarticletitle{Learning Important Features Through Propagating
  Activation Differences}. In \bibinfo{booktitle}{\emph{Proceedings of the 34th
  International Conference on Machine Learning}}, Vol.~\bibinfo{volume}{70}.
  \bibinfo{publisher}{PMLR}, \bibinfo{address}{International Convention Centre,
  Sydney, Australia}, \bibinfo{pages}{3145--3153}.
\newblock
\urldef\tempurl%
\url{http://proceedings.mlr.press/v70/shrikumar17a.html}
\showURL{%
\tempurl}


\bibitem[\protect\citeauthoryear{Simonyan, Vedaldi, and Zisserman}{Simonyan
  et~al\mbox{.}}{2013}]%
        {Simonyan2013DeepIC}
\bibfield{author}{\bibinfo{person}{Karen Simonyan}, \bibinfo{person}{Andrea
  Vedaldi}, {and} \bibinfo{person}{Andrew Zisserman}.}
  \bibinfo{year}{2013}\natexlab{}.
\newblock \showarticletitle{Deep Inside Convolutional Networks: Visualising
  Image Classification Models and Saliency Maps}.
\newblock \bibinfo{journal}{\emph{CoRR}} (\bibinfo{year}{2013}).
\newblock
\urldef\tempurl%
\url{http://arxiv.org/abs/1312.6034}
\showURL{%
\tempurl}


\bibitem[\protect\citeauthoryear{{SQLFlow}}{{SQLFlow}}{2019}]%
        {sqlflow}
\bibfield{author}{\bibinfo{person}{{SQLFlow}}.}
  \bibinfo{year}{2019}\natexlab{}.
\newblock \bibinfo{title}{SQLFlow: Bridging Data and AI}.
\newblock \bibinfo{howpublished}{\url{https://sqlflow.org}}.
\newblock
\newblock
\shownote{[Online; accessed 10-October-2019].}


\bibitem[\protect\citeauthoryear{Sundararajan, Taly, and Yan}{Sundararajan
  et~al\mbox{.}}{2017}]%
        {integratedgradients}
\bibfield{author}{\bibinfo{person}{Mukund Sundararajan}, \bibinfo{person}{Ankur
  Taly}, {and} \bibinfo{person}{Qiqi Yan}.} \bibinfo{year}{2017}\natexlab{}.
\newblock \showarticletitle{Axiomatic Attribution for Deep Networks}. In
  \bibinfo{booktitle}{\emph{Proceedings of the 34th International Conference on
  Machine Learning}}, Vol.~\bibinfo{volume}{70}. \bibinfo{publisher}{PMLR},
  \bibinfo{address}{International Convention Centre, Sydney, Australia},
  \bibinfo{pages}{3319--3328}.
\newblock
\urldef\tempurl%
\url{http://proceedings.mlr.press/v70/sundararajan17a.html}
\showURL{%
\tempurl}


\bibitem[\protect\citeauthoryear{Tan, Zhang, Elmeleegy, and Srivastava}{Tan
  et~al\mbox{.}}{2017}]%
        {Tan2017ReverseEA}
\bibfield{author}{\bibinfo{person}{Wei~Chit Tan}, \bibinfo{person}{Meihui
  Zhang}, \bibinfo{person}{Hazem Elmeleegy}, {and} \bibinfo{person}{Divesh
  Srivastava}.} \bibinfo{year}{2017}\natexlab{}.
\newblock \showarticletitle{Reverse Engineering Aggregation Queries}.
\newblock \bibinfo{journal}{\emph{Proc. VLDB Endow.}} \bibinfo{volume}{10},
  \bibinfo{number}{11} (\bibinfo{date}{Aug.} \bibinfo{year}{2017}),
  \bibinfo{pages}{1394–1405}.
\newblock
\showISSN{2150-8097}
\urldef\tempurl%
\url{https://doi.org/10.14778/3137628.3137648}
\showDOI{\tempurl}


\bibitem[\protect\citeauthoryear{Tanaka, Ikami, Yamasaki, and Aizawa}{Tanaka
  et~al\mbox{.}}{2018}]%
        {DBLP:conf/cvpr/TanakaIYA18}
\bibfield{author}{\bibinfo{person}{Daiki Tanaka}, \bibinfo{person}{Daiki
  Ikami}, \bibinfo{person}{Toshihiko Yamasaki}, {and} \bibinfo{person}{Kiyoharu
  Aizawa}.} \bibinfo{year}{2018}\natexlab{}.
\newblock \showarticletitle{Joint Optimization Framework for Learning With
  Noisy Labels}. In \bibinfo{booktitle}{\emph{2018 {IEEE} Conference on
  Computer Vision and Pattern Recognition, {CVPR} 2018, Salt Lake City, UT,
  USA, June 18-22, 2018}}. \bibinfo{publisher}{{IEEE} Computer Society},
  \bibinfo{pages}{5552--5560}.
\newblock
\urldef\tempurl%
\url{https://doi.org/10.1109/CVPR.2018.00582}
\showDOI{\tempurl}


\bibitem[\protect\citeauthoryear{Tran and Chan}{Tran and Chan}{2010}]%
        {Tran2010HowTC}
\bibfield{author}{\bibinfo{person}{Quoc~Trung Tran} {and}
  \bibinfo{person}{Chee-Yong Chan}.} \bibinfo{year}{2010}\natexlab{}.
\newblock \showarticletitle{How to ConQueR Why-Not Questions}. In
  \bibinfo{booktitle}{\emph{Proceedings of the 2010 ACM SIGMOD International
  Conference on Management of Data}} (Indianapolis, Indiana, USA)
  \emph{(\bibinfo{series}{SIGMOD ’10})}. \bibinfo{publisher}{Association for
  Computing Machinery}, \bibinfo{address}{New York, NY, USA},
  \bibinfo{pages}{15–26}.
\newblock
\showISBNx{9781450300322}
\urldef\tempurl%
\url{https://doi.org/10.1145/1807167.1807172}
\showDOI{\tempurl}


\bibitem[\protect\citeauthoryear{Varma, Hancock, Suri, Dunnman, and Ré}{Varma
  et~al\mbox{.}}{2018}]%
        {dawnblog}
\bibfield{author}{\bibinfo{person}{Paroma Varma}, \bibinfo{person}{Braden
  Hancock}, \bibinfo{person}{Sahaana Suri}, \bibinfo{person}{Jared Dunnman},
  {and} \bibinfo{person}{Chris Ré}.} \bibinfo{year}{2018}\natexlab{}.
\newblock \bibinfo{title}{Debugging Training Data for Software 2.0}.
\newblock
  \bibinfo{howpublished}{\url{https://dawn.cs.stanford.edu/2018/08/30/debugging2/}}.
\newblock
\newblock
\shownote{[Online; accessed 10-October-2019].}


\bibitem[\protect\citeauthoryear{Wang, Dong, and Meliou}{Wang
  et~al\mbox{.}}{2015}]%
        {Wang2015DataXA}
\bibfield{author}{\bibinfo{person}{Xiaolan Wang}, \bibinfo{person}{Xin~Luna
  Dong}, {and} \bibinfo{person}{Alexandra Meliou}.}
  \bibinfo{year}{2015}\natexlab{}.
\newblock \showarticletitle{Data X-Ray: A Diagnostic Tool for Data Errors}. In
  \bibinfo{booktitle}{\emph{Proceedings of the 2015 ACM SIGMOD International
  Conference on Management of Data}} (Melbourne, Victoria, Australia)
  \emph{(\bibinfo{series}{SIGMOD ’15})}. \bibinfo{publisher}{Association for
  Computing Machinery}, \bibinfo{address}{New York, NY, USA},
  \bibinfo{pages}{1231–1245}.
\newblock
\showISBNx{9781450327589}
\urldef\tempurl%
\url{https://doi.org/10.1145/2723372.2750549}
\showDOI{\tempurl}


\bibitem[\protect\citeauthoryear{Wang, Meliou, and Wu}{Wang
  et~al\mbox{.}}{2017}]%
        {qfix}
\bibfield{author}{\bibinfo{person}{Xiaolan Wang}, \bibinfo{person}{Alexandra
  Meliou}, {and} \bibinfo{person}{Eugene Wu}.} \bibinfo{year}{2017}\natexlab{}.
\newblock \showarticletitle{QFix: Diagnosing Errors through Query Histories}.
  In \bibinfo{booktitle}{\emph{Proceedings of the 2017 ACM International
  Conference on Management of Data}} (Chicago, Illinois, USA)
  \emph{(\bibinfo{series}{SIGMOD ’17})}. \bibinfo{publisher}{Association for
  Computing Machinery}, \bibinfo{address}{New York, NY, USA},
  \bibinfo{pages}{1369–1384}.
\newblock
\showISBNx{9781450341974}
\urldef\tempurl%
\url{https://doi.org/10.1145/3035918.3035925}
\showDOI{\tempurl}


\bibitem[\protect\citeauthoryear{Wu and Madden}{Wu and Madden}{2013}]%
        {scorpion}
\bibfield{author}{\bibinfo{person}{Eugene Wu} {and} \bibinfo{person}{Samuel
  Madden}.} \bibinfo{year}{2013}\natexlab{}.
\newblock \showarticletitle{Scorpion: Explaining Away Outliers in Aggregate
  Queries}.
\newblock \bibinfo{journal}{\emph{Proc. VLDB Endow.}} \bibinfo{volume}{6},
  \bibinfo{number}{8} (\bibinfo{date}{June} \bibinfo{year}{2013}),
  \bibinfo{pages}{553–564}.
\newblock
\showISSN{2150-8097}
\urldef\tempurl%
\url{https://doi.org/10.14778/2536354.2536356}
\showDOI{\tempurl}


\bibitem[\protect\citeauthoryear{Wu, Flokas, Wu, and Wang}{Wu
  et~al\mbox{.}}{2020}]%
        {Rain-Full}
\bibfield{author}{\bibinfo{person}{Weiyuan Wu}, \bibinfo{person}{Lampros
  Flokas}, \bibinfo{person}{Eugene Wu}, {and} \bibinfo{person}{Jiannan Wang}.}
  \bibinfo{year}{2020}\natexlab{}.
\newblock \showarticletitle{Complaint-driven Training Data Debugging for Query
  2.0}.
\newblock \bibinfo{journal}{\emph{CoRR}} (\bibinfo{year}{2020}).
\newblock
\urldef\tempurl%
\url{https://sfu.ca/~youngw/files/Rain-arXiv.pdf}
\showURL{%
\tempurl}


\bibitem[\protect\citeauthoryear{Xu, Zhang, Friedman, Liang, and Van~den
  Broeck}{Xu et~al\mbox{.}}{2018}]%
        {semanticloss}
\bibfield{author}{\bibinfo{person}{Jingyi Xu}, \bibinfo{person}{Zilu Zhang},
  \bibinfo{person}{Tal Friedman}, \bibinfo{person}{Yitao Liang}, {and}
  \bibinfo{person}{Guy Van~den Broeck}.} \bibinfo{year}{2018}\natexlab{}.
\newblock \showarticletitle{A Semantic Loss Function for Deep Learning with
  Symbolic Knowledge}. In \bibinfo{booktitle}{\emph{Proceedings of the 35th
  International Conference on Machine Learning}}, Vol.~\bibinfo{volume}{80}.
  \bibinfo{publisher}{PMLR}, \bibinfo{address}{Stockholmsmässan, Stockholm
  Sweden}, \bibinfo{pages}{5502--5511}.
\newblock
\urldef\tempurl%
\url{http://proceedings.mlr.press/v80/xu18h.html}
\showURL{%
\tempurl}


\bibitem[\protect\citeauthoryear{Zeiler and Fergus}{Zeiler and Fergus}{2014}]%
        {Zeiler2013VisualizingAU}
\bibfield{author}{\bibinfo{person}{Matthew~D. Zeiler} {and}
  \bibinfo{person}{Rob Fergus}.} \bibinfo{year}{2014}\natexlab{}.
\newblock \showarticletitle{Visualizing and Understanding Convolutional
  Networks}. In \bibinfo{booktitle}{\emph{Computer Vision - {ECCV} 2014 - 13th
  European Conference, Zurich, Switzerland, September 6-12, 2014, Proceedings,
  Part {I}}}, Vol.~\bibinfo{volume}{8689}. \bibinfo{publisher}{Springer},
  \bibinfo{pages}{818--833}.
\newblock
\urldef\tempurl%
\url{https://doi.org/10.1007/978-3-319-10590-1\_53}
\showDOI{\tempurl}


\bibitem[\protect\citeauthoryear{Zeng, Liu, Chen, and Zhao}{Zeng
  et~al\mbox{.}}{2015}]%
        {Zeng2015DistantSF}
\bibfield{author}{\bibinfo{person}{Daojian Zeng}, \bibinfo{person}{Kang Liu},
  \bibinfo{person}{Yubo Chen}, {and} \bibinfo{person}{Jun Zhao}.}
  \bibinfo{year}{2015}\natexlab{}.
\newblock \showarticletitle{Distant Supervision for Relation Extraction via
  Piecewise Convolutional Neural Networks}. In
  \bibinfo{booktitle}{\emph{Proceedings of the 2015 Conference on Empirical
  Methods in Natural Language Processing, {EMNLP} 2015, Lisbon, Portugal,
  September 17-21, 2015}}. \bibinfo{publisher}{The Association for
  Computational Linguistics}, \bibinfo{pages}{1753--1762}.
\newblock
\urldef\tempurl%
\url{https://doi.org/10.18653/v1/d15-1203}
\showDOI{\tempurl}


\bibitem[\protect\citeauthoryear{Zhang, Zhu, and Wright}{Zhang
  et~al\mbox{.}}{2018}]%
        {duti}
\bibfield{author}{\bibinfo{person}{Xuezhou Zhang}, \bibinfo{person}{Xiaojin
  Zhu}, {and} \bibinfo{person}{Stephen~J. Wright}.}
  \bibinfo{year}{2018}\natexlab{}.
\newblock \showarticletitle{Training Set Debugging Using Trusted Items}. In
  \bibinfo{booktitle}{\emph{Proceedings of the Thirty-Second {AAAI} Conference
  on Artificial Intelligence}}. \bibinfo{publisher}{{AAAI} Press},
  \bibinfo{pages}{4482--4489}.
\newblock
\urldef\tempurl%
\url{https://www.aaai.org/ocs/index.php/AAAI/AAAI18/paper/view/16155}
\showURL{%
\tempurl}


\end{thebibliography}

\clearpage
\appendix
\let\oldcref\cref
\section{Ambiguity \& \naive}
In this section, we will describe a setting where \naive is unlikely to identify the correct training errors because the complaint is ambiguous. Specifically, we will formally prove that very few solutions to the SQL step of \naive can lead to a training error discovery in the influence step.

For the needs of our setting, we will focus on debugging a binary logistic regression model $M$. Its training set $T$ consists of data that is drawn from a clean distribution as well as a single noisily labeled example $\pmb{t}$. Let $l'\in\{0,1\}$ be the noisy label of $\pmb{t}$. For convenience, let the feature vector of $\pmb{t}$ be orthogonal to the other records in $T$, i.e. its inner product with all other feature vectors is zero. Symmetrically, the \qrecs distribution contains $n$ records with all but $m$ being orthogonal to $\pmb{t}$, just like the clean distribution. The remaining $m$ records can be arbitrary.   

Let the query $Q$ count the number of records in the \qdats where $M.predict(r)=1-l'$, the opposite of the $\pmb{t}$'s incorrect label. The user complaint is that the query result should be $k$ when the current result is $0$. We use \naive with an influence analysis step based on \cite{influence}.
\begin{theorem}\label{thm:ambiguity}
Assuming that the ILP solver picks uniformly at random from the satisfying solution space, then for fixed $m,k$ the probability that \naive assigns $\pmb{t}$ a non-zero score in the influence analysis step converges to $0$ as $n \to \infty$.
\end{theorem}

The intuition of the proof is straightforward. Given the orthogonality condition, the predictions of the $n-m$ \qrecs coming from the clean distribution would be the same regardless if $\pmb{t}$ existed or not. Unless the ILP assignment picks at least one of the remaining $m$ records, influence analysis at the second step will always assign a zero score for $\pmb{t}$. As $n$ increases with $k$ and $m$ fixed the probability of picking even one of the $m$ records decreases to 0.

For each of the ILP solutions that do not favor the recovery of $\pmb{t}$, we can always construct clean \trecs with positive scores that are ranked above $\pmb{t}$. Injecting as many as we want for each ILP solution, we can guarantee that \naive ranks $\pmb{t}$ arbitrarily low. We will conclude this section with the proof of the main theorem.
\begin{proof}
Let $\pmb{\theta}$ be the parameters of the logistic regression problem. We can write
\begin{equation*}
   \pmb{\theta} = \pmb{\theta}_{\textrm{noise}} + \pmb{\theta}_{\textrm{clean}}
\end{equation*}
where $\pmb{\theta}_{\textrm{noise}}$ is the projection of $\pmb{\theta}$ on the direction of the feature vector of $\pmb{t}$ and the second term is the orthogonal residue. We will call $\pmb{v}_{\textrm{noise}}$ the feature vector of $\pmb{t}$ and $\pmb{v}_i$ and $y_i$ the feature vectors and labels of the clean data. Let $\ell$ be the sample loss function of $M$ 
\begin{equation*}
    L(\pmb{\theta}) = \sum_{i=1}^{T_c}\ell(\pmb{v}_i, y_i, \pmb{\theta}) + \ell(\pmb{v}_{\textrm{noise}}, l', \pmb{\theta}) + \lambda \norm{\pmb{\theta}}^2
\end{equation*}
where the first term corresponds to the loss of the clean data and the second term corresponds to the loss of $\pmb{t}$. $\ell$ takes the feature vector and projects it to $\pmb{\theta}$ and then applies the sigmoid function and then the log loss. Let $f$ denote the function implementing the steps after the projection
\begin{align*}
    \ell(\pmb{\theta}, \pmb{v}_i, y_i) &= f(\pmb{\theta} \cdot \pmb{v}_i, y_i) = f(\pmb{\theta}_{\textrm{clean}} \cdot \pmb{v}_i, y_i)\\
    \ell(\pmb{\theta}, \pmb{v}_{\textrm{noise}}, l') &= f(\pmb{\theta} \cdot \pmb{v}_{\textrm{noise}}, l') = f(\pmb{\theta}_{\textrm{noise}} \cdot \pmb{v}_{\textrm{noise}}, l')
\end{align*}
That is the clean distribution loss depends only on $\pmb{\theta}_{\textrm{clean}}$ and the loss on $\pmb{t}$ depends only on $\pmb{\theta}_{\textrm{noise}}$.
Thus we essentially have two loss functions that depend on disjoint variables
\begin{align*}
    L(\pmb{\theta}) &= \sum_{i=1}^{T_c}f(\pmb{\theta}_{\textrm{clean}} \cdot \pmb{v}_i, y_i)+ \lambda \norm{\pmb{\theta}_{\textrm{clean}}}^2\\
    &+ f(\pmb{\theta}_{\textrm{noise}} \cdot \pmb{v}_{\textrm{noise}}, l') + \lambda \norm{\pmb{\theta}_{\textrm{noise}}}^2\\
    &= L_1(\pmb{\theta}_{\textrm{clean}}) + L_2(\pmb{\theta}_{\textrm{noise}})
\end{align*}
Essentially we have two distinct optimization problems
\begin{align*}
    \pmb{\theta}_{\textrm{clean}}^* &= \argmin_{\pmb{\theta}_{\textrm{clean}}} L_1(\pmb{\theta}_{\textrm{clean}})\\
    \pmb{\theta}_{\textrm{noise}}^* &= \argmin_{\pmb{\theta}_{\textrm{noise}}} L_2(\pmb{\theta}_{\textrm{noise}})
\end{align*}

Observe that the existence of $\pmb{t}$ does not affect the value of $\pmb{\theta}_{\textrm{clean}}^*$. Additionally, predictions on \qrecs that have feature vectors that are orthogonal to $\pmb{t}$ depend only on $\pmb{\theta}_{\textrm{clean}}^*$. Thus complaints on these \qrecs cannot be resolved by deleting $\pmb{t}$. For these complaints, $\pmb{t}$ would be assigned a zero score by the influence step of \naive. 

The feature vectors of the $m$ \qrecs are the only ones that could have a non-zero inner product with $\pmb{t}$. Thus, out of all the satisfying solutions of the ILP, only ones that assign the label of $1-l'$ to at least one of the $m$ \qrecs has any hope of giving $\pmb{t}$ a non-zero score in the influence analysis step. Observe that there are $\binom{n}{k}$ solutions to the ILP and there are $\binom{n-m}{k}$ assignments that do not pick any of the $m$ points. The probability of picking such an assignment is converging to 1.
\begin{align*}
   \lim_{n \to \infty} \frac{\binom{n-m}{k}}{\binom{n}{k}} &= \lim_{n \to \infty} \frac{(n-m)!(n-k)! }{(n-m-k)!n!}\\
   &= \lim_{n \to \infty} \prod_{i=0}^{k-1} \frac{n - i - m }{n + i} = 1
\end{align*}
The probability of assigning $\pmb{t}$ a non-zero score goes to 0.
\end{proof}

\section{\holistic Relaxation Examples}
In this section, we will provide additional examples of how \holistic handles multi-class classification models and aggregate comparisons in SQL queries.

\stitle{Multi-class models} Relaxing SQL queries that use multi-class classification models is also supported. As an example, let us consider the MNIST dataset that we describe in \cref{sec:exp}. We can design a classifier that takes one image, represented by a $28 \times 28$ grid of pixels, and yields a number from $0$ to $9$ corresponding to the digit displayed. We may want to use this model in an optical character recognition application that takes a handwritten multi-digit number, segments it into small images each containing a single digit and uses the classifier to figure out the numerical value of the whole number. Let us assume that the segmentation has occurred and that we have the sequence of $N$ images stored in a table $\textsf{DIGITS}$ in an attribute called $\textsf{image}$. Along with each image we have a field $\textsf{position}$ indicating the digit position from the right. The numeric value of the number is represented in SQL by the following query
\begin{equation*}
{\footnotesize
    \textsf{SELECT SUM(POWER(10, position)*predict(image)) FROM DIGITS}.
}
\end{equation*}

Let $\pmb{\theta}$ be the parameters of our model and $p_{ij}(\pmb{\theta})$ be the probability assigned by the classifier that the image at position $i$ is digit $j$. Then the relaxation of the query output is the following quantity
\begin{equation*}
\sum_{i=1}^{N} 10^{i-1}\sum_{j=0}^9 j \cdot  p_{ij}(\pmb{\theta}) 
\end{equation*}

\stitle{Aggregate comparisons} SQL queries are allowed to use comparisons in their selection and join predicates. Unfortunately, the relaxation rules for \holistic as described in \cref{sec:solution} do not directly support comparison operators. Regardless of its complexity, every comparison has an equivalent logical formula involving only $\texttt{AND}, \texttt{OR}, \texttt{NOT}$ that \holistic supports. Finding such a logical formula can be non-trivial when aggregate values are  compared. SQL can express aggregate comparisons through a $\textsf{HAVING}$ clause. 

For example, let us revisit the optical character recognition application of the previous subsection. We want to express a predicate selecting numbers that are greater or equal than $95$. Let $x_{i,j}$ be the boolean value expressing that the digit at position $i$ from the right is classified as being $j$. For simplicity we focus on the case of two-digit numbers. Then the aggregate comparison is equivalent to the following formula
\begin{equation*}
    x_{2,9} \texttt{ AND } \left(\bigvee_{9 \geq \ell \geq 5}  x_{1,\ell} \right)
\end{equation*}

In general, finding the equivalent formula can be a computationally expensive procedure and the resulting formula can have a large amount of terms slowing down the influence analysis step. Identifying ways to relax comparisons directly is a promising direction for future work. It is important to note that comparisons that are part of the complaint itself do not require special care. They can be handled directly based on the techniques described in \cref{sec:solution}.
\section{The value of complaints}
In this section, we will describe a setting where ordering \trecs based on loss or loss sensitivity ranks training corruptions at the bottom. At the same time, an appropriately selected complaint is sufficient to rank all corrupted \trecs at the top.

For the needs of our setting, we will focus on debugging a binary logistic regression model $M$. Our training set $T$ is a mixture of clean and corrupted \trecs. Clean records have been perfectly labelled whereas the corrupted ones have had theirs inverted. Corrupted \trecs labelled as being in class $1$ are truly in class $0$ and vice versa. 

For simplicity, we are going the two following assumptions. First, the feature vectors of the clean \trecs are all orthogonal to the ones in the corrupted distribution. That is for each pair of records from the two distributions, the corresponding feature vectors have zero inner product. Second, the feature vectors for all corrupted training records are parallel. That is for each pair of feature vectors $\pmb{v}_i$ and $\pmb{v}_j$ from the corrupted \trecs, there is a $\kappa_{ij} \in \mathbb{R}$ such that $\pmb{v}_i = \kappa_{ij} \pmb{v}_j$. Third, we are going to assume that the corrupted \trecs are linearly separable, i.e. there is a linear classifier that can correctly specify the labels of the corrupted \trecs.

Let us discuss the two ways we can use the model loss to rank \trecs. The first one is to use the loss value of each \trec. The \loss baseline discussed in the experiments ranks \trecs with higher loss at the top. The second one ranks \trecs based on loss sensitivity. Specifically, \cite{influence} considers the effect of the removal of each \trec on its own loss. This is the \infloss baseline. For each \trec $\pmb{z}$ it computes a loss sensitivity
\begin{equation*}
    - \nabla_{\pmb{\theta}} \ell(\pmb{z}, \pmb{\theta}^*) H_{\pmb{\theta}^*}^{-1} \nabla_{\pmb{\theta}} \ell(\pmb{z}, \pmb{\theta}^*).
\end{equation*}
These scores are negative or zero since the Hessian of logistic regression, and thus its inverse, is positive definite. Large negative values indicate that when the training record is removed, its own loss tends to increase rapidly. These \trecs are ranked at the top by \infloss. The following holds
\begin{theorem}
As the number of corrupted \trecs goes to infinity, the loss and loss sensitivity of corrupted \trecs goes to zero.
\end{theorem}
Observe that $0$ is the minimum value of the loss and maximum value of the loss sensitivity. Thus both approaches rank corrupted \trecs at the bottom. As a first step let us prove this theorem.
\begin{proof}
Let $\pmb{\theta}$ be the parameters of the logistic regression problem. We can once again write
\begin{equation*}
   \pmb{\theta} = \pmb{\theta}_{\textrm{noise}} + \pmb{\theta}_{\textrm{clean}}.
\end{equation*}
$\pmb{\theta}_{\textrm{noise}}$ is the projection of $\pmb{\theta}$ on the direction of the feature vector of a corrupted \trec. It does not matter which one since all are parallel. $\pmb{\theta}_{\textrm{clean}}$ is the orthogonal residue. 

We can apply the same techniques as in the proof of \oldcref{thm:ambiguity} to get two independent optimization problems. Let $K$ be the number of corrupted points, $\pmb{v}_i$ the feature vectors and $y_i$ the corresponding corrupted labels. Using the $f$ function from the proof of \oldcref{thm:ambiguity}
\begin{equation*}
    \pmb{\theta}_{\textrm{noise}}^* = \argmin_{\pmb{\theta}_{\textrm{noise}}} \left(\sum_{i=1}^K f(\pmb{v}_i \cdot \pmb{\theta}_{\textrm{noise}}, y_i) + \lambda \norm{\pmb{\theta}_{\textrm{noise}}}^2 \right) 
\end{equation*}
As a first step we want to prove that
\begin{equation*}
    \lim_{K \to \infty} f(\pmb{v}_i \cdot \pmb{\theta}_{\textrm{noise}}^*, y_i) = 0
\end{equation*}
for all pairs of $\pmb{v}_i$ and $y_i$ from the corrupted \trecs. This states that the loss of the corrupted records goes to 0, our first claim. Let $\sigma$ be the sigmoid function. By first order conditions we have that $\pmb{\theta}_{\textrm{noise}}^*$ satisfies
\begin{equation*}
    \sum_{i=1}^K \left(\sigma(\pmb{v}_i \cdot \pmb{\theta}_{\textrm{noise}}^*) -y_i\right) \pmb{v}_i + 2 \lambda \cdot \pmb{\theta}_{\textrm{noise}}^* = \pmb{0}.
\end{equation*}
We have assumed that the corrupted \trecs are linearly separable. Thus there exists a vector $\pmb{u}$ that linearly separates the data with margin $1$. Let us multiply with this vector the equation above.
\begin{equation*}
    \sum_{i=1}^K \left(\sigma(\pmb{v}_i \cdot \pmb{\theta}_{\textrm{noise}}^*) -y_i\right) \pmb{v}_i \cdot \pmb{u} + 2  \lambda \cdot\pmb{\theta}_{\textrm{noise}}^* \cdot \pmb{u} = 0.
\end{equation*}
In turn we have that
\begin{equation*}
    \sum_{i=1}^K \frac{\left(\sigma(\pmb{v}_i \cdot \pmb{\theta}_{\textrm{noise}}^*) -y_i\right) \pmb{v}_i \cdot \pmb{u}}{\pmb{\theta}_{\textrm{noise}}^* \cdot \pmb{u} } = - 2  \lambda.
\end{equation*}
Given that $\pmb{u}$ has margin $1$, we have that
\begin{equation*}
    (2y_i-1) \cdot \pmb{v}_i \cdot \pmb{u} \geq 1
\end{equation*}
and thus for both $y_i$ equal to $0$ and $1$, all summation terms of the previous equation need to have the same sign. As the number of terms $K$ increases, the terms cannot be bounded away from $0$ while the sum remains finite.  Thus $\pmb{\theta}_{\textrm{noise}}^*$ is such that either the numerator goes to zero or the denominator goes to infinity or both. In either case 
\begin{equation*}
    \lim_{K \to \infty} \pmb{\theta}_{\textrm{noise}}^* \cdot \pmb{u} = \infty
\end{equation*}
Given that all $\pmb{v}_i$ are parallel to $\pmb{u}$ and $\pmb{\theta}_{\textrm{noise}}^*$, the loss of the samples going to $0$ follows immediately. 

Similarly, the gradients of the losses go to a $0$ norm. For the loss sensitivity scores, we have
\begin{equation*}
    -\nabla_{\pmb{\theta}} \ell(\pmb{z}, \pmb{\theta}^*) H_{\pmb{\theta}^*}^{-1} \nabla_{\pmb{\theta}} \ell(\pmb{z}, \pmb{\theta}^*) \geq - \norm{\nabla_{\pmb{\theta}} \ell(\pmb{z}, \pmb{\theta}^*)}^2 \lambda_{\textrm{max}}(H_{\pmb{\theta}^*}^{-1})
\end{equation*}
where $\lambda_{\textrm{max}}(H_{\pmb{\theta}^*}^{-1})$ the biggest eigenvalue of the inverse Hessian. The minimum eigenvalue of the Hessian is at least $2\lambda$. Thus the inverse Hessian has bounded eigenvalues by $\frac{1}{2\lambda}$. 
\begin{equation*}
    - \nabla_{\pmb{\theta}} \ell(\pmb{z}, \pmb{\theta}^*) H_{\pmb{\theta}^*}^{-1} \nabla_{\pmb{\theta}} \ell(\pmb{z}, \pmb{\theta}^*) \geq - \norm{\nabla_{\pmb{\theta}} \ell(\pmb{z}, \pmb{\theta}^*)}^2 \frac{1}{2\lambda}
\end{equation*}
With the gradient norms going to $0$, the loss sensitivity scores of all corrupted \trecs need to going to $0$ as well.
\end{proof}

Thus for a large enough number of corrupted \trecs and the clean ones fixed, we can force the corrupted \trecs to the bottom of the rankings for both \loss and \infloss. What remains to discuss is how an appropriate complaint can bring these records to the top of the ranks. 

Complaints on \qrecs with feature vectors parallel to the ones of the corrupted \trecs are particularly interesting. For these complaints, the clean \trecs receive $0$ influence scores following the same discussion as in \oldcref{thm:ambiguity}. It thus remains to find one such complaint that assigns positive scores to all corrupted \trecs. 

Even identifying one of the mispredicted \qrecs that have the property above will do. Let $\pmb{z}_q = (\pmb{v}_q, y_q)$ be the identified record with its correct label. The influence score of each corrupted \trec  $\pmb{z}_i = (\pmb{v}_i, y_i)$ is
\begin{equation*}
     - \nabla_{\pmb{\theta}} \ell(\pmb{z}_q, \pmb{\theta}^*) H_{\pmb{\theta}^*}^{-1} \nabla_{\pmb{\theta}} \ell(\pmb{z}_i, \pmb{\theta}^*).
\end{equation*}
We have that
\begin{align*}
    \nabla_{\pmb{\theta}} \ell(\pmb{z}_q, \pmb{\theta}^*) &= \left(\sigma(\pmb{v}_q \cdot \pmb{\theta}_{\textrm{noise}}^*) -y_q\right) \pmb{v}_q\\
    \nabla_{\pmb{\theta}} \ell(\pmb{z}_i, \pmb{\theta}^*) &= \left(\sigma(\pmb{v}_i \cdot \pmb{\theta}_{\textrm{noise}}^*) -y_i\right) \pmb{v}_i.
\end{align*}
$\pmb{v}_q$ and $\pmb{v}_i$ are parallel but $\pmb{z}_q$ is mispredicted while $\pmb{z}_i$ is correctly predicted. Simple algebra shows that the gradients have opposite directions. That is there is a $k>0$ such that
\begin{align*}
   \nabla_{\pmb{\theta}} \ell(\pmb{z}_q, \pmb{\theta}^*) = -k  \nabla_{\pmb{\theta}} \ell(\pmb{z}_i, \pmb{\theta}^*)
\end{align*}
Since the Hessian of logistic regression and thus its inverse is positive definite, we have
\begin{align*}
     -& \nabla_{\pmb{\theta}} \ell(\pmb{z}_q, \pmb{\theta}^*) H_{\pmb{\theta}^*}^{-1} \nabla_{\pmb{\theta}} \ell(\pmb{z}_i, \pmb{\theta}^*) =\\
     &\quad \quad k \nabla_{\pmb{\theta}} \ell(\pmb{z}_i, \pmb{\theta}^*) H_{\pmb{\theta}^*}^{-1} \nabla_{\pmb{\theta}} \ell(\pmb{z}_i, \pmb{\theta}^*) >0.
\end{align*}
Thus the influence scores of all corrupted \trecs are positive and the complaint ranks all of them at the top.
\section{Debugging on NN}
In this section, we evaluate \naive, \loss, and \holistic on a convolutional neural network (CNN) model.  We execute the COUNT query $Q_5$ on the  MNIST dataset, and we corrupt the training set by flipping $50\%$ of the $1$ digit images to be labeled $7$. The CNN model consists of 3-layers (convolution, max pooling, dense with RELU activation). \cite{influence} showed empirically that the influence function analysis works even for neural networks, which are non-convex, including CNN architectures. We also include the logistic regression model for comparison.  

This section reports the \auc for the three approaches.  We find that \holistic degrades slightly when debugging the CNN model. This agrees with the findings of \cite{influence}, where the influence analysis scores were shown to be less accurate for non-convex than convex models. Recent work on influence analysis \cite{sgdcleansing} has provided improved approaches that are more accurate on non-convex models. \sys is compatible with \cite{sgdcleansing} and can continue to leverage any improvements on influence analysis by the ML community.  

\begin{figure}[htpb]
    \centering
    \includegraphics[width=.8\columnwidth]{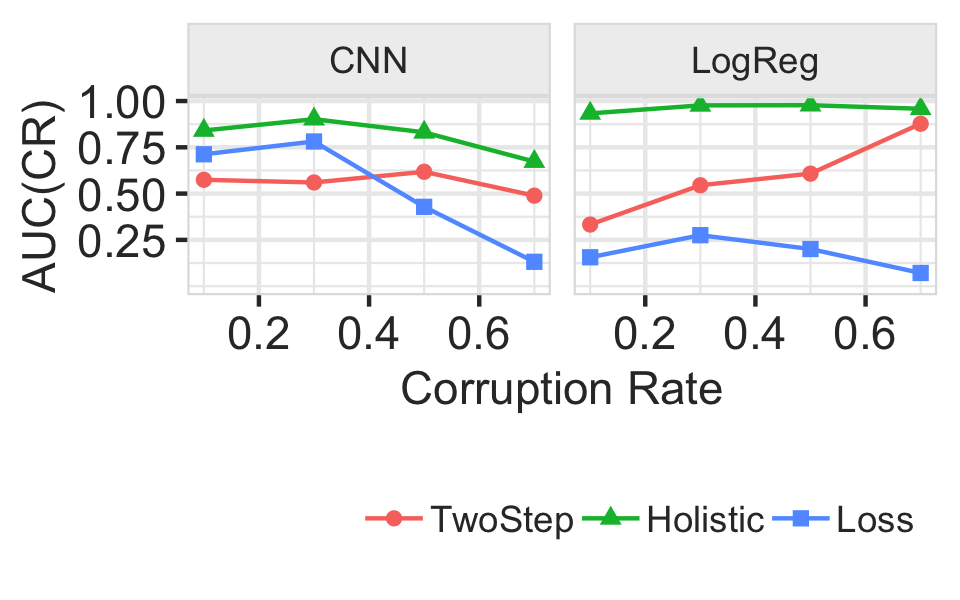}
    \caption{\auc when using CNN and logistic regression models.}
    \label{fig:mnist-logreg-nn}
\end{figure}

\begin{figure}[!htbp]
    \centering
    \includegraphics[width=\columnwidth]{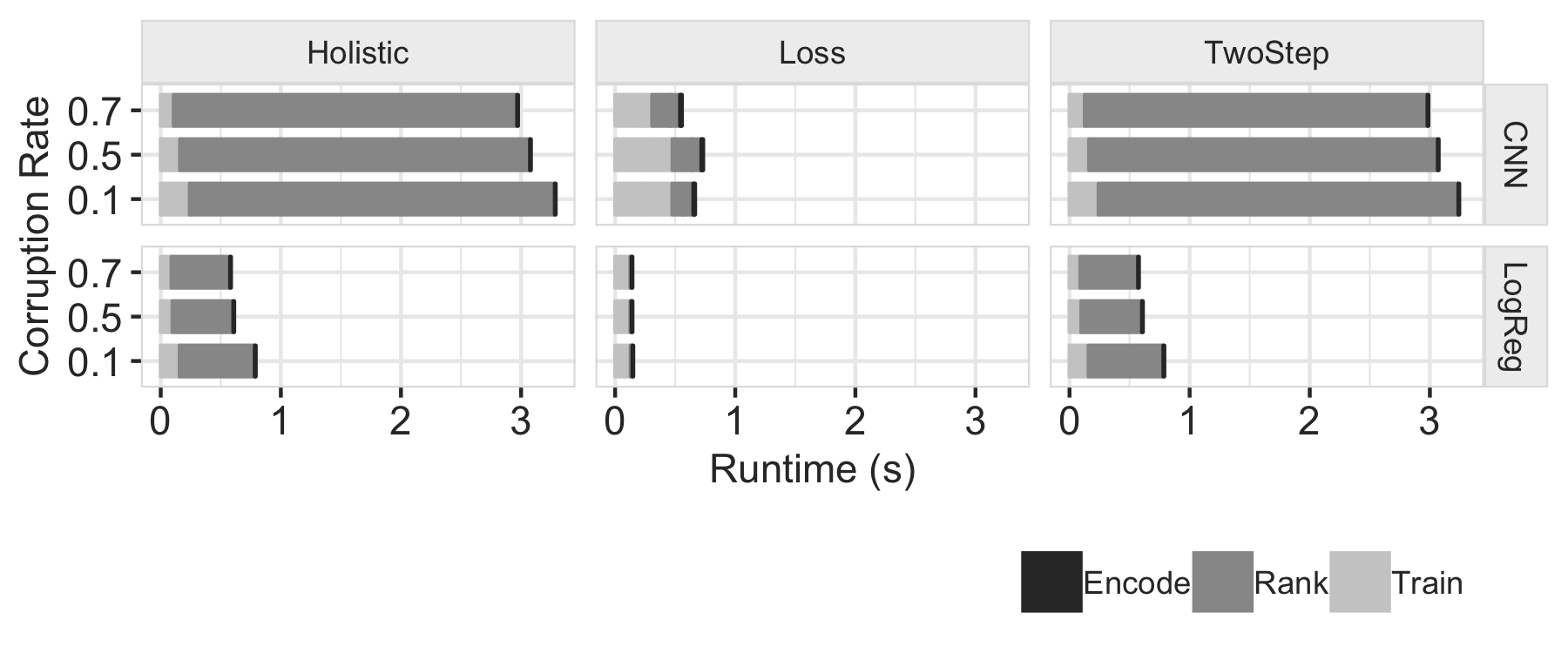} 
    \caption{Per-iteration runtimes for debugging CNN and logistic regression models for different corruption rates. }
    \label{fig:mnist-logreg-nn-runtime}
\end{figure}

\Cref{fig:mnist-logreg-nn-runtime} shows that per-iteration runtimes for \loss is dominated by retraining costs. We note that the models are trained incrementally in each iteration i.e. the previous values of the weights are used as initializations for the next debugging iteration. Thus, retraining costs can vary across methods depending on which points are removed. Intuitively removing points with high loss may result in significant model changes and thus can lead to higher retraining costs, explaining the higher retraining time for \loss in \Cref{fig:mnist-logreg-nn-runtime} when compared to \naive and \holistic.

In contrast,  \naive and \holistic are dominated by calculating the Hessian vector products required by the conjugate gradient approach of \cite{hessianfree}. Even if the conjugate gradient approach is much faster than naively computing the inverse Hessian, its cost grows linearly with the number of parameters of the model. \cite{influence} suggested doing the influence analysis on neural networks by considering only the weights of the last layer as parameters, treating all the previous layer as a fixed feature transformation. We leave studying the effects of this optimization on debugging runtimes as future work.

{\it\ititle{Takeaways:}
    \holistic supports queries that use neural network models.  It performs well under low and moderate corruption rates and dominates \naive and \loss.  However, each iteration takes $\approx 3$ seconds due to the cost of the ranking (hessian inverse) step, which is particularly costly for neural network models.  
}

\end{document}